\documentclass[a4paper,11pt]{article}
\pdfoutput=1 
\usepackage{jheppub}

\usepackage{subcaption}

\usepackage{amsfonts,mathtools}
\usepackage{physics} 
\usepackage{bbold} 
\usepackage{slashed} 
\usepackage{siunitx} 
\usepackage{pifont} 

\usepackage{enumitem} 
\usepackage[bottom]{footmisc}

\usepackage{hyperref}

\newcommand{\hc}{\text{h.c.}} 
\newcommand{\eq}{\text{eq}} 
\DeclareMathOperator{\diag}{diag} 

\arxivnumber{2510.24843}
\title{\boldmath Impact of dim-6 $\nu$SMEFT operators on low-scale leptogenesis}

\author[a,b]{Kaori Fuyuto,}
\author[c]{Julia Harz,}
\author[c,1]{and Sascha Weber\note{Corresponding author.}}

\affiliation[a]{KEK Theory Center, IPNS, KEK, Tsukuba 305–0801, Japan}
\affiliation[b]{Theoretical Division, Los Alamos National Laboratory, Los Alamos, NM 87545, USA}
\affiliation[c]{PRISMA$^{++}$ Cluster of Excellence \& Mainz Institute for Theoretical Physics, FB 08 - Physics, Mathematics and Computer Science, Johannes Gutenberg-Universität Mainz, Staudingerweg 9, 550099 Mainz, Germany}

\emailAdd{kfuyuto@post.kek.jp}
\emailAdd{julia.harz@uni-mainz.de}
\emailAdd{wesascha@uni-mainz.de}

\preprint{KEK-TH-2770, LA-UR-25-23538, MITP-25-068}
\abstract{We investigate the impact of higher-dimensional operators on low-scale leptogenesis (LG) via oscillations of right-handed neutrinos within the neutrino-extended Standard Model Effective Field Theory ($\nu$SMEFT) and discuss the connection to neutrinoless double beta decay ($0\nu\beta\beta$). 
Focusing on a dimension-six, lepton number conserving operator, we explore how new interactions can significantly alter the production and equilibration dynamics of right-handed neutrinos. We derive the relevant quantum kinetic equations incorporating both renormalizable and non-renormalizable interactions and perform a comprehensive numerical analysis for benchmark scenarios in both the oscillatory and overdamped regimes. 
Our results reveal that even in the absence of explicit lepton number violation by the operator, it can enhance or suppress the baryon asymmetry of the universe (BAU) by several orders of magnitude, depending on the EFT scale. 
We further connect these effects to predictions for $0\nu\beta\beta$ decay, demonstrating that the same operator can lead to enhanced decay rates, potentially within reach of the next generation of experiments. 
Our findings indicate that the observation of $0\nu\beta\beta$ could rule out a large part of the parameter space for successful low-scale LG within the $\nu$SMEFT, implying low RHN masses and low reheating temperatures. }


\begin{document}

\maketitle
\flushbottom

\section{Introduction}

The Standard Model (SM) successfully describes the known fundamental interactions, but clear evidence points to physics beyond it. Neutrino oscillations, for example, imply non-zero neutrino masses \cite{Esteban:2020cvm,deSalas:2020pgw}, which the SM cannot accommodate. A minimal and well-motivated extension introduces right-handed neutrinos (RHNs), allowing neutrinos to acquire Dirac masses. Being gauge singlets, RHNs naturally admit Majorana mass terms. The interplay between Dirac and Majorana masses leads to a set of Majorana mass eigenstates, with RHN-dominated states potentially much heavier than the active neutrinos. This setup realizes the type-I seesaw mechanism \cite{Minkowski:1977sc,Yanagida:1979as,Gell-Mann:1979vob,Mohapatra:1979ia,Schechter:1980gr}.\footnote{In this context, RHNs are also called sterile neutrinos (since they are gauge singlets) or heavy neutral leptons (since they are much heavier than the active neutrinos).}

Another indication of physics beyond the Standard Model (BSM) is the baryon asymmetry of the Universe (BAU). Observations of the cosmic microwave background (CMB) and predictions from Big Bang nucleosynthesis (BBN) point to a baryon-to-photon ratio of $\eta_B = 6.2 \times 10^{-10}$ \cite{Planck:2018vyg}, which the SM fails to explain. A well-established mechanism for generating the BAU is leptogenesis (LG) \cite{Fukugita:1986hr}, which can be naturally realized through the dynamics of RHNs. In standard LG scenarios, a lepton asymmetry is generated via SM lepton number violating (LNV) interactions involving RHNs and subsequently converted into a baryon asymmetry via SM $(B+L)$-violating sphaleron processes \cite{Kuzmin:1985mm}.

Depending on the RHN mass scale, different LG scenarios emerge: thermal LG \cite{Fukugita:1986hr,Davidson:2002qv} and resonant LG \cite{Liu:1993tg,Flanz:1994yx,Flanz:1996fb,Covi:1996wh,Covi:1996fm,Pilaftsis:1997jf,Buchmuller:1997yu,Pilaftsis:2003gt} operate via freeze-out dynamics, while LG via RHN oscillations -- also known as the Akhmedov-Rubakov-Smirnov (ARS) mechanism\footnote{This mechanism is also referred to as freeze-in LG or baryogenesis via RHN oscillation.} \cite{Akhmedov:1998qx,Asaka:2005pn} -- proceeds through freeze-in. Although these regimes differ in their dynamics and applicable temperature ranges, they are all continuously connected within the same underlying seesaw framework \cite{Klaric:2020phc,Klaric:2021cpi}.

Probing RHNs is difficult due to their gauge singlet nature and the seesaw relation, which implies either very heavy masses or suppressed Yukawa couplings. However, their characteristic LNV interactions offer a promising avenue: observables that are highly suppressed or absent in the SM may be significantly enhanced in the presence of RHNs. A prime example is neutrinoless double beta ($0\nu\beta\beta$) decay, a rare nuclear process that would signify both LNV and the Majorana nature of neutrinos. While essentially unobservable in the SM, $0\nu\beta\beta$ decay is predicted in a wide range of BSM scenarios involving RHNs. Experimental searches have been conducted using different isotopes e.g.~$^{136}$Xe in the KamLAND-Zen experiment \cite{KamLAND-Zen:2022tow, KamLAND-Zen:2024eml} and $^{76}$Ge in LEGEND \cite{LEGEND:2025jwu}. The most stringent bound from the isotope $^{136}$Xe currently comes from KamLAND-Zen, improving the half-life limit to $T^{0\nu}_{1/2}(^{136}{\rm Xe}) > 3.8 \times 10^{26}$ yr \cite{KamLAND-Zen:2024eml}, a factor of 1.7 stronger than its previous result \cite{KamLAND-Zen:2022tow}. The LEGEND collaboration recently announced its first half-life limit, $T^{0\nu}_{1/2} (^{76}{\rm Ge}) > 1.9 \times 10^{26}$ yr \cite{LEGEND:2025jwu}, the strongest constraint for the isotope $^{76}{\rm Ge}$. Future experiments are expected to significantly extend this reach. KamLAND2-Zen aims to probe lifetimes of order $10^{27}$ yr, while LEGEND is projected to reach sensitivities of $T^{0\nu}_{1/2} \sim 10^{28}$ yr. These milestones will allow exploration of key regions in parameter space, including the standard inverted neutrino mass hierarchy.

While RHNs provide an elegant explanation for both neutrino masses and the baryon asymmetry of the Universe, considering them in isolation is likely an idealization. In realistic ultraviolet (UV) completions, additional fields or interactions are often required, not only to generate RHN masses or explain flavor structures, but potentially also to address other open problems of the Standard Model. These extended models can alter the predictions for both LG and low-energy observables such as $0\nu\beta\beta$ decay. To study such effects in a systematic and model-independent way, it is useful to employ an effective field theory (EFT) approach. While for the SM, the Standard Model Effective Field Theory (SMEFT) \cite{Buchmuller:1985jz,Grzadkowski:2010es} provides a powerful framework, RHNs which are lighter than the cutoff scale, must be kept as explicit degrees of freedom -- leading to the so-called neutrino-extended SMEFT ($\nu$SMEFT) \cite{delAguila:2008ir,Bhattacharya:2015vja,Liao:2016qyd}. Examples of UV completions giving rise to additional operators in the $(\nu)$SMEFT are leptoquark models \cite{Dorsner:2016wpm}, left-right-symmetric models \cite{Pati:1974yy,Mohapatra:1974gc,Mohapatra:1980yp}, gauged baryon and lepton number models \cite{Langacker:1980js,Hewett:1988xc,Faraggi:1990ita}, and Grand Unified Theories \cite{Bando:1998ww}.

The effect of SMEFT operators on LG has been studied in \cite{Deppisch:2013jxa,Deppisch:2015yqa,Deppisch:2017ecm}, including implications on neutrino masses and $0\nu\beta\beta$ decay. While $0\nu\beta\beta$ decay in the $\nu$SMEFT has been studied extensively in \cite{Dekens:2020ttz}, a corresponding study of LG within the $\nu$SMEFT framework remains absent. Interestingly, Ref.~\cite{Dekens:2020ttz} found that a specific lepton number conserving (LNC) dimension-six operator in the $\nu$SMEFT can significantly enhance the $0\nu\beta\beta$ decay rate when RHNs have masses in the MeV-\si{\giga\electronvolt} range. Motivated by this result, we focus on this mass range, for which the dominant LG production occurs via freeze-in.

The standard freeze-in LG scenario has been established and refined in \cite{Akhmedov:1998qx,
Asaka:2005an,Asaka:2005pn,Shaposhnikov:2008pf,Canetti:2010aw,Canetti:2012kh,
Asaka:2011wq,
Hernandez:2015wna,Hernandez:2016kel,Hernandez:2022ivz,Sandner:2023tcg,
Anisimov:2010gy,Besak:2012qm,Ghiglieri:2017gjz,Ghiglieri:2018wbs,
Drewes:2012ma,Drewes:2016jae,Drewes:2016gmt,Antusch:2017pkq,Abada:2018oly,Abada:2017ieq,Klaric:2020phc,Klaric:2021cpi,
Shuve:2014zua,Hambye:2016sby,Hambye:2017elz,
Drewes:2017zyw} and various extensions have been explored, including specific ultraviolet (UV) completions \cite{Shuve:2014zua,Caputo:2018zky,Flood:2021qhq,Fischer:2021nha,Astros:2024yee} and effective field theory approaches with higher-dimensional operators \cite{Asaka:2017rdj}. However, the scenario we consider here differs in several important ways from these other extensions. First, previous studies have focused primarily on LNV interactions \cite{Caputo:2018zky,Flood:2021qhq,Fischer:2021nha,Astros:2024yee}, while we focus exclusively on a LNC dimension-six operator. Second, Refs.~\cite{Shuve:2014zua,Caputo:2018zky,Flood:2021qhq,Fischer:2021nha,Astros:2024yee} generally focused on additional light particles, while we are interested in heavy particles in this work (such that they can be captured in the EFT approach). Third, \cite{Asaka:2017rdj} examined only the impact of higher-dimensional operators on initial conditions of RHNs, while we study their influence on the full dynamical evolution of RHNs throughout the LG process.

In this work, we focus on a single LNC $\nu$SMEFT operator. As discussed in ~\cite{Dekens:2020ttz}, the operator can give a dominant contribution to $0\nu\beta\beta$ process. A key result of our work is that also for freeze-in LG the final BAU can be altered by several orders of magnitude depending on the EFT operator scale. The BAU can be suppressed, as might be expected for a freeze-in process with additional interactions, but somewhat surprisingly the BAU can also be enhanced, even though the additional operator is LNC. The suppression happens usually for small EFT scales where they can lead to observable signatures in $0\nu\beta\beta$ decay. We show how an observation of $0\nu\beta\beta$ decay in this case can essentially exclude large parts of the freeze-in parameter space. 
Simultaneously observing $0\nu\beta\beta$ decay and requiring successful freeze-in LG could point us towards small RHN masses and low reheating temperatures. Moreover, we show how the enhancement of the BAU can be used to alleviate some of the mass degeneracy for the RHNs, which is typical for freeze-in LG.

This paper is structured as follows. In Section~\ref{sec:nuSMEFT} we introduce the $\nu$SMEFT and the dimension-six LNC operator we consider in this work. Section~\ref{sec:0vbb} is devoted to the study of $0\nu\beta\beta$ decay. In Section~\ref{sec:LG} we set up the quantum kinetic equations (QKEs), necessary in the freeze-in LG framework, including higher-dimensional operators. We estimate how the solution qualitative behaves and solve the QKEs for the final BAU numerically for a few representative benchmark points. In Section~\ref{sec:results} we show how our results depend on the reheating temperature and connect to $0\nu\beta\beta$ decay experiments. We also show how to reduce the mass degeneracy between the RHNs before concluding in Section~\ref{sec:conclusion}.

\section{The \texorpdfstring{$\nu$SMEFT}{vSMEFT} Lagrangian}\label{sec:nuSMEFT}

The $\nu$SMEFT consists of the usual SM fields augmented by a set of RHNs with masses below the EFT scale(s). In this work we will focus on the addition of two RHNs $\nu_{R,I}, I \in\{1,2\}$.\footnote{While the formulas we present in this work are easily generalizable for three or more RHNs, the number of free parameters grows quickly. Here we stick to the two-RHN case, as is common in many LG via oscillation studies. Qualitatively, our results are also present in cases with three or more RHN flavors.} Including higher-dimensional operators the Lagrangian can be written as
\begin{align}\label{eq:lagrangian}
    \mathcal{L} &= \mathcal{L}_{\mathrm{SM}} + \overline{\nu_{R,I}} (i\slashed{\partial}) \nu_{R,I} + \left(- F_{\alpha I} \overline{L_{\alpha}} \Tilde{H} \nu_{R,I} - \frac{1}{2} M_{I} \overline{\nu_{R,I}^c}\nu_{R,I} + \hc \right) \notag \\
    &\quad + \mathcal{L}^{(5)} + \mathcal{L}_{\nu_{R}}^{(5)} + \mathcal{L}^{(6)} + \mathcal{L}_{\nu_{R}}^{(6)} + \mathcal{L}^{(7)} + \mathcal{L}_{\nu_{R}}^{(7)} + \dots \, . 
\end{align}
Here, $\mathcal{L}_{\mathrm{SM}}$ is the SM Lagrangian, $F_{\alpha I}$ is the Yukawa coupling between the SM lepton doublet $L_{\alpha}$ of flavor $\alpha \in \{e,\mu,\tau\}$ and the RHNs $\nu_{R,I}$, $\Tilde{H}=i\sigma_2H^{*}$ is the conjugate Higgs doublet with a vacuum expectation value $v = \expval{H^0} = \SI{174}{\giga\electronvolt}$ and the superscript $c$ denotes charge conjugation\footnote{We define charge conjugation for a Dirac or Majorana field via $\Psi^c = C \overline{\Psi}^T$ with the usual charge conjugation Matrix $C=i\gamma^0\gamma^2$ and use the convention $\Psi_{L,R}^c=(\Psi_{L,R})^c$ throughout.}. We use $M_I$ for the Majorana masses of the RHNs $\nu_R$.

In Eq.~\eqref{eq:lagrangian} the $\mathcal{L}^{(n)}$ include all Lorentz and gauge invariant operators with mass dimension dim-$n$ involving only SM fields and $\mathcal{L}_{\nu_{R}}^{(n)}$ include the remaining operators with mass dimension dim-$n$ involving at least one RHN $\nu_R$. Lists of possible operators in the SMEFT and $\nu$SMEFT can be found in~\cite{Grzadkowski:2010es,Lehman:2014jma,Liao:2016hru,Liao:2016qyd,Cirigliano:2018yza,Dekens:2020ttz}.

\subsection{Neutrino parameters}

After electroweak symmetry breaking (EWSB) the mass terms of the neutrino sector are generally given by
\begin{align}
    {\cal L}_m=-\frac{1}{2}\overline{N^c}M_{\nu}N
+\hc,\hspace{1cm}
M_{\nu}=
\begin{pmatrix}
    M_L & M_D^*\\
    M_D^{\dagger} & M_R^{\dagger}
\end{pmatrix},
\end{align}
where $N=(\nu_{L\alpha},\nu_{RI}^c)^T$ and $M_L,M_D$ and $M_R$ can receive contributions from the renormalizable interactions, i.e.\ the first line of Eq.~\eqref{eq:lagrangian}, and higher-dimensional operators, i.e. the second line of Eq.~\eqref{eq:lagrangian}. In the case of only renormalizable interactions one has $M_L=0$, $M_D=vF$ and $M_R = M$. In this work, we focus on higher-dimensional operators which do not contribute to the mass matrix $M_{\nu}$ at tree-level, so we limit the following discussion to only the renormalizable interactions, commenting on the loop-level contribution of the higher-dimensional operators in Section~\ref{sec:loop_neutrino_masses}.

Details on the diagonalization of the mass matrix $M_{\nu}$ are given in Appendix~\ref{app:yuakawa}, here we summarize the main points. The mass matrix can be diagonalized by a $5\times 5$ unitary matrix $U$ leading to
\begin{align}
    {\cal L}_m = \frac{1}{2}\bar{\nu}M_{\nu}^{\rm diag}\nu,
\end{align}
with the diagonal mass matrix $M_{\nu}^{\rm diag} = U^TM_{\nu}U$ and the Majorana mass eigenstates $\nu=N_m+N_m^c$ where $N_m=U^{\dagger}N$, explicitly
\begin{align}
    M^{\rm diag}_{\nu}={\rm diag}(m_1,m_2,m_3,m_4,m_5).
\end{align}
In the seesaw limit $M_{D} \ll M_{R}$, which we will take throughout this work, the lightest three states approximately correspond to the active neutrinos and the two heaviest ones to the RHNs. In the case of two RHNs with normal ordering (NO) the lightest state will be massless $m_1 = 0$ and it follows that $m_2 = (\Delta m_{\mathrm{sol}}^2)^{1/2}, m_3 = (\Delta m_{\mathrm{atm}}^2)^{1/2}$ from the known solar and atmospheric mass squared differences $\Delta m_{\mathrm{sol}}^2$ and $\Delta m_{\mathrm{atm}}^2$, respectively \cite{Esteban:2024eli}. The heavy masses satisfy $m_{4,5} \approx M_{1,2}$. Moreover, in the seesaw limit it is natural to write the unitary matrix $U$ in terms of active neutrinos (a) and sterile neutrinos (s), i.e.\ RHNs, via \cite{Donini:2012tt, Huang:2013kma, Hernandez:2014fha}
\begin{align}
    U=
    \begin{pmatrix}
        U_{aa} & U_{as}\\
        U_{sa} & U_{ss}
    \end{pmatrix}.
\end{align}
The $3\times3$ matrix $U_{aa}$ describes the change from the mass basis of the active neutrinos to the flavor basis of the active neutrinos and is given in the seesaw limit by $U_{aa} \approx U_{\mathrm{PMNS}}$ with the Pontecorvo–Maki–Nakagawa–Sakata (PMNS) matrix $U_{\mathrm{PMNS}}$ given in Appendix~\ref{app:yuakawa}. Similarly, the $2\times 2$ matrix $U_{ss}$ describes the basis change of the sterile states and the $3 \times 2$ ($2\times 3$) matrix $U_{as}$ ($U_{sa}$) describes the active-sterile mixing. The overall size of the active-sterile mixing is given by $U^2 = \sum_{\alpha,I} (U_{as})_{\alpha I} (U_{as})_{\alpha I}^{*} = \Tr(U_{as} U_{as}^{\dagger})$. This quantity is especially useful to characterize low-scale leptogenesis since the thermally averaged rates are proportional to $\Tr(FF^{\dagger}) \propto U^2$ (see Section~\ref{sec:LG}).

\begin{figure}[t]
    \centering
    \includegraphics[width=0.6\linewidth]{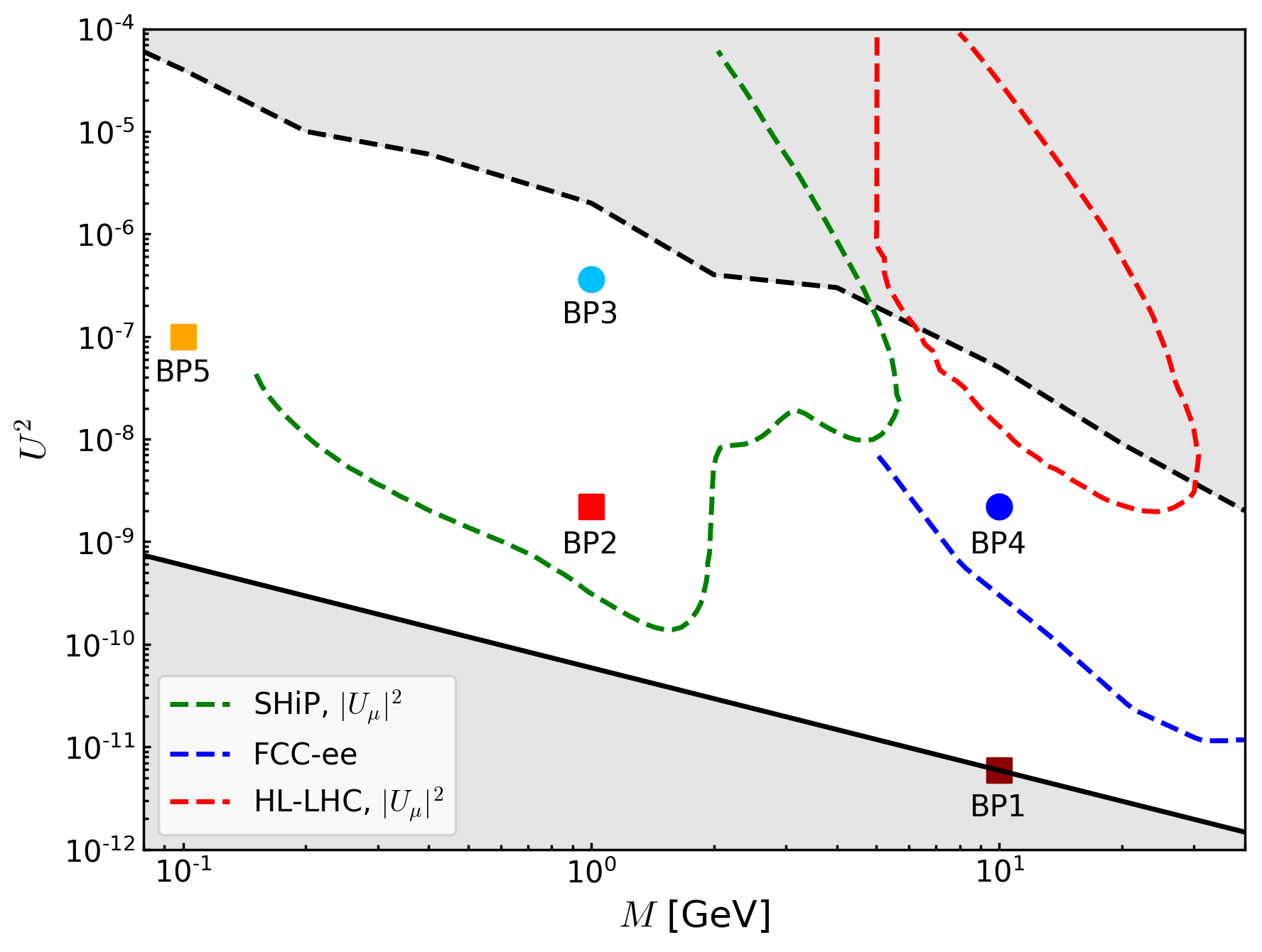}
    \caption{Benchmark points (BPs) used in this work (see Table~\ref{tab:BPs}). We use different shapes to show which BPs are in the oscillatory regime (\ding{110}) or overdamped regime (\ding{108}) of leptogenesis via oscillations (see Section~\ref{sec:scales}). The lower gray region of the parameter space is excluded from requiring that the seesaw mechanism reproduces the active neutrino masses (see Appendix~\ref{app:yuakawa}). In the upper gray region one cannot reproduce the correct baryon asymmetry given our approximations and assumptions in Section~\ref{sec:LG}. Note that the allowed region agrees approximately with the state-of-the-art computations, see e.g. \cite{Klaric:2020phc,Hernandez:2022ivz}. Furthermore, we show the expected sensitivities of SHiP (green), HL-LHC (red) and FCC-ee (blue), taken from~\cite{Klaric:2020phc}.}
    \label{fig:BPs}
\end{figure}

For low-scale leptogenesis, where the RHNs are usually close in mass to reproduce the correct BAU\footnote{The asymmetry generation rate is proportional to the Yukawa coupling and the oscillation timescale $z_{\mathrm{osc}}$ (see Section~\ref{sec:scales}), but the washout rate is also proportional to the Yukawa couplings. For example, in the oscillatory regime (see Section~\ref{sec:scales}) the Yukawa couplings are small Yukawa enough such that washout rates are negligible and one needs to tune the mass splitting to achieve the longest possible oscillation timescale for a sufficient asymmetry generation \cite{Shuve:2014zua}.}, it is more natural to express the masses via $m_{4,5} \approx M \pm \Delta M/2$ with the average mass $M = (M_{1}+M_{2})/2$ and the (small) mass splitting $\Delta M = M_2-M_1$. For a given $M$ and $\Delta M$ the complex entries of the Yukawa coupling $F_{\alpha I}$ are not independent, since one has to reproduce the active neutrino masses $m_{1,2,3}$ as well as the PMNS matrix $U_{\mathrm{PMNS}}$. The independent free parameters of the neutrino sector can be taken as the average RHN mass $M$, the RHN mass difference $\Delta M$, a Majorana phase $\alpha_{31}$ and one complex angle $\omega$ if one takes the convenient Casas-Ibarra (CI) parametrization (see Appendix~\ref{app:yuakawa}). In this work we will focus on a few interesting Benchmark points (BPs) to show the qualitative behavior. We consider the NO case for the active neutrino masses, the values used for the remaining free parameters are given in Table~\ref{tab:BPs} and we show the BPs in the $M$-$U^2$ plane in Fig.~\ref{fig:BPs}. We focus on the mass range shown in Fig.~\ref{fig:BPs} due to cosmological constraints from BBN and to focus on the ARS mechanism for leptogenesis (LG). For $M\lesssim \SI{1e-1}{\giga\electronvolt}$ the sterile neutrinos do not decay before BBN and for $M\gtrsim\SI{1e1}{\giga\electronvolt}$ the contribution of the resonant LG production \cite{Klaric:2020phc,Klaric:2021cpi} as well as the LNV interactions of order $M/T$ become important for the ARS mechanism. For later convenience, we also indicate in Table~\ref{tab:BPs} and Fig.~\ref{fig:BPs} in which LG regime each BP falls. There are two regimes relevant for the ARS mechanism, the oscillatory regime and the overdamped regime, which are explained in detail in Section~\ref{sec:scales}. 
\begin{table}[t]
    \centering
    \begin{tabular}{|c|ccccc|cc|}
    \hline
         & $M$ & $\Delta M / M$ & $\alpha_{31}$ & $\Re \omega$ & $\Im \omega$ & $U^2$ & Regime \\
        \hline
        BP1 & \SI{10}{\giga\electronvolt} & $10^{-8}$ & 5.75 & $\pi/4$ & 0 & \num{5.9e-12} & osc.\\
        BP2 & \SI{1}{\giga\electronvolt} & $10^{-5}$ & 8 & $3\pi/4$ & 2.16 & \num{2.2e-9} & osc. \\
        BP3 & \SI{1}{\giga\electronvolt} & $\num{0.5e-6}$ & 8.53 & $\pi/4$ & 4.71 & \num{3.6e-7} & ov. \\
        BP4 & \SI{10}{\giga\electronvolt} & $10^{-7}$ & 7.24 & $\pi/4$ & 3.31 & \num{2.2e-9} & ov. \\
        \hline
        BP5 & \SI{0.1}{\giga\electronvolt} & $10^{-7}$ & 7.20 & $3\pi/4$ & 2.91 & \num{9.9e-8} & osc. \\
        \hline
    \end{tabular}
    \caption{Benchmark points (BPs) used in this work. BP1 is taken from \cite{Asaka:2011wq} and BP2 and BP3 from \cite{Drewes:2016gmt}. We changed $\alpha_{31}$ for all of them to reproduce the correct baryon asymmetry for a vanishing dim-6 operator using the latest \textit{NuFit-6.0} data \cite{Esteban:2024eli}. We also show the active-sterile mixing $U^2$ and in which LG regime each BP falls. For the LG via oscillations considered in this work one differentiates between the oscillatory (osc.) regime and the overdamped (ov.) regime (see Section~\ref{sec:scales}).}
    \label{tab:BPs}
\end{table}

\subsection{Lepton number conservation and violation}\label{sec:LN}

$0\nu\beta\beta$ decay as well as LG rely on the lepton number violating (LNV) interactions. However, the extension of the standard model lepton number $L_{\rm SM}$ to include a charge for the RHNs is not unique, i.e.~we can choose how to assign lepton number to the RHNs. Following \cite{Abada:2018oly} we differentiate between two useful notions of an extended lepton number: generalized lepton number $\Bar{L}$ and effective lepton number $\Tilde{L}$, which we will explain in the following.  
The neutrino (Dirac) Yukawa interaction violates $L_{\rm SM}$. Further one can introduce the combination $\Bar{L}=L_{\rm SM}+L_{\nu_R}$ (see Tab.~\ref{tab:charge_assignments}), which is violated when both the Dirac and Majorana mass term are present. One way to observe the violation of $\Bar{L}$ is to look at processes involving SM lepton number in the initial or final states only, e.g.~via $0\nu\beta\beta$ decay searches.

However, for special cases of the neutrino Yukawa coupling matrix $F$ and the RHN Majorana mass matrix $M$ there are $\Bar{L}$ assignments for the RHNs such that exact $\Bar{L}$ conservation is realized.\footnote{Actually, taking into account all other SM interactions, on a non-perturbative level, only $B-\Bar{L}$ is conserved.} In these cases the RHNs either i) decouple, ii) have vanishing Majorana masses or iii) arrange themselves in pairs that form (pseudo-)Dirac spinors \cite{Abada:2018oly}. An interesting region of the RHN parameter space is where these conditions are only approximately satisfied, leading to an approximate $\Bar{L}$NC (a$\Bar{L}$NC). This allows for large active-sterile mixing without leading to too large active neutrino masses, making RHNs easier to detect. This corresponds to large values of $U^2$ in Fig.~\ref{fig:BPs}, e.g.~BP3 is a representative example of a$\Bar{L}$NC.
\begin{table}[t]
    \centering
    \begin{tabular}{|ccc|}
    \hline
        spinor & $L_{\nu_R}$-charge & $L_{\rm SM}$-charge \\
        \hline
        $\nu_{R,s}=\frac{1}{\sqrt{2}}(\nu_{R,1}+i\nu_{R,2})$ & $+1$ & 0 \\
        $\nu_{R,w}=\frac{1}{\sqrt{2}}(\nu_{R,1}-i\nu_{R,2})$ & $-1$ & 0\\
    \hline
    \end{tabular} \qquad
    \begin{tabular}{|ccc|}
    \hline
        spinor & $L_{N}$-charge & $L_{\rm SM}$-charge \\
        \hline
        $P_{+}N_i, \quad \overline{N}_{i}P_{-}$ & $+1$ & 0\\
        $P_{-}N_i, \quad \overline{N}_{i}P_{+}$ & $-1$ & 0\\
    \hline
    \end{tabular}
    \caption{Charge assignments for the RHNs used in this work (see also \cite{Abada:2018oly}). The SM fields are charged as usual under $L_{\rm SM}$ and are uncharged under the new lepton numbers, $L_{\nu_R}({\rm SM})=L_{N}({\rm SM})=0$. From these we can define the combinations $\Bar{L}=L_{\rm SM}+L_{\nu_R}$ and $\Tilde{L} = L_{\rm SM} + L_{N}$ for all fields.   
    We assign $L_{\nu_R}$-charge to the specific flavor combinations $\nu_{R,s}$ and $\nu_{R,w}$, such that $\nu_{R,s}$ ($\nu_{R,w}$) couples strongly (weakly) to the SM leptons in the case of a$\Bar{L}$NV. The different RHN helicity states $P_{\pm}N_i$ are approximately conserved at high temperatures, which allows for the definition of the $L_{N}$-charge. Here $P_{\pm}$ stands for the projector onto different helicity states.}
    \label{tab:charge_assignments}
\end{table}

Moving on to early Universe and LG via oscillation, even in the case of a$\Bar{L}$NC, $\Bar{L}$ is significantly violated by the oscillations among the heavy neutrinos \cite{Abada:2018oly}. However, for the temperature range of LG via oscillation, $T \gg M$, the different helicity states of the RHN $N$ basically act as “particle” and “antiparticle” states, such that an effective lepton number $\Tilde{L} = L_{\rm SM} + L_{N}$ (see Tab.~\ref{tab:charge_assignments}) \cite{Abada:2018oly} is conserved by all interactions in the plasma up to correction of order $M/T \ll 1$. Note that LG via oscillation can work even without including these corrections, i.e.~for exact $\Tilde{L}$NC. Then any asymmetry in the SM sector $L_{\rm SM}\neq0$ will be compensated by an asymmetry in the RHN sector $L_{N}=-L_{\rm SM}$, but part of $L_{\rm SM}\neq0$ is converted to a baryon asymmetry by the EW sphalerons. We will explain LG via oscillation in more detail in Section~\ref{sec:LG}. In the LG via oscillation literature it is precisely $\Tilde{L}$NC which is meant when speaking of ``LNC interactions". In this work we only focus on the $\Tilde{L}$NC case, neglecting all $M/T$ corrections. For the calculation in Section~\ref{sec:LG}, we explicitly use a basis for the RHNs where the $\Tilde{L}$NC is apparent. The explicit charge assignments to the RHNs are summarized in Tab.~\ref{tab:charge_assignments}.

\subsection{\texorpdfstring{$\nu$SMEFT}{vSMEFT} Operators} \label{sec:ScalarOp}

Additional interactions mediated by the higher-dimensional operators can change whether lepton number is conserved or not. In this work we will focus on the single\footnote{In general, renormalization group running will lead to other operators being generated. Here we focus on this single operator and leave a more detailed study of general operators as well as their running for future work.} $\nu$SMEFT operator
\begin{equation}\label{eq:our_dim6}
    \mathcal{L}_{\nu_{R}}^{(6)} \supset \mathcal{O}_6 = \frac{G_{\alpha I}}{\Lambda^2} (\overline{L_{\alpha}}\nu_{R,I})(\overline{u_R}Q) \, .
\end{equation}
Here $\Lambda$ characterizes the EFT cut-off scale and we choose $G_{\alpha I}$ as the flavor dependent components of the Wilson coefficient. We leave in Eq.~\eqref{eq:our_dim6} the quark flavor structure of $G$ implicit, allowing for couplings to all three generations of right-handed up-quarks $u_{R}$ and quark doublets $Q$. For $0\nu\beta\beta$ decay and the ARS mechanism the (lepton) flavor structure of the operator will be important, which is encoded by different values for the entries in $G$. We choose this operator because it explicitly involves a RHN $\nu_{R}$ and conserves $\Tilde{L}$ for large temperatures, similarly as the neutrino Yukawa interaction. The operator also allows for a$\Bar{L}$NC for special structures of $G$. One example of this is $G_{\alpha I}=F_{\alpha I}/\sqrt{\Tr(F^\dagger F)}$, which we will take throughout this work, but other choices are possible as well. In this way, if $F$ has a structure which leads to a$\Bar{L}$NC, $G$ will inherit this structure leading to overall a$\Bar{L}$NC. With an overall a$\Bar{L}$NC, large mixing angles for the RHNs, as is the case e.g.~for BP3, remain technical natural. Moreover, with this choice, the only additional free parameter, beside the ones in the renormalizable neutrino sector, is given by the scale $\Lambda$. Later, we will show our results in terms of $\Lambda$, but one should keep in mind that the corresponding Wilson coefficient satisfies $G_{\alpha I} < 1$, in particular $G_{\alpha I} \sim \order{0.1}$ in the absence of hierarchies in the Yukawa coupling $F$. We reiterate that the designation ``LNC” for the operator refers primarily to $\tilde{L}$NC, consistent with the conventions used in the LG via oscillation literature. 

A recast of existing constraints on RHNs translates into bounds on the operator in Eq.~\eqref{eq:our_dim6}~\cite{Fernandez-Martinez:2023phj}. While Ref.~\cite{Fernandez-Martinez:2023phj} considers the case of a single RHN, their results can be straightforwardly generalized to the present scenario with two RHNs by assuming that the two mass eigenstates contribute additively to the processes providing the strongest constraints. With $G_{\alpha I} = F_{\alpha I}/\sqrt{\Tr(F^\dagger F)}$ the flavor dependence for each BP is fixed and the experimental limits can be interpreted as constraints on the scale~$\Lambda$. Explicitly, the bounds for the BPs considered in this work are summarized in Tab.~\ref{tab:BP_bounds} and are mostly determined by the mass $M$.
\begin{table}[t]
    \centering
    \begin{tabular}{|c|c|c|}
    \hline
         & $M$ & $\Lambda$ \\
        \hline
        BP1 & \SI{10}{\giga\electronvolt} & $>\SI{2.7}{\tera\electronvolt}$ \\
        BP2 & \SI{1}{\giga\electronvolt} & $>\SI{17.1}{\tera\electronvolt}$  \\
        BP3 & \SI{1}{\giga\electronvolt} & $>\SI{17.3}{\tera\electronvolt}$  \\
        BP4 & \SI{10}{\giga\electronvolt} & $>\SI{2.9}{\tera\electronvolt}$ \\
        \hline
        BP5 & \SI{0.1}{\giga\electronvolt} & $>\SI{43.9}{\tera\electronvolt}$ \\
        \hline
    \end{tabular}
    \caption{Bounds on the new-physics scale $\Lambda$ associated with the operator in Eq.~\eqref{eq:our_dim6} for the benchmark points (BPs) considered in this work. The limits are obtained by rescaling the bounds of Ref.~\cite{Fernandez-Martinez:2023phj}. For each BP, values of $\Lambda$ below the quoted bound are experimentally excluded.}
    \label{tab:BP_bounds}
\end{table}

\subsection{Loop contribution to neutrino masses}\label{sec:loop_neutrino_masses}

\begin{figure}[t]
\centering
\includegraphics[width=4.5cm]{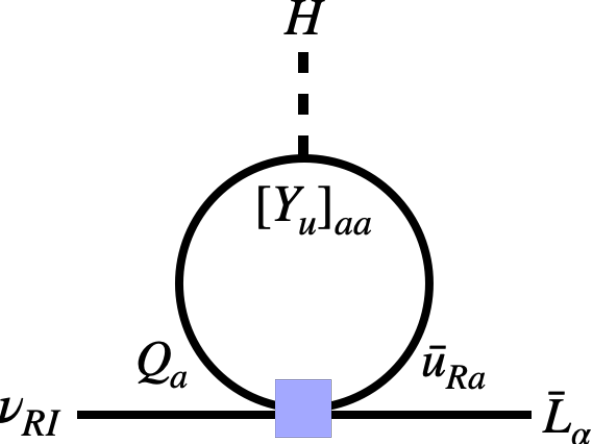}
\caption{The Yukawa interaction arising from the operator ${\cal O}_{6}$ denoted by the blue square.}
\label{fig:rge}
\end{figure}

Any higher-dimensional operator will in general contribute to the neutrino masses. From a given upper limit on the neutrino masses we can estimate a lower limit on the scale of the higher-dimensional operator. These contribution can even be sizable at loop level for high cut-off scales. SMEFT operators, like the Weinberg operator, give in general a contribution to the Majorana mass term of the LH neutrinos. (In these cases one could estimate their contribution following~\cite{deGouvea:2007qla,Deppisch:2017ecm,Gargalionis:2020xvt} using dimensional analysis). Instead, the operator in Eq.~\eqref{eq:our_dim6} contributes to the Yukawa coupling (Dirac mass term) via a loop level contribution, as depicted in Fig.~\ref{fig:rge}. Performing an analogous calculation as in~\cite{Jenkins:2013zja} we find
\begin{align}\label{eq:EFT-running}
\mu\frac{d}{d\mu}F_{\alpha I}=\frac{1}{(4\pi)^2}\frac{m^2_H}{\Lambda^2}(G_{\alpha I})_{aa}[Y_u]_{aa},
\end{align}
with the Higgs mass $m_H\simeq 125~$GeV and the up-type quark Yukawa coupling $Y_u~(a=u,c,t)$. Working in the leading log approximation, we obtain
\begin{align}
F_{\alpha I}(\mu_0)&=F_{\alpha I}(\Lambda)-\frac{1}{(4\pi)^2}\frac{m^2_H}{\Lambda^2}[G_{\alpha I}]_{aa}[Y_u]_{aa}\log\left(\frac{\Lambda}{\mu_0} \right)\nonumber\\
&\equiv F_{\alpha I}(\Lambda)-\Delta F_{\alpha I}.
\end{align}
From this we can infer a correction to the Yukawa coupling $F$ of the order
\begin{align}\label{eq:running-contribution}
\Delta F_{\alpha I}\sim 
\num{1.8e-12} \left(\frac{\SI{1e5}{\giga\electronvolt}}{\Lambda}\right)^2 \left(\frac{[Y_u]_{aa}}{\num{1.3e-5}}\right)[G_{\alpha I}]_{aa} ,
\end{align}
for $\Lambda\sim\order{\SI{1e5}{\giga\electronvolt}}$, which represents the lowest value of $\Lambda$ we will consider in this work, and $\mu_{0}\sim\order{\SI{0.1}{\giga\electronvolt}}$ corresponding to the scale of $0\nu\beta\beta$ decay. For our choice $G = F/\sqrt{\Tr(F^{\dagger}F)}$ one can focus on the overall scale of the contribution which is given by the trace. For our BPs we have $F \sim \sqrt{\Tr(F^{\dagger}F)} \gtrsim \num{1e-7}$, and comparing this value to Eq.~\eqref{eq:running-contribution} we conclude that the correction to the neutrino Yukawa coupling $F$ is negligible if the dim-6 operator involves the first generation ($[Y_u]_{11} \sim \num{1.3e-5}$), which is the quark generation probed in $0\nu\beta\beta$ decay experiments. Thus we safely can use the same values for $F$ at the $0\nu\beta\beta$ decay scale and higher scales relevant for LG, leaving a more detailed investigation of renormalization group effects to a future study.

Note that the exact contribution to the Yukawa couplings and masses can crucially depend on the structure of the UV completion \cite{Fridell:2024pmw}. In the following, we show that the above calculation holds in simple UV completions. There are two simple ways to generate this operator with a single additional heavy field. One is a model with a heavy Higgs $H^{*}(1,2,1/2)$ and the other one is a model with a vector leptoquark $U_1(3,1,2/3)$, following the notation of~\cite{Dorsner:2016wpm}. Focusing on the leptoquark, the relevant Lagrangian reads
\begin{align}\label{eq:LQ-lagrangian}
    \mathcal{L}_{\mathrm{LQ}} &\supset g_1^{LL} U_{1,\mu} \overline{Q}\gamma^{\mu} L + g_1^{\overline{RR}} U_{1,\mu} \overline{u_R} \gamma^{\mu} \nu_{R} + g_1^{RR} U_{1,\mu} \overline{d_R} \gamma^{\mu} e_{R} . 
\end{align}
For a heavy leptoquark one can integrate out $U_1$ and after using the Fierz transformation~\cite{Nieves:2003in} $(\overline{L}\nu_{R})(\overline{u_R}Q) = -1/2 (\overline{L}\gamma_{\mu} Q)(\overline{u_R} \gamma^{\mu} \nu_{R})$ one obtains
\begin{align}\label{eq:matching-couplings}
    \frac{G_{\alpha I}}{\Lambda^2} &= \left[\frac{2g_1^{LL*}g_1^{\overline{RR}}}{m_{\mathrm{LQ}}^2}\right]_{\alpha I} .
\end{align}
In this UV-complete model one gets a similar contribution to the running as given in Eq.~\eqref{eq:running-contribution}, but the situation can be different when multiple additional fields are involved \cite{Fridell:2024pmw}. Depending on the value of the UV coupling, additional operators could be generated giving further detection possibilities.

\section{\texorpdfstring{$0\nu\beta\beta$ decay}{0vbb decay}}\label{sec:0vbb}

Theoretical analyses of $0\nu\beta\beta$ require to deal with a large scale separation from the BSM scale $\Lambda$ through the nuclear scale $\sim$ MeV, which involves hadrons and nuclei that become degrees of freedom below the GeV scale. Employing chiral perturbation theory ($\chi$PT) \cite{Weinberg:1978kz, Gasser:1983yg, Jenkins:1990jv, Bernard:1995dp}, the formulation has been significantly developed over the last few decades; starting from the application of $\chi$PT to processes induced by dimension-9 operators \cite{Prezeau:2003xn}, the analysis has been completed up to dimension 9 operators within the framework of the SMEFT \cite{Cirigliano:2017djv, Cirigliano:2018yza}. In recent years, the study was extend to $\nu$SMEFT, demonstrating a significant contribution from $\nu$SMEFT operators \cite{Dekens:2020ttz}. Following the $\nu$SMEFT analysis therein, we will calculate $0\nu\beta\beta$ half-lives induced by the operator of interest and compare them to the latest experimental limits.

Including the operator in Eq.~\eqref{eq:our_dim6} and rotating to the neutrino mass basis (see the previous section and Appendix~\ref{app:yuakawa}) modify the charged currents relevant for $0\nu\beta\beta$ processes to
\begin{align}
    {\cal L}_{\rm cc}=\frac{2G_F}{\sqrt{2}}\sum_{i=1}^5\Bigg\{\bar{u}_L\gamma^{\mu}d_L\bar{e}_L\gamma_{\mu}\left[C^{(6)}_{\rm VLL}\right]_{ei}\nu_i + \bar{u}_Rd_L\bar{e}_L \left[C^{(6)}_{\rm SLR}\right]_{ei}\nu_i\Bigg\}, \label{cc_opt}
\end{align}
where
\begin{align}\label{eq:0vbb_wilson_coeff}
    \left[C^{(6)}_{\rm VLL}\right]_{ei}=-2V_{ud}U_{ei},\hspace{1cm}
    \left[C^{(6)}_{\rm SLR}\right]_{ei}=2\left(\frac{v}{\Lambda}\right)^2\sum_{I=4,5}G_{eI}U_{Ii}^*,
\end{align}
with $V_{ud}$ being the entry of the Cabibbo–Kobayashi–Maskawa matrix connecting the up- and down-quark. {Note that $U_{ei}\,(i=1,2,3)$ is given in terms of the neutrino oscillation data (see Appendix~\ref{app:yuakawa}), and  Tab.~\ref{tab:BPs_0vbb} states the values of $U_{eI}$ and $\sum_{I=4,5} G_{eI}U^{*}_{IJ}\,(J=4,5)$ for our BPs.\footnote{{In the current setup with two RHNs, $U_{I1}^*=0$ and $U^*_{I2},U^*_{I3}$ are at most ${\cal O}(10^{-4})$ for $I=4,5$. On the other hand, $U_{IJ}\simeq \delta_{IJ}\, (I,J=4,5)$, such that the dominant contribution to $\left[C^{(6)}_{\rm SLR}\right]_{ei}$ in Eq.~\eqref{eq:0vbb_wilson_coeff} comes from $i=4,5$.}}}
\begin{table}[t]
    \centering
    \begin{tabular}{|c|cc|}
    \hline
         & $ U_{e4}$ & $ U_{e5}$ \\
        \hline
        BP1 & $6.2\cdot 10^{-8}+5.9\cdot 10^{-7}\,i$ & $6.2\cdot 10^{-8}-1.3\cdot 10^{-7}\,i$ \\
        BP3 &  $8.2\cdot 10^{-6}+4.395\cdot 10^{-5}\,i$ & $4.393\cdot 10^{-5} - 8.2\cdot 10^{-6}\,i$ \\
        \hline
        BP5 & $4.27\cdot 10^{-5} - 4.2\cdot 10^{-6}\,i$ & $-4.1\cdot 10^{-6} - 4.30\cdot 10^{-5}\, i$\\
        \hline
    \end{tabular}
    \begin{tabular}{|c|cc|}
    \hline
          & $\sum_{I=4,5}G_{eI}U^{*}_{I4}$ & $\sum_{I=4,5}G_{eI}U^*_{I5}$ \\
        \hline
        BP1 & $0.025+0.24\,i$ & $0.025-0.055\,i$ \\
        BP3 & $0.01357 + 0.0728\, i$ & $0.0727 - 0.01359\, i$ \\
        \hline
        BP5& $0.135 - 0.0134\, i$ & $-0.0130 - 0.136 \,i$ \\
        \hline
    \end{tabular}
    \caption{The values of $U_{eI}$ and $\sum_{I=4,5}G_{eI}U^*_{IJ}~(J=4,5)$ relevant for the RHN contributions to $0\nu\beta\beta$ decay for BP1, BP3, and BP5.}
    \label{tab:BPs_0vbb}
\end{table}
Using these coefficient we follow the analysis in \cite{Dekens:2020ttz} to obtain the half-life
\begin{align}
    T^{0\nu}_{1/2}=\Bigg[g_A^4G_{01}\left|{\cal A}_L(m_i)+{\cal A}_{L}^{(\nu)}(m_i)\right|^2 \Bigg]^{-1},
\end{align}
with the nucleon axial coupling $g_A=1.27$ \cite{ParticleDataGroup:2024cfk} and phase space factor $G_{01}=1.5~(0.22) \times 10^{-14}$ yr$^{-1}$ for $^{136}$Xe ($^{76}$Ge) \cite{Horoi:2017gmj}. Here, ${\cal A}_L$ and ${\cal A}_L^{(\nu)}$ represent amplitudes of long- and short-distance contributions, respectively. The latter contribution corresponds to the exchange of hard neutrinos (with the neutrino four momentum $q^0\sim {\bf q}\sim 1~{\rm GeV}$), which was first pointed out in \cite{Cirigliano:2018hja, Cirigliano:2019vdj}.
\begin{figure}[t]
\centering
\includegraphics[width=12cm]{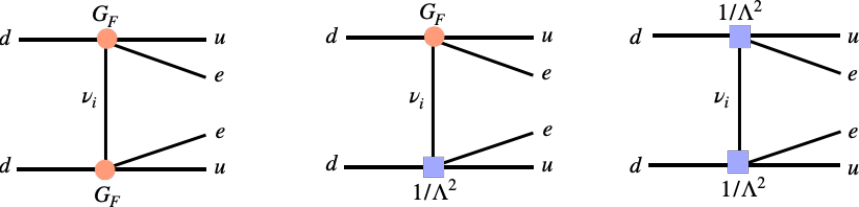}
\caption{$0\nu\beta\beta$ processes generated by the operators in Eq. (\ref{cc_opt}). The left diagram is the standard process arising from the weak interaction (orange bullet), while the middle and right ones originate from the non-standard $\nu_R$ interaction of $(\bar{L}\nu_R)(\bar{u}_RQ)$ (blue square).}
\label{fig:DBD_diagrams}
\end{figure}
The amplitudes are given by
\begin{align}
    {\cal A}_L(m_i)=&-\frac{m_i}{4m_e}\Bigg\{\Bigg({\cal M}_V(m_i)+{\cal M}_A(m_i)\Bigg)\left[C_{\rm VLL}^{(6)}\right]_{ei}^2 
    +\Bigg({\cal M}_{PS}(m_i)\frac{B^2}{m^2_{\pi}} +{\cal M}_S(m_i)\Bigg)\left[C_{\rm SLR}^{(6)}\right]_{ei}^2\nonumber\\
    &\hspace{1.5cm}+\frac{2B}{m_i}{\cal M}_{PS}(m_i)\left[C_{\rm VLL}^{(6)}\right]_{ei}\left[C_{\rm SLR}^{(6)}\right]_{ei}\Bigg\},\label{eq:AL}\\
    {\cal A}_L^{(\nu)}(m_i)=&-\frac{m_i}{m_e}m^2_{\pi}\Bigg\{\frac{2}{F_{\pi}^2m^2_{\pi}}g^{\pi\pi}_{\rm S1}(m_i){\cal M}_{PS,sd}\left[C^{(6)}_{\rm SLR}\right]_{ei}^2\nonumber\\
    &\hspace{1.5cm}+\frac{1}{2g_A^2}M_{F,sd}\Bigg(g^{NN}_{\nu}(m_i)\left[C^{(6)}_{\rm VLL}\right]_{ei}^2+g^{NN}_{\rm S1}(m_i)\left[C^{(6)}_{\rm SLR}\right]_{ei}^2 \Bigg) \Bigg\} \label{eq:ALnu},
\end{align}
with the quark condensate $B=2.7$ GeV and the pion decay constant $F_{\pi}=92.2$ MeV. ${\cal M}_A~(A=V,A,PS,S)$, ${\cal M}_{PS,sd}$ and $M_{F,sd}$ are nuclear matrix elements (NMEs), and $g_{\nu}^{NN}$ and $g_{\rm S1}^{\pi\pi/NN}$ are low-energy constants (LECs). We employ the interpolation formulae for the LECs derived in \cite{Dekens:2020ttz}. The recent theoretical development in the LECs can be found in \cite{Cirigliano:2020dmx, Cirigliano:2021qko, Richardson:2021xiu, VanGoffrier:2024lmo, Richardson:2024ihs} for a light Majorana exchange and \cite{Dekens:2024hlz, Cirigliano:2024ccq} for sterile neutrino processes.

In Fig.~\ref{fig:DBD_diagrams}, we illustrate three  representative processes; the left diagram corresponds to the standard process induced by weak interactions $[C^{(6)}_{\rm VLL}]^2$, while the diagrams in the middle and on the right are the processes induced through non-standard $\nu_R$ interactions $[C^{(6)}_{\rm VLL}][C^{(6)}_{\rm SLR}]$ and $[C^{(6)}_{\rm SLR}]^2$, respectively.  As described in \cite{Blennow:2010th, Dekens:2020ttz}, if $m_{4,5}\ll {\bf q}\sim {\cal O}(100)$ MeV, which is assumed to be a typical momentum transfer between nucleons, the amplitude of the standard process is naively expressed by the expansion parameter $m^2_i/{\rm q}^2$
\begin{align}
    {\cal A}_{\rm standard}(m_i) \propto \sum_{i}\frac{m_i}{{\rm q}^2+m^2_i}U_{ei}^2 \sim \sum_{i}\frac{m_i}{{\rm q}^2}U^2_{ei}\left(1+\frac{m^2_i}{{\rm q}^2}+\cdots\right),
\end{align}
where the leading-order term vanishes due to $\sum_{i}m_iU^2_{ei}=0$ (which follows from Eq.~\eqref{eq:neutrino_mass_diagonalization}). Thus, the first nonzero contribution arises from the next-leading order term $m^3_iU^2_{ei}$, which results in the suppressed amplitudes in the light mass region.\footnote{A recent study found that the process scales as $U^2_{ei}m^2_i$ or $U^2_{ei}m^3_i\log m_i$ more precisely \cite{Dekens:2024hlz}.}

{As discussed in Sec.~\ref{sec:LN}, the choice of $G=F/\sqrt{\Tr(F^{\dagger}F)}$, which is motivated by an a$\Bar{L}$N symmetry, can give large active-sterile mixing angles. On the other hand, the specific structure leads to a suppression of the dim-6 operator contribution of the RHNs to the $0\nu\beta\beta$ decay amplitude. Focusing on their contributions to the non-standard process with $[C_{\rm SLR}^{(6)}]^2$ (right diagram in Fig.~\ref{fig:DBD_diagrams}), we can see
\begin{align}
    {\cal A}_{\rm dim\,6}\propto \sum_{I=4,5} \, \frac{m_I}{{\rm q^2}+m^2_{I}}~[C_{\rm SLR}^{(6)}]^2_{eI}\sim \left(\frac{v}{\Lambda}\right)^4\frac{M}{{\rm q}^2+M^2}\left(G_{e4}^2+G_{e5}^2 \right) + \order{\Delta M}.
\end{align}
As seen in Tab.~\ref{tab:BPs_0vbb}, there could be a cancellation between the two RHN contributions if $G_{e5}\simeq-iG_{e4}$. While $\sum_{I=4,5}G_{eI}^2\sim {\cal O}(10^{-2})$ for BP1, it is ${\cal O}(10^{-6\,(4)})$ for BP3~(BP5), a feature of the a$\Bar{L}$N symmetry.}

\begin{figure}[t]
\centering
\includegraphics[width=7cm]{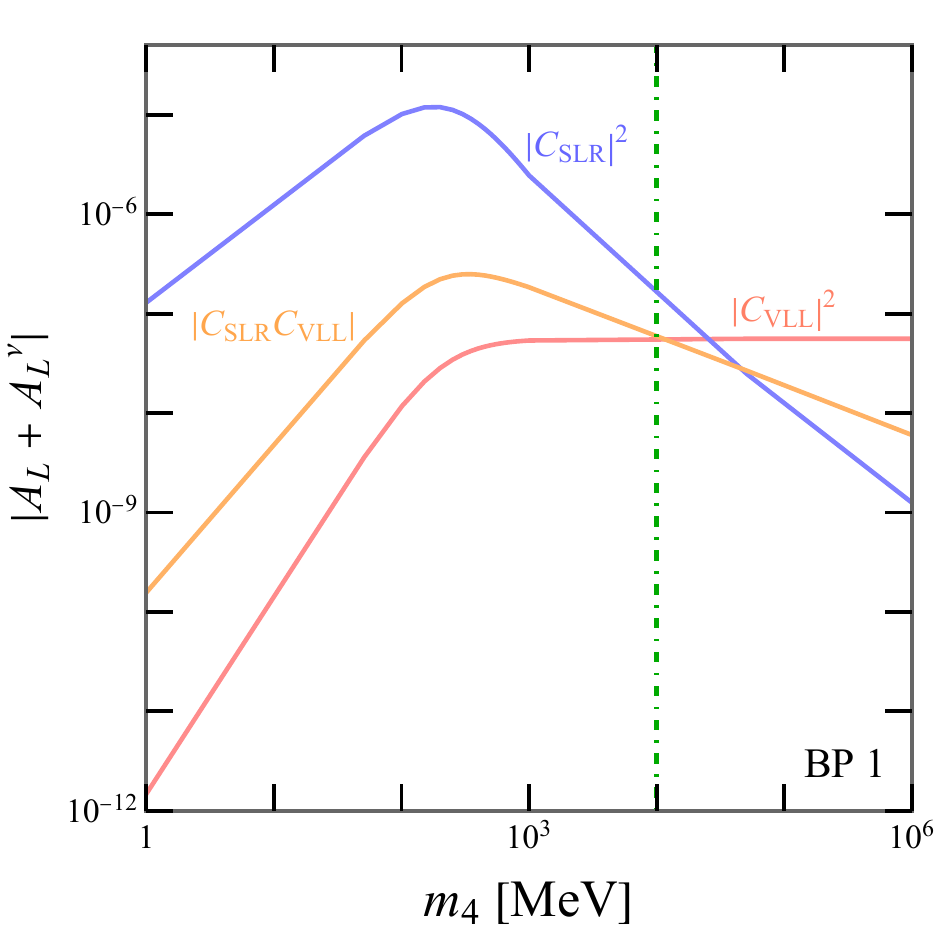}\hspace{0.5cm}
\includegraphics[width=6.8cm]{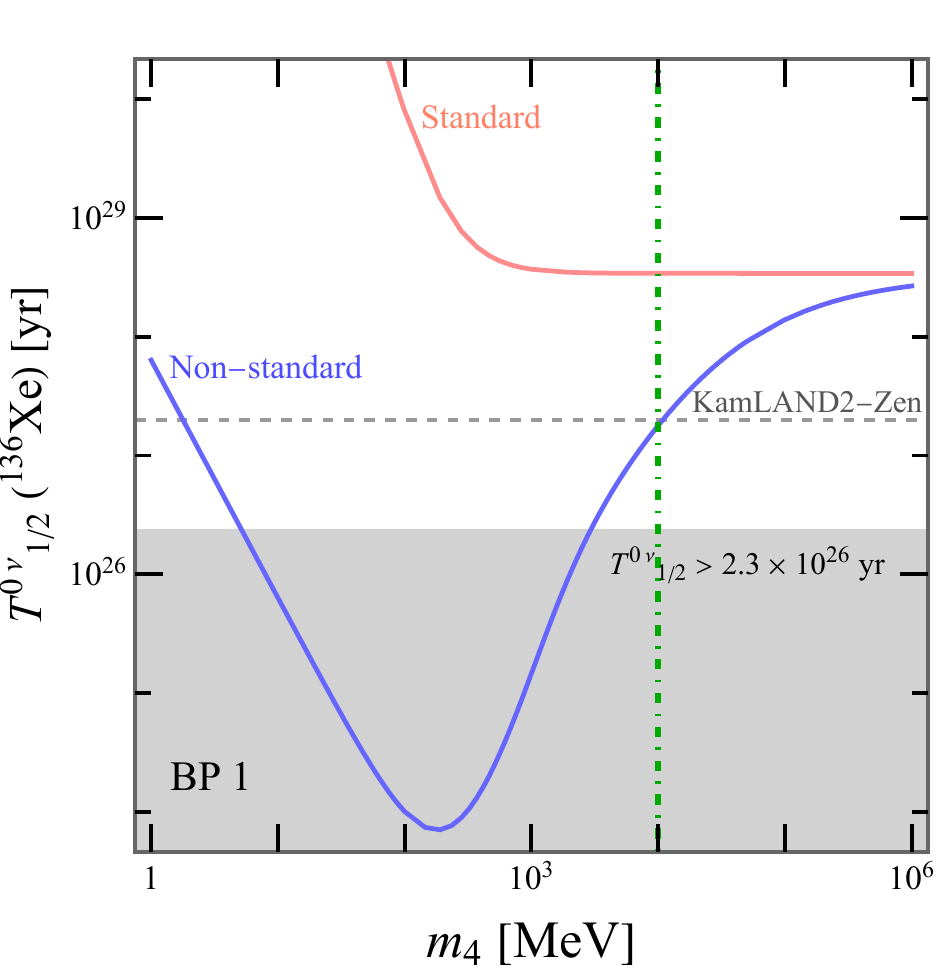}
\caption{BP1~~(Left) contributions to the total amplitude from the standard $|C_{\rm VLL}|^2$ (pink), non-standard $|C_{\rm SLR}|^2$ (blue), and mixed terms $|C_{\rm VLL}C_{\rm SLR}|$ (orange) at {$\Lambda=20$ TeV} as a function of $m_4$. We take $m_5=m_4+\Delta M$ with $\Delta M$ fixed as given in Tab.~\ref{tab:BPs}. The vertical green dash-dotted line indicates the specific value of $m_4$ used for BP1. (Right) the half-lives of $0\nu\beta\beta$ decay against $m_4$. The pink line presents the prediction from the standard interactions while the blue line includes the non-standard contributions. The excluded region is represented by the gray region, and the gray dashed line represents the expected sensitivity of $T^{0\nu}_{1/2}(^{136}{\rm Xe})=2.0\times 10^{27}~$yr at KamLAND2-Zen.}
\label{fig:DBD_halflife_BP1}
\end{figure}

Assuming NO ($m_i<m_{i+1}$) and taking BP1, BP3 and BP5 (except varying $M\approx m_4$) we discuss the predictions of $0\nu\beta\beta$ decay in the following. We further take $m_5 = m_4 +\Delta M$, with $\Delta M$ fixed corresponding to the BPs listed in Tab.~\ref{tab:BPs} (see also Tab.~\ref{tab:BPs_0vbb} for the RHN mixing and couplings of dim-6 operator relevant for $C_{\rm VLL}$ and $C_{\rm SLR}$).
Fig.~\ref{fig:DBD_halflife_BP1} presents the case of BP1 with {$\Lambda=20$ TeV}. The left panel shows contributions from the standard $(C_{\rm VLL}^2)$, non-standard $(C_{\rm SLR}^2)$, and mixed $(C_{\rm VLL}C_{\rm SLR})$ terms to the amplitude in Eqs.~\eqref{eq:AL} and \eqref{eq:ALnu}. The dash-dotted line corresponds to the actual value of $m_4$ used for BP1. It is seen that the non-standard interaction dominates the process up to $m_4\sim 10$ GeV, and is decoupled as $m_4$ becomes heavier. In the right panel of Fig.~\ref{fig:DBD_halflife_BP1}, the pink and blue lines correspond to the prediction of $T^{0\nu}_{1/2}(^{136}{\rm Xe})$ from the standard $(C_{\rm VLL}\neq 0,~C_{\rm SLR}=0)$ and the non-standard $(C_{\rm VLL},~C_{\rm SLR}\neq 0)$ cases, respectively. The presence of the non-standard interaction results in much shorter half-lives compared to the standard one. As expected, if we take larger $\Lambda$, the non-standard case approaches the standard one. The current experimental bound is depicted by the gray region,\footnote{Our analysis takes the experimental result published in \cite{KamLAND-Zen:2022tow}.} and the gray dashed line represents the future prospect of $T^{0\nu}_{1/2}(^{136}{\rm Xe})=2.0\times 10^{27}~$yr at KamLAND2-Zen that can probe the region of {$m_4\sim {\cal O}(1-10^4)$ MeV}. For the actual value $m_4=\SI{10}{\giga\electronvolt}$ used for BP1, indicated by the dash-dotted line, {we see that the half-life for $\Lambda=20$ TeV is just around $2\times 10^{27}$ yr, therefore, it could be accessible in the next-generation search.}
\begin{figure}[t]
\centering
\includegraphics[width=7cm]{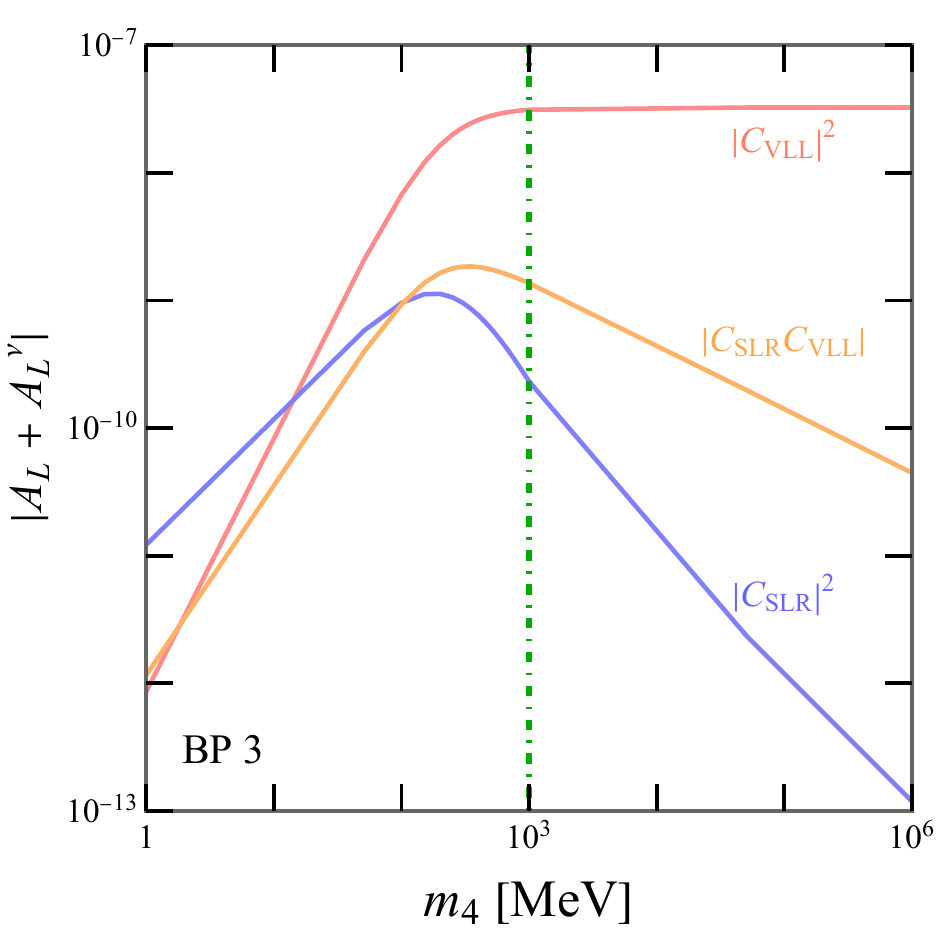}\hspace{0.5cm}
\includegraphics[width=6.8cm]{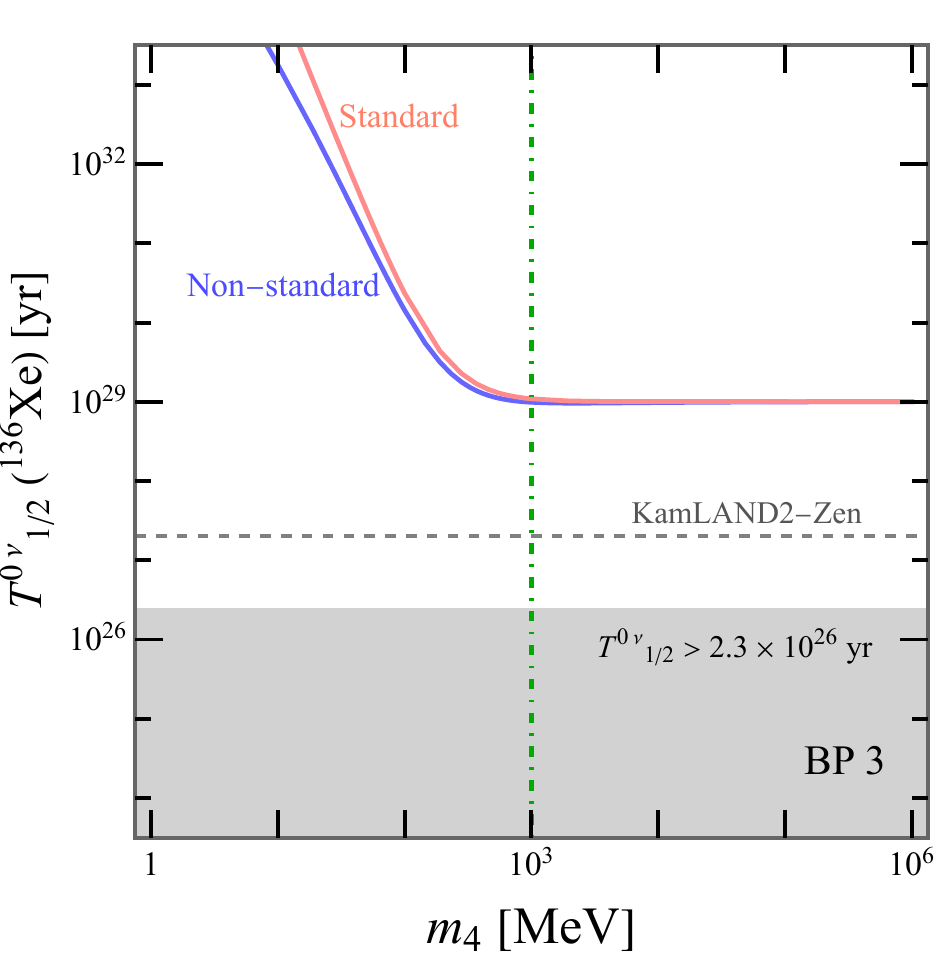}
\caption{BP3~~(Left) contributions to the total amplitude from the standard $|C_{\rm VLL}|^2$ (pink), non-standard $|C_{\rm SLR}|^2$ (blue), and mixed terms $|C_{\rm VLL}C_{\rm SLR}|$ (orange) at {$\Lambda=20$} TeV as a function of $m_4$. We take $m_5=m_4+\Delta M$ with $\Delta M$ fixed as given in Tab.~\ref{tab:BPs}. (Right) the half-lives of $0\nu\beta\beta$ against $m_4$. The pink line presents the prediction from the standard interactions while the blue line includes the non-standard contributions. The current experimental bound is represented by the gray region, and the dashed line corresponds to the expected sensitivity of $T^{0\nu}_{1/2}(^{136}{\rm Xe})=2.0\times 10^{27}~$yr at KamLAND2-Zen. The value of $m_4$ used for BP3 is indicated by the vertical dash-dotted line.}
\label{fig:DBD_halflife_BP3}
\end{figure}
\begin{figure}[t]
\includegraphics[width=7cm]{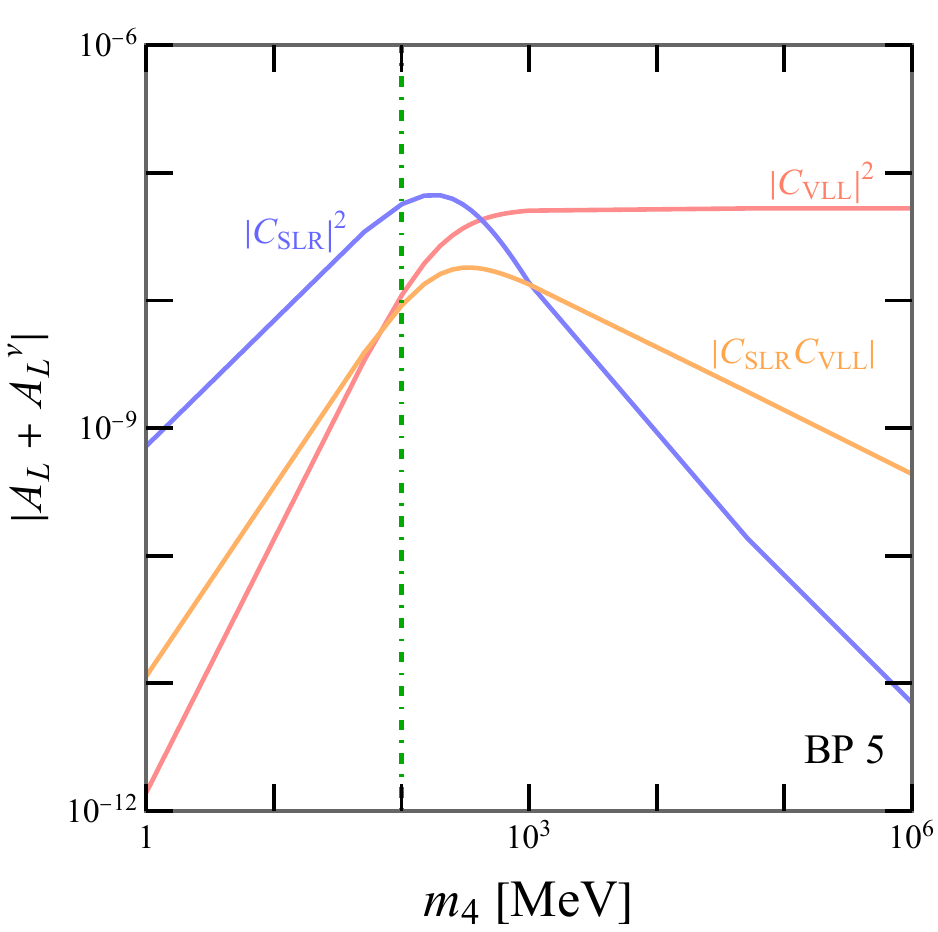}\hspace{0.5cm}
\includegraphics[width=6.8cm]{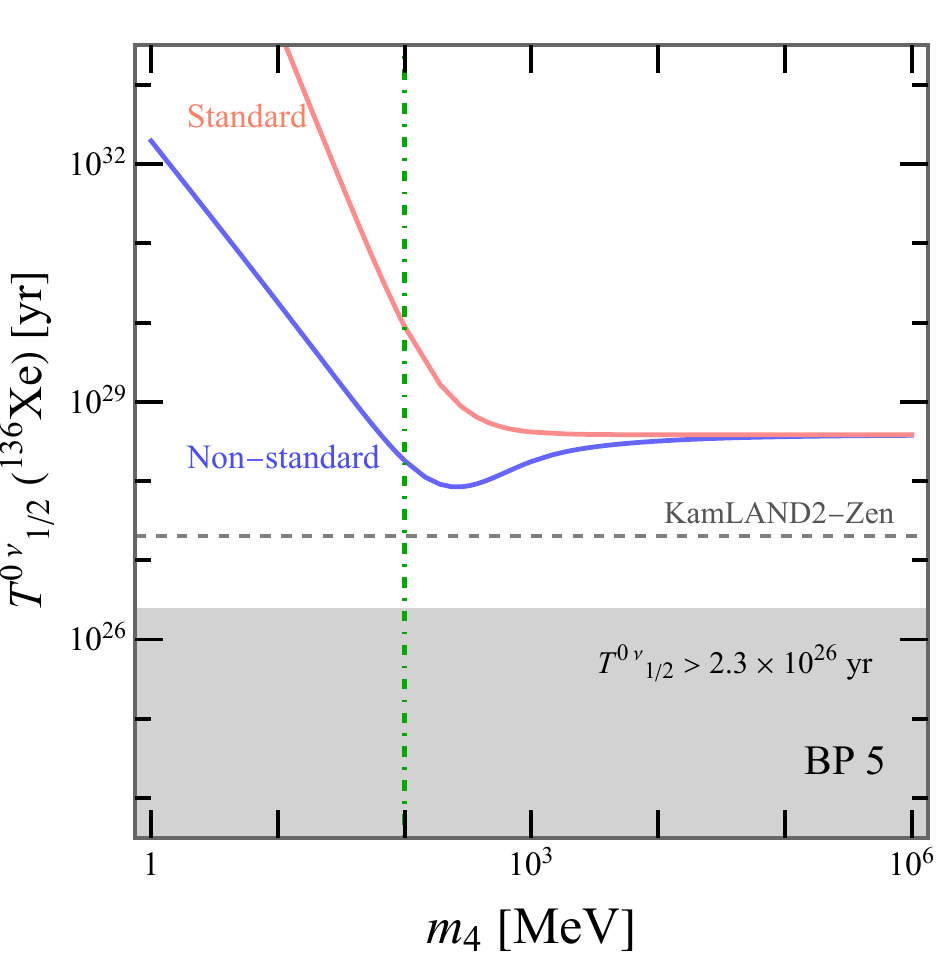}
\caption{BP5~~(Left) contributions to the total amplitude from the standard $|C_{\rm VLL}|^2$ (pink), non-standard $|C_{\rm SLR}|^2$ (blue), and mixed terms $|C_{\rm VLL}C_{\rm SLR}|$ (orange) as a function of $m_4$. We take $m_5=m_4+\Delta M$ with $\Delta M$ fixed as given in Tab.~\ref{tab:BPs}. (Right) the half-lives of $0\nu\beta\beta$ against $m_4$. Also see the caption in Fig.~\ref {fig:DBD_halflife_BP3}. {Note that, although we take $\Lambda=20$ TeV for illustration, this scale is already excluded for BP5, as indicated in Tab.~\ref{tab:BP_bounds}.}}
\label{fig:DBD_halflife_BP5}
\end{figure}

BP3 and BP5 are presented in Fig.~\ref{fig:DBD_halflife_BP3} and \ref{fig:DBD_halflife_BP5}.
{One difference in BP3 from the BP1 case is a relatively strong cancellation between the RHN contributions, which originates from the a$\Bar{L}$N symmetry of BP3 as seen from the structure of $G$ indicated in Tab.~\ref{tab:BPs_0vbb}, leading to a suppression of the non-standard processes. Therefore, we see from the right panel that the $0\nu\beta\beta$ prediction is almost the same as that from the standard contribution. For BP5, although the non-standard contribution can still dominate the process, 
it is unlikely for next-generation searches to reach the predicted half-lives.}
{Note that in Figs.~\ref{fig:DBD_halflife_BP1}, \ref{fig:DBD_halflife_BP3} and \ref{fig:DBD_halflife_BP5}, we are taking $\Lambda= 20$ TeV as a representative value {(compare with Tab.~\ref{tab:BP_bounds} for bounds on $\Lambda$ obtained by recasting the results in \cite{Fernandez-Martinez:2023phj})}, and the contributions from $[C_{\rm SLR}^{(6)}]^2$ and $[C_{\rm SLR}^{(6)}][C_{\rm VLL}^{(6)}]$ scale as $1/{\Lambda^4}$ and $1/{\Lambda^2}$, respectively.} We will explore
the $\Lambda$ dependence together with the implication of requiring successful LG in a later section.

We conclude that, {none of the standard cases can be observed in the next generation $0\nu\beta\beta$ decay experiments. Vice versa, a positive detection of $0\nu\beta\beta$ decay in this mass range would indicate additional physics beyond the renormalizable interactions with RHNs, for example it could point towards the dim-6 $\nu$SMEFT operator $\mathcal{O}_6$ in Eq.~\eqref{eq:our_dim6}, {BP1 as seen in Fig.~\ref{fig:DBD_halflife_BP1}}.} In Section~\ref{sec:results} we show the implications of such a detection on low-scale leptogenesis, for which we derive the relevant equations in the next section.

\section{Leptogenesis}\label{sec:LG}

For RHN masses in the \si{\giga\electronvolt} range, the dominant mechanism for generating a lepton asymmetry is via RHN oscillations, commonly referred to as the ARS mechanism~\cite{Akhmedov:1998qx}. This process falls under the category of freeze-in production: assuming that RHNs have a negligible initial abundance after inflation\footnote{See \cite{Asaka:2017rdj} for non-vanishing initial abundances.}, they are slowly populated through interactions with the SM thermal bath, due to their feeble Yukawa couplings. If part of the RHN sector remains out of equilibrium until the electroweak scale, the corresponding Sakharov condition is naturally satisfied. As the RHNs are produced, they can undergo coherent oscillations between different flavors. These oscillations, together with flavor-dependent CP-violating phases in the Yukawa couplings, generate asymmetries among the different SM lepton flavors~\cite{Shuve:2014zua}. These flavor asymmetries are partially washed out by inverse processes, with rates that also depend on the lepton flavor due to the same Yukawa couplings. As a result, a net lepton asymmetry accumulates in the SM sector, which is balanced by an equal and opposite asymmetry in the RHN sector\footnote{Neglecting lepton-number-violating effects of order $M/T$.}. Since EW sphaleron transitions couple only to SM leptons, they partially convert the SM lepton asymmetry into a baryon asymmetry. Once the temperature drops below the electroweak scale, sphaleron processes freeze out, effectively preserving the generated BAU. At later times, even if the RHN sector equilibrates and therefore erases the lepton asymmetry in the SM sector, the BAU remains fixed. For a more detailed discussion, see \cite{Akhmedov:1998qx,
Asaka:2005an,Asaka:2005pn,Shaposhnikov:2008pf,Canetti:2010aw,Canetti:2012kh,
Asaka:2011wq,
Hernandez:2015wna,Hernandez:2016kel,Hernandez:2022ivz,Sandner:2023tcg,
Anisimov:2010gy,Besak:2012qm,Ghiglieri:2017gjz,Ghiglieri:2018wbs,
Drewes:2012ma,Drewes:2016jae,Drewes:2016gmt,Antusch:2017pkq,Abada:2018oly,Abada:2017ieq,Klaric:2020phc,Klaric:2021cpi,
Shuve:2014zua,Hambye:2016sby,Hambye:2017elz,
Drewes:2017zyw}.

\subsection{Quantum Kinetic Equations}\label{sec:QKE}

Because of the coherent oscillation between the RHNs one needs to go beyond the usual semi-classical Boltzmann equation approach. Similar to the active neutrino oscillations in the early Universe this is a quantum effect and one needs to use so-called quantum kinetic equations (QKEs).
The QKEs can be derived using the density matrix formalism \cite{Sigl:1993ctk}. The object of interest is the so called density matrix $\rho$, which is a matrix in flavor space and whose diagonal elements $\rho_{ii}$ correspond to the phase space distribution function of the particle~$i$. The equation governing the time evolution of the density matrix $\rho = \rho(k)$ reads~\cite{Asaka:2011wq}
\begin{equation}\label{eq:QKE}
    \left(\pdv{}{t} - H k \pdv{}{k}\right)\rho = -i \left[H_{\mathrm{eff}},\rho\right] - \frac{1}{2} \left\{\Gamma^d,\rho\right\} + \frac{1}{2} \left\{\Gamma^p,1-\rho\right\} \, .
\end{equation}
Here $H=T^2/M_{P}^{*}$ is the Hubble rate with $M_{P}^{*} = \sqrt{45/(4\pi^3g_{*}(T))} M_{\rm Planck} = \SI{7.12e17}{\giga\electronvolt}$, $[H_{\mathrm{eff}}]_{ij} = E_{ij} + V_{ij}$ is the effective Hamiltonian, where $E_{ij} = \delta_{ij}\sqrt{k^2+m_i^2}$ corresponds to the energy of particle $i$ with mass $m_i$, and $V_{ij}$ is the effective potential induced by medium effects. Furthermore, $\Gamma^d$ and $\Gamma^p$ are the destruction and production rates, respectively. If there are no correlations between different particles one can treat their density matrices independently. The states in a process like $X + A \to B + C$ are described by density matrices $\rho_X, \rho_A,\rho_B,\rho_C$, where e.g.\ $X_i$ are the RHN flavors labeled by $i$, $A_{a}$ the different SM lepton flavors labeled by $a$ and $B,C$ some final states. In this case one can write the destruction rate $\Gamma^d$ entering the QKE of $\rho_{X}(k)$ as
\begin{align}\label{eq:destruction_rates_formula}
    \left[\Gamma^d(k)\right]_{ij} &= \frac{1}{2E_k} \int \frac{\mathrm{d}^3\vec{p}_A}{(2\pi)^3 2E_A} \frac{\mathrm{d}^3\vec{p}_B}{(2\pi)^3 2E_B} \frac{\mathrm{d}^3\vec{p}_C}{(2\pi)^3 2E_C} (2\pi)^4\delta^{(4)}\left(k+p_A-p_B-p_C\right) \notag \\
    &\quad\times \sum_{a_1,a_2} \mathcal{M}^{\dagger}(X_i+A_{a_1}\to B + C)\left[\rho_{A}(p_A)\right]_{a_1a_2}\mathcal{M}(X_{j}+A_{a_2}\to B + C)\,,
\end{align}
and one has a similar expression for $\Gamma^p$. In the above expression we have neglected Pauli-Blocking and Bose-Enhancement. In the limit where one can neglect all correlations, i.e.\ there are no off-diagonal elements, one recovers for Eq.~\eqref{eq:destruction_rates_formula} the usual collision term and for Eq.~\eqref{eq:QKE} the usual Boltzmann equation.

\begin{figure}[t]
    \centering
    \begin{subfigure}{0.3\textwidth}
        \centering
        \includegraphics{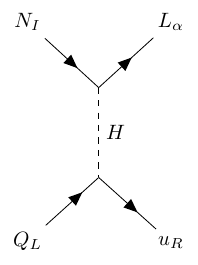} \\
        \includegraphics{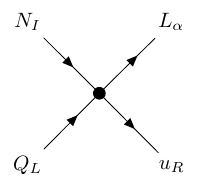}
        \caption{}
    \end{subfigure}
    \hfill
    \begin{subfigure}{0.3\textwidth}
        \centering
        \includegraphics{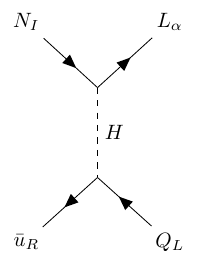} \\
        \includegraphics{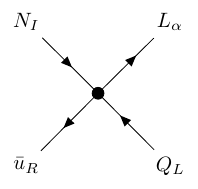}
        \caption{}
    \end{subfigure}
    \hfill
    \begin{subfigure}{0.3\textwidth}
        \centering
        \raisebox{0.7cm}{
        \includegraphics{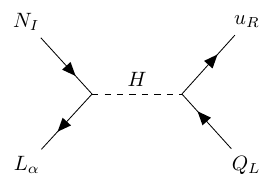}
        } \\
       \includegraphics{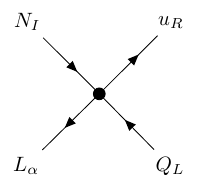}
        \caption{}
    \end{subfigure}
    \caption{Diagrams contributing to the destruction rates $\Gamma^d$ and production rates $\Gamma^p$ from the scatterings mediated by dim-4 interactions (top) and from the scatterings mediated by the dim-6 operator $\mathcal{O}_6$ (bottom). The diagrams were created using Ti\textit{k}Z-Feynman~\cite{Ellis:2016jkw}.}
    \label{fig:quark_scatterings}
\end{figure}

In the following, we specialize the equations above to the case of leptogenesis via neutrino oscillations. In this case we are interested in the evolution of the density matrices of the right-handed neutrinos $\rho_N$, the leptons $\rho_L = \rho_{\nu} + \rho_{e}$ as well as their helicity conjugate and antiparticle density matrices $\rho_{\bar{N}}$ and $\rho_{\Bar{L}}$ respectively. It is convenient to describe the RHNs in terms of helicity states, since this makes $\tilde{L}$NC explicit, as discussed in Section~\ref{sec:LN}. A detailed derivation of the QKE used in ARS and the corresponding rates from different processes can be found in e.g.\ \cite{Ghiglieri:2017gjz}. Here we include the additional interactions of the higher-dimensional operator $\mathcal{O}_6$. In the previous studies \cite{Akhmedov:1998qx,
Asaka:2005an,Asaka:2005pn,Shaposhnikov:2008pf,Canetti:2010aw,Canetti:2012kh,
Asaka:2011wq,Asaka:2017rdj,
Hernandez:2015wna,Hernandez:2016kel,Hernandez:2022ivz,Sandner:2023tcg,
Anisimov:2010gy,Besak:2012qm,Ghiglieri:2017gjz,Ghiglieri:2018wbs,
Drewes:2012ma,Drewes:2016jae,Drewes:2016gmt,Antusch:2017pkq,Abada:2018oly,Abada:2017ieq,Klaric:2020phc,Klaric:2021cpi,
Shuve:2014zua,Hambye:2016sby,Hambye:2017elz,
Drewes:2017zyw} the RHNs only interacted with the SM bath via the Yukawa coupling $F$ and thus all rates were proportional to $F^2$. Here we derive the equations for the general case where the coupling $G$ of $\mathcal{O}_6$ given in Eq.~\eqref{eq:our_dim6} can have a different flavor structure than the Yukawa coupling $F$.
To find simple analytic expressions and to be able to differentiate the new effects of the higher-dimensional operator we make some simplifying assumptions. First, we will use Maxwell-Boltzmann statistics $\rho^{\eq}(k) = \exp(-E_k/T) \mathbb{1}$ for particles in equilibrium and neglect Pauli-Blocking and Bose-Enhancement. While there has been over the years tremendous progress in the precise calculation in the destruction and production rates \cite{Anisimov:2010gy,Besak:2012qm,Ghiglieri:2018wbs}, including quark and gauge scattering as well as $1\leftrightarrow 2$ processes with enhancements by multiple soft scatterings, the Landau-Pomeranchuk-Migdal (LPM) effect~\cite{Landau:1953um,Migdal:1956tc}, we will only include the analytically simple case of the quark scatterings depicted in Fig.~\ref{fig:quark_scatterings} (top row).
These approximations do not change our result qualitatively (see e.g.~\cite{Hernandez:2016kel}) and we leave an exact quantitative study for future work. We will also neglect spectator effects \cite{Drewes:2016gmt} as well as the contribution of the higher-dimensional operator to the effective Hamiltonian and for simplicity take $V = T^2/(8E_k) F^{\dagger} F$ \cite{Asaka:2011wq}. As is common, we will take the momentum average $\expval{\dots}$ as given in Eq.~\eqref{eq:thermal_average} over Eq.~\eqref{eq:QKE} using the ansatz
\begin{equation}\label{eq:density_matrix_RHN_momentum_ansatz}
    \rho_N(t,k) = R_N(t) \rho^{\eq}(k),  \qquad \rho_{\bar{N}}(t,k) = R_{\bar{N}}(t) \rho^{\eq}(k) \ ,
\end{equation}
and study the evolution of $R_{N},R_{\bar{N}}$. In \cite{Asaka:2011wq,Ghiglieri:2018wbs} it has been shown that this leads to a good approximation for the final baryon yield and we expect this to be still valid when including more interactions that drive the system to equilibrium. We explicitly neglect the $\Tilde{L}$NV interactions in the following to show that the effect of the dim-6 operator in Eq.~\eqref{eq:our_dim6} is a $\Tilde{L}$NC one. Since we consider RHN masses $M_{I}\leq\SI{10}{\giga\electronvolt}$, such that for temperatures $T>T_{\rm EW}$ $\Tilde{L}$NV interactions are suppressed by $M_{I}^2/T^2$ compared to the $\Tilde{L}$NC interactions \cite{Drewes:2016gmt,Ghiglieri:2017gjz}, we expect the BAU to only change mildly with the inclusion of these terms \cite{Antusch:2017pkq}. Under this assumption, one can explicitly show, that the evolution is $\Tilde{L}$NC (see Section~\ref{sec:LN}) even when including the dim-6 operator, meaning
\begin{equation}
    \left(\pdv{}{t} - H k \pdv{}{k}\right)\left[\sum_{I} (N_{N_I} - N_{\bar{N}_I}) + \sum_{\alpha} (N_{L_\alpha} - N_{\bar{L}_\alpha})\right] = 0\ , 
\end{equation}
where $N_{i}$ is the number of particles of species $i$. This confirms that the effect of the higher-dimensional operator in Eq.~\eqref{eq:our_dim6} on the resulting baryon asymmetry is a completely $\Tilde{L}$NC one.

With these assumption one can calculate the relevant rates and simplify the QKEs. Details of the derivation can be found in Appendix~\ref{app:rates}, here we state the resulting equations in the style of \cite{Granelli:2023vcm,Hernandez:2022ivz}. Introducing $z=T_{\mathrm{EW}}/T$ and expanding in the small chemical potential\footnote{We use $\rho_L(k) = N_D \rho^{\eq}(k) \diag(e^{\mu_{\nu_{e}}},e^{\mu_{\nu_{\mu}}},e^{\mu_{\nu_{\tau}}})$ with the $SU(2)_L$ factor $N_D = 2$.} $\mu_{\alpha}$ with $\alpha \in \{\nu_{e},\nu_{\mu},v_{\tau}\}$, the QKEs read
\begin{align}
    z H \dv{R_N}{z} &= -i\left[\expval{H_{\mathrm{eff}}},R_N\right] - \frac{1}{2} \sum_{X,Y\in\{F,G\}} \expval{\gamma_N^{(0),XY}} \left\{X^{\dagger}Y,R_N-1\right\} \notag \\
    &\quad + \sum_{X,Y\in\{F,G\}} \expval{\gamma_N^{(1),XY}} X^{\dagger}\mu Y - \frac{1}{2} \sum_{X,Y\in\{F,G\}} \expval{\gamma_N^{(2),XY}} \left\{X^{\dagger}\mu Y,R_N\right\} \label{eq:QKE_LG_1} \\
    z H \dv{R_{\bar{N}}}{z} &= -i\left[\expval{H_{\mathrm{eff}}},R_{\bar{N}}\right] - \frac{1}{2} \sum_{X,Y\in\{F,G\}} \expval{\gamma_N^{(0),XY}} \left\{X^{T}Y^{*},R_{\bar{N}}-1\right\} \notag \\
    &\quad - \sum_{X,Y\in\{F,G\}} \expval{\gamma_N^{(1),XY}} X^{T}\mu Y^{*} + \frac{1}{2} \sum_{X,Y\in\{F,G\}} \expval{\gamma_N^{(2),XY}} \left\{X^{T}\mu Y^{*},R_{\bar{N}}\right\} \label{eq:QKE_LG_2}\\
    z H \dv{\mu_{\alpha}}{z} &= \frac{1}{2N_D} \sum_{X,Y\in\{F,G\}} \expval{\gamma_N^{(0),XY}} \left[YR_NX^{\dagger}-Y^{*}R_{\bar{N}}X^{T}\right]_{\alpha\alpha} \notag \\
    &\quad - \frac{1}{N_D} \sum_{X,Y\in\{F,G\}} \expval{\gamma_N^{(1),XY}} \left[YX^{\dagger}\right]_{\alpha\alpha} \mu_{\alpha} \notag \\
    &\quad + \frac{1}{2N_D} \sum_{X,Y\in\{F,G\}} \expval{\gamma_N^{(2),XY}} \left[YR_NX^{\dagger}+Y^{*}R_{\bar{N}}X^{T}\right]_{\alpha\alpha} \mu_{\alpha} \ ,\label{eq:QKE_LG_3}
\end{align}
with the initial conditions $R_{N}(z_{\mathrm{rh}})=R_{\bar{N}}(z_{\mathrm{rh}})=\mu_{\alpha}(z_{\mathrm{rh}})=0$ at $z_{\rm rh}=T_{\rm EW}/T_{\rm rh}$ where we consider $T_{\rm rh}$ as the reheating temperature. In Section~\ref{sec:scales} we comment on the dependence of the solution on $T_{\rm rh}$ and show the explicit dependence of the BAU on $T_{\rm rh}$ in Section~\ref{sec:results}. We consider the bath to be only radiation dominated after $T_{\rm rh}$ such that $H=T^2/M_{P}^{*}$ is the Hubble rate with $M_{P}^{*} \approx \SI{7.12e17}{\giga\electronvolt}$. We have also explicitly factored out the flavor structure of the rates via $\expval{\Gamma} \sim \sum_{X,Y\in\{F,G\}} \expval{\gamma^{XY}}X^{\dagger}Y$ where $X$ and $Y$ indicate the flavor dependent matrices $F$ or $G$ of the respective processes whose interactions rate are expressed by $\expval{\gamma^{(i),XY}}$ with $i$ labeling the terms with different powers in $\mu$ and $R_N$. Note that when including only dim-4 interactions the only coupling between the RHNs and the SM sector is via the Yukawa coupling $F$ so all rates will be proportional to the Yukawa coupling squared, e.g.\ $\expval{\Gamma} \sim \expval{\gamma} F^{\dagger} F$. With the introduction of the dim-6 operator we introduce a new coupling $G\neq F$, which in general has a different structure compared to the RHN Yukawa coupling. Thus we have to sum in Eqs.~\eqref{eq:QKE_LG_1}-\eqref{eq:QKE_LG_3} over $F$ and $G$ to include both contributions and their interference. The explicit rates are given in Appendix~\ref{app:rates}. While the results above hold for general values of $G$, we simplify in the following the flavor structure for our numerical calculations in the subsequent sections to reduce the number of free parameters as follows. While for the dim-4 interactions the scatterings to the top quark dominate due to the large top Yukawa coupling, the dominating quark flavor for the scattering via the dim-6 operator depends on the underlying new physics. For our numerical calculations we limit ourselves to a coupling to the top quark to asses the potential impact of the interference between the dim-4 interactions and the dim-6 operator and assume $G = F/\sqrt{\Tr(F^{\dagger} F)}$ for the leptonic part.
In this way the only free parameter of the dim-6 coupling is the overall scale $\Lambda$. For our BPs we find that the interference terms of the thermally averaged rates always plays a subdominant role (see Appendix~\ref{sec:interference}). While we limit ourselves to the third generation couplings, our results can be generalized to arbitrary quark flavors by 
simply rescaling $\Lambda$ according to the corresponding interaction (see Appendix~\ref{sec:interference}). In this way, the results could similarly account for e.g. flavor democratic couplings or a coupling to only up quarks, directly relating to the $0\nu\beta\beta$ decay rates.\footnote{Note that, if the dim-6 operator couples to top quarks, the top quark Yukawa coupling needs to be used for the running estimate in Eq.~\eqref{eq:running-contribution}, for which the loop contribution is right at the boundary of becoming relevant. However, only first generation coupling influences the $0\nu\beta\beta$ decay results. For the following LG results, some consideration must be taken when one interprets them for couplings to the third generation and not the first. We leave a more detailed analysis of running effects for future work.} This rescaling would have not been possible with dominant interference terms, as in here another dependence on the coupling $G$ would have been present.

The QKEs~\eqref{eq:QKE_LG_1}-\eqref{eq:QKE_LG_3} describe how oscillations generate a non-zero SM lepton asymmetry in $\mu_{\alpha}$. Neglecting spectator effects, the EW sphalerons convert this asymmetry to a baryon asymmetry. Here we assume an instantaneous shut-off of the sphaleron interactions at $T_{\mathrm{EW}} = \SI{131.7}{\giga\electronvolt}$ \cite{DOnofrio:2014rug} such that the final baryon asymmetry is given by 
\cite{Asaka:2011wq}
\begin{equation}
    Y_{\Delta B} = -\frac{28}{79} Y_{\Delta L} = -\frac{28}{79} \left(\frac{45 N_D}{\pi^4 g_{*s}} \sum_{\alpha} \mu_{\alpha}(z_{\mathrm{EW}})\right) \ .
\end{equation}
A non-zero baryon asymmetry can only be generated if the Sakharov conditions are satisfied. The solution of the QKEs~\eqref{eq:QKE_LG_1}-\eqref{eq:QKE_LG_3} can result in a non-zero SM lepton number, i.e.\ $\sum_{\alpha} \mu_{\alpha}(z_{\mathrm{EW}})\neq0$, if the RHN Yukawa interactions are CP violating and at least one of the RHNs is out of equilibrium. The role of the SM baryon number violating EW sphalerons is then to convert this SM lepton asymmetry to the BAU. The experimentally observed BAU is given by~\cite{Planck:2018vyg}
\begin{equation}
    Y_{\Delta B}^{\mathrm{exp}} = (8.66 \pm 0.05)\times 10^{-11} \ .
\end{equation}
Before comparing the numerical solutions of the QKEs~\eqref{eq:QKE_LG_1}-\eqref{eq:QKE_LG_3} to this value, it is instructive to look at the involved thermalization rates to understand the time evolution of the system.

\subsection{Thermalization rates}\label{sec:scales}

The resulting dynamics crucially depend on the interplay of different scales in the system. These scales are set by the different rates $\expval{\Gamma}$ in the QKEs~\eqref{eq:QKE_LG_1}-\eqref{eq:QKE_LG_3}, which scale parametrically as (see Appendix~\ref{app:rates})
\begin{align}\label{eq:rate_scalings}
    \expval{\Gamma_{\mathrm{osc}}} \sim \frac{M \Delta M}{T}, \qquad \expval{\Gamma_4} \sim T [F^{\dagger} F], \qquad \expval{\Gamma_6} \sim \frac{T^5}{\Lambda^4} [G^{\dagger} G], \qquad H \sim \frac{T^2}{M_{P}^{*}} \ .
\end{align}
Here $\expval{\Gamma_{\mathrm{osc}}} = \expval{H_{\mathrm{eff}}}_{22}-\expval{H_{\mathrm{eff}}}_{11}$ characterizes the oscillation rate between the RHNs $N_1$ and $N_2$ with mass $M$ and mass splitting $\Delta M$, and $\expval{\Gamma_4}$ and $\expval{\Gamma_6}$ are the rates of the dim-4 interactions and our new dim-6 operator, respectively.
Once the oscillation rate becomes comparable to the Hubble rate, $\expval{\Gamma_{\mathrm{osc}}} \sim H$, the RHNs start to oscillate and asymmetries in the lepton flavors start to build up, which will later lead to a SM lepton asymmetry and a corresponding equal and opposite asymmetry in the RHNs. From Eq.~\eqref{eq:rate_scalings} one can see that the oscillations become faster over time, i.e.\ when the Universe expands and the temperature decreases. Comparing the rates $\expval{\Gamma_4}$ and $\expval{\Gamma_6}$ with the Hubble rate, we see that the rate of the dim-4 interactions increases over time and the rate of the dim-6 operator decreases. In general the dim-4 rate will be small initially and not be relevant until $\expval{\Gamma_4} \sim H$ and the dim-6 rate will be large at early times and become negligible once $\expval{\Gamma_6} \sim H$. Thus the dim-4 rate is IR dominated and the dim-6 rate UV dominated, similar to IR and UV freeze-in \cite{Hall:2009bx,Elahi:2014fsa}. Due to the UV dependence of the dim-6 operator rate, we will be sensitive to the earliest time in the evolution of the thermal plasma. In this work, we assume an instantaneous reheating of the plasma after an inflationary period and start all our evolution from a reheating temperature $T_{\mathrm{rh}}$. To be consistent with our effective operator description we require $\Lambda>T_{\mathrm{rh}}$, applicable for $G\sim\order{0.1-1}$, at all times.

The rates in Eq.~\eqref{eq:rate_scalings} depend on the temperature and thus on time. To characterize the full time evolution of the system it is useful to integrate the rates over the time variable $z = T_{\mathrm{EW}}/T$ and normalize the rates $\expval{\Gamma}$ to the Hubble rate $H$. We define the integrated rates \cite{Hernandez:2022ivz}
\begin{equation}
     \overline{\Gamma}_i(z) = \left( \int_{z_{\mathrm{rh}}}^{z} \frac{\mathrm{d}z'}{z'} \frac{\expval{\Gamma_i}}{H}\right) \ .
\end{equation}
There will be a corresponding integrated rate $\overline{\Gamma}_i$ for each $\expval{\Gamma_i}$ in Eq.~\eqref{eq:rate_scalings}. Approximately one full oscillation between the two RHNs has occurred once $\overline{\Gamma}_{\mathrm{osc}}=1$ \cite{Flood:2021qhq} and we define the time when this happens as $z_{\mathrm{osc}} = T_{\mathrm{EW}}/T_{\mathrm{osc}}$.\footnote{For two RHNs there is only one mass difference and thus only one oscillation timescale.} Explicitly one has
\begin{equation}\label{eq:osc_timescale}
    \overline{\Gamma}_{\mathrm{osc}}(z_{\mathrm{osc}})=1 \implies z_{\mathrm{osc}} \approx \left(\frac{6 T_{\mathrm{EW}}^3}{M \Delta M M_{P}^{*}}\right)^{1/3} \ ,
\end{equation}
where we have dropped the contribution proportional to $z_{\mathrm{rh}}$ on the right-hand side of the right equation valid for $z_{\mathrm{osc}}\gg z_{\mathrm{rh}}$. The integrated rates $\overline{\Gamma}_{4}$ and $\overline{\Gamma}_{6}$ describe the production of RHNs by the dim-4 interactions and dim-6 operator, respectively. They have to be large enough such that there are RHNs in the thermal plasma to perform oscillations but not too large otherwise the RHNs reach equilibrium too early. Moreover, $\overline{\Gamma}_{4}$ and $\overline{\Gamma}_{6}$ are flavor dependent via the couplings $F$ and $G$. Let us first focus solely on the dim-4 rate $\overline{\Gamma}_{4}$ corresponding to the standard ARS case. The physical relevant rates are characterized by the eigenvalues of $\overline{\Gamma}_{4}$, corresponding to independent linear combinations of the RHNs, called modes \cite{Drewes:2016gmt,Hernandez:2022ivz}. Potential hierarchical Yukawa couplings $F$ would be correspondingly be present between the rates for these modes. In the case of no hierarchies the rates for these modes, i.e.\ the eigenvalues of $\overline{\Gamma}_{4}$, will be of the same order as the flavor invariant quantity $\Tr(\overline{\Gamma}_{4})$. The RHNs equilibrate once $\Tr(\overline{\Gamma}_{4}) \simeq 1$ and we can define the time when this happens as $z_{\mathrm{eq}}^{\operatorname{dim-4}}$ via
\begin{equation}\label{eq:eq_timescale}
    \Tr(\overline{\Gamma}_{4}(z_{\eq}^{\operatorname{dim-4}}))=1 \implies z_{\eq}^{\operatorname{dim-4}} \approx \frac{128\pi^3}{3N_D N_C h_t^2} \frac{T_{\mathrm{EW}}}{\Tr(F^{\dagger}F)M_{P}^{*}} \ ,
\end{equation}
where we have dropped the contribution proportional to $z_{\mathrm{rh}}$ on the right-hand side of the right equation valid for $z_{\eq}^{\operatorname{dim-4}} \gg z_{\mathrm{rh}}$. 

If $z_{\mathrm{osc}} \gg z_{\eq}^{\operatorname{dim-4}}$ the RHNs have already equilibrated when the RHNs would start to oscillate and the BAU is exponentially suppressed. The requirement can be relaxed in case of hierarchical eigenvalues of $\overline{\Gamma}_{4}$ \cite{Hernandez:2022ivz}. If one eigenvalue is much smaller than $\Tr(\overline{\Gamma}_{4})$, the corresponding \textit{slow} mode equilibrates much later at a time $z_{\eq,\mathrm{slow}}^{\operatorname{dim-4}} \gg z_{\eq}^{\operatorname{dim-4}}$. If $z_{\mathrm{osc}} \ll z_{\eq,\mathrm{slow}}^{\operatorname{dim-4}}$ an asymmetry can be generated. Thus one can differentiate two regimes \cite{Drewes:2016gmt,Hernandez:2022ivz}
\begin{align}
    \text{oscillatory regime: } \qquad z_{\mathrm{osc}} &\ll z_{\eq}^{\operatorname{dim-4}} \ ,\\
    \text{overdamped regime: } \qquad z_{\mathrm{osc}} &\gg z_{\eq}^{\operatorname{dim-4}} \ .
\end{align}
In the oscillatory regime the RHNs undergo a large number of coherent oscillations before reaching equilibrium. While these oscillations become more rapid over time, the BAU is most efficiently generated during the first few oscillations, and later oscillations average out. In the overdamped regime, for which a hierarchy is necessary, at least one RHN reaches equilibrium before performing a complete flavor oscillation leading to an overdamped behavior of the oscillations. In Fig.~\ref{fig:BPs} we indicate which of our BPs falls into which regime considering only the dim-4 interactions. Note that $\Tr(F^{\dagger}F) \simeq U^2M^2/v^2$ and thus $z_{\eq}^{\operatorname{dim-4}}$ is large for small $U^2$. Therefore, the oscillatory regime usually corresponds to small values of $U^2$ and the overdamped regime to large values of $U^2$.

\begin{figure}[t]
\begin{subfigure}[b]{0.5\textwidth}
    \includegraphics[width=\textwidth]{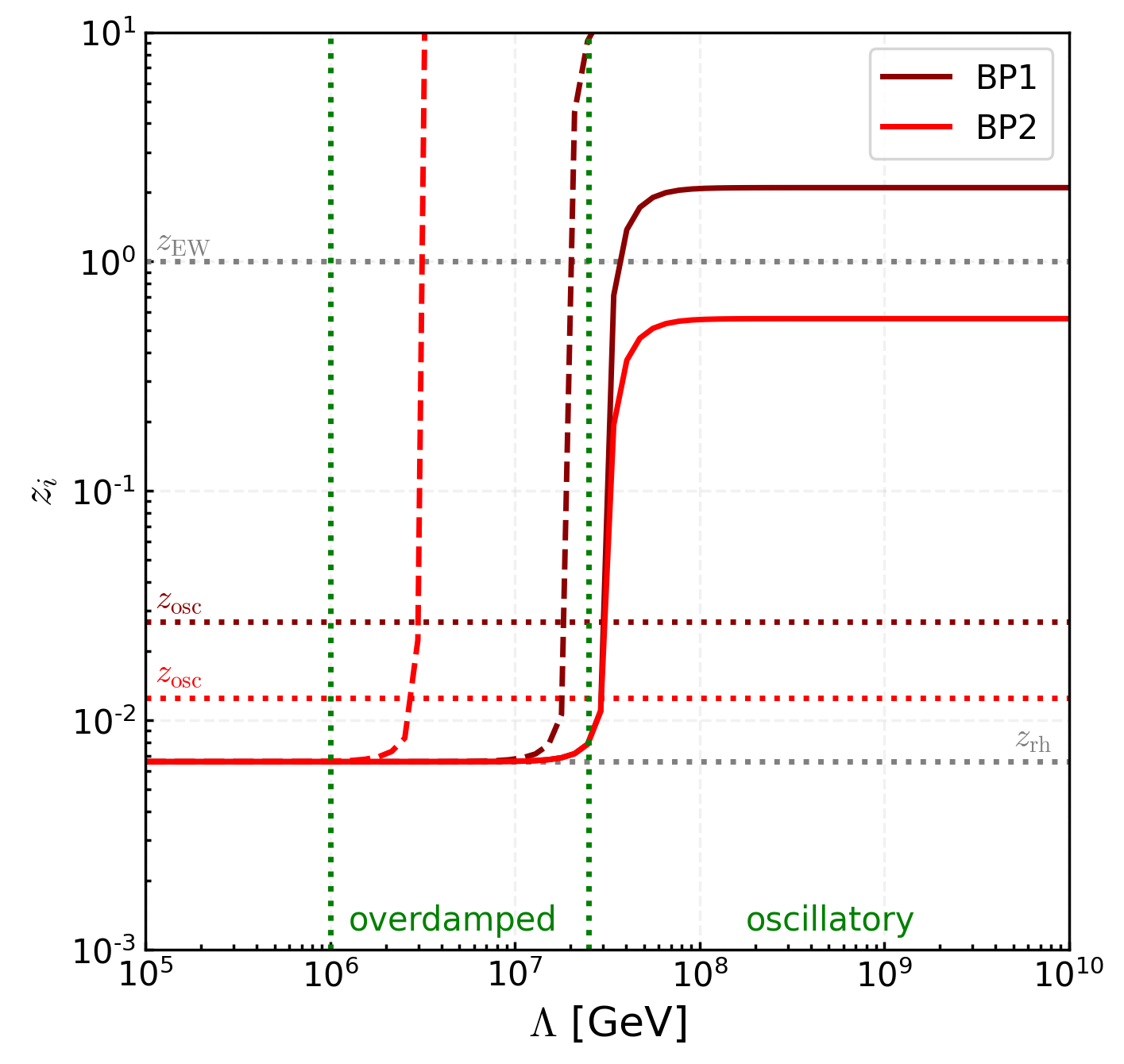}
\end{subfigure}
\hfill
\begin{subfigure}[b]{0.5\textwidth}
    \includegraphics[width=\textwidth]{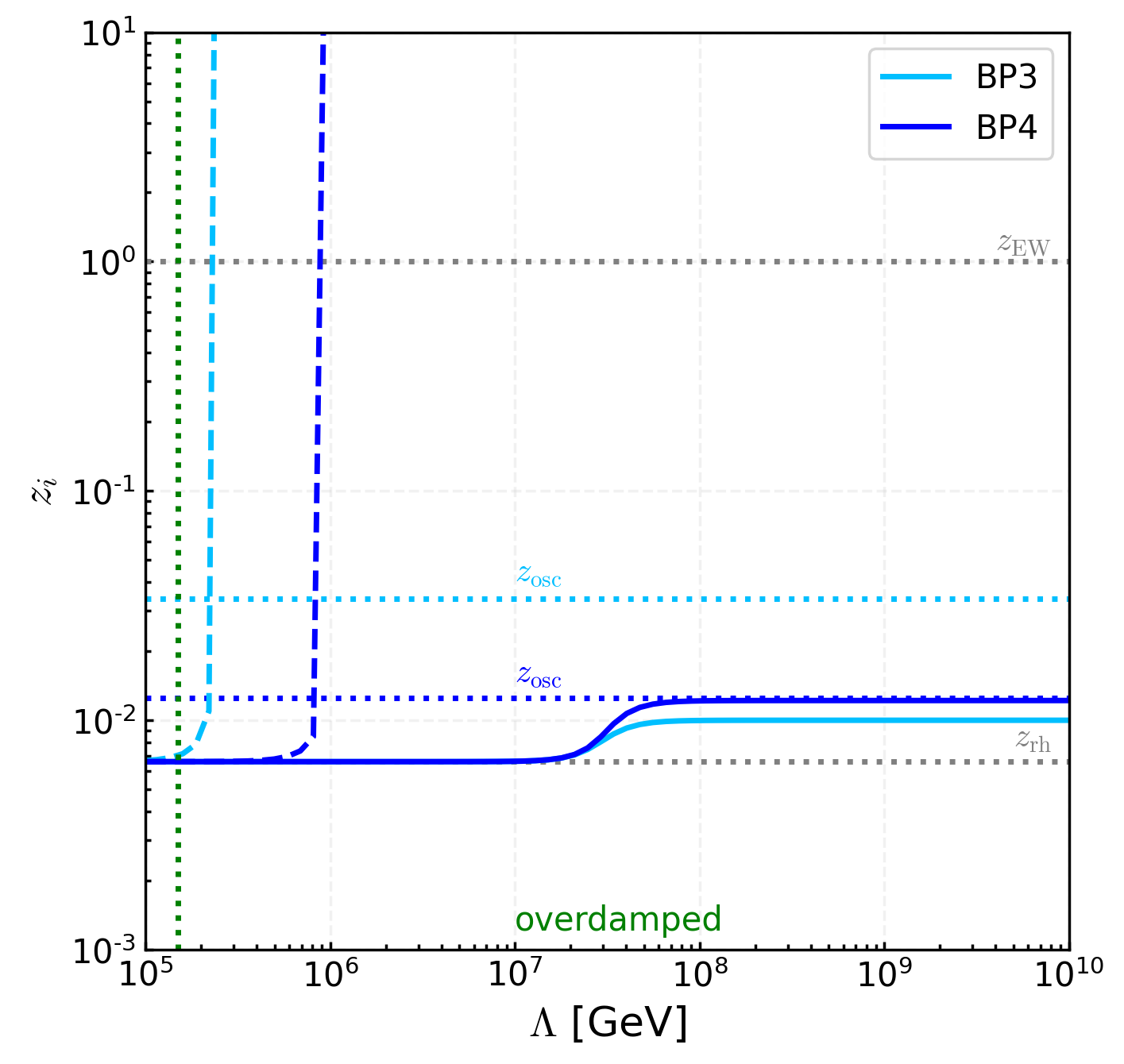}
\end{subfigure}
\caption{Various timescales $z_i$ as a function of the dim-6 operator scale $\Lambda$ of the operator $\mathcal{O}_6$ in Eq.~\eqref{eq:our_dim6} with $G= F / \sqrt{\Tr(F^{\dagger}F)}$ for different benchmark points (BPs) given in Fig.~\ref{fig:BPs}. Left (right) for the BPs corresponding to the oscillatory (overdamped) regime in the limit $\Lambda\to\infty$. Solid lines are the overall equilibration timescales $z_{\eq}$ and dashed lines the equilibration timescales $z_{\eq,\mathrm{slow}}$ for the slow mode. In colored dotted lines we show the corresponding oscillation timescales $z_{\mathrm{osc}}$ and in gray dotted the reheating timescale $z_{\mathrm{rh}}$ ($T_{\mathrm{rh}} = \SI{2e4}{\giga\electronvolt}$) and the EW timescale $z_{\rm EW}$. We also show in green if the evolution is oscillatory ($z_{\rm osc} < z_{\rm eq} $) or overdamped ($z_{\rm osc} > z_{\rm eq} $). Note that when $z_{\eq,\mathrm{slow}} < z_{\rm osc}$ the BAU is exponentially suppressed.}
\label{fig:eq_timescales}
\end{figure}

The above discussion is similarly applicable when including the interactions mediated via the dim-6 operator. The total integrated rate $\overline{\Gamma}_{\mathrm{tot}} = \overline{\Gamma}_{4} + \overline{\Gamma}_{6}$ defines an equilibration scale $z_{\eq}$ for $\Tr(\overline{\Gamma}_{\mathrm{tot}})$ and $z_{\eq,\mathrm{slow}}$ for the slow mode. Note that $z_{\rm osc}$ is independent of the interactions and remains the same as in Eq.~\eqref{eq:osc_timescale}. We show both $z_{\eq}$ and $z_{\eq,\mathrm{slow}}$ as a function of the dim-6 operator scale $\Lambda$ in Fig.~\ref{fig:eq_timescales}.
The oscillation timescales $z_{\mathrm{osc}}$, depicted by colored dotted lines in Fig.~\ref{fig:eq_timescales}, are independent of the scale $\Lambda$ as can be seen from Eq.~\eqref{eq:osc_timescale}. We choose a reheating temperature $T_{\mathrm{rh}} = \SI{2e4}{\giga\electronvolt}$ such that $z_{\mathrm{rh}} < z_{\mathrm{osc}}$ for all four BPs. In the limit $\Lambda \to \infty$ the dim-6 operator decouples and one recovers the standard ARS case with $z_{\eq} \to z_{\eq}^{\operatorname{dim-4}}$ given in Eq.~\eqref{eq:eq_timescale} which takes a different value depending on the BP. We see that for BP1 and BP2 in the left panel of Fig.~\ref{fig:eq_timescales} $z_{\eq}^{\operatorname{dim-4}} > z_{\mathrm{osc}}$ and thus these BPs lie in the oscillatory regime and for BP3 and BP4 in the right panel of Fig.~\ref{fig:eq_timescales} $z_{\eq}^{\operatorname{dim-4}} < z_{\mathrm{osc}}$ and thus these BPs lie in the overdamped regime.\footnote{For BP4 $z_{\mathrm{osc}}$ and $z_{\eq}^{\operatorname{dim-4}}$ are very close, so one could also classify this point in an intermediate regime. As we will see in the next section, BP4 behaves more like points in the overdamped regime.} When decreasing $\Lambda$ one observes that $z_{\eq}$ stays constant until a critical value of $\Lambda$ at which $z_{\eq}$ rapidly decreases to $z_{\mathrm{rh}}$. The reason for this rapid change is the $\Lambda^{-4}$ scaling of the dim-6 operator rate (see Eq.~\eqref{eq:rate_scalings}) and the value $z_{\mathrm{rh}}$ is reached because of the UV dominance. As $z_{\eq}$ is determined from $\Tr(\overline{\Gamma}_{\mathrm{tot}})$ and we take for our BPs $G = F / \sqrt{\Tr(F^{\dagger} F)}$ the dim-6 operator rate is proportional to $\Tr(G^{\dagger}G)=1$ and thus independent of $G$, i.e.\ the specific BP, and only depends on $\Lambda$. Therefore, this critical value of $\Lambda \approx \SI{3e7}{\giga\electronvolt}$ is the same for all four BPs. For each BP we can also consider the slow mode characterized by $z_{\eq,\mathrm{slow}}$ given by the dashed lines in Fig.~\ref{fig:eq_timescales}. The curves of $z_{\eq,\mathrm{slow}}$ have the same shape as the ones of $z_{\eq}$, e.g.\ they are constant for large values of $\Lambda$ with $z_{\eq,\mathrm{slow}} \geq \num{1e1}$. In particular $z_{\eq,\mathrm{slow}} > z_{\mathrm{osc}} > z_{\eq}$ for BP3 and BP4 in the right panel of Fig.~\ref{fig:eq_timescales} which is necessary to produce the BAU in the overdamped regime. The value of $z_{\eq,\mathrm{slow}}$ are constant for large $\Lambda$ until they quickly decrease below a critical value of $\Lambda$, which we denote as $\Lambda_{\rm crit}$. Since the eigenvalues for the different BPs are different, $\Lambda_{\rm crit}$ is different for each BP. Note that there is a hierarchy between $z_{\rm eq}$ and $z_{\rm eq, slow}$ also for the BPs in the left panel of Fig.~\ref{fig:eq_timescales}, especially for BP2. In the standard ARS case ($\Lambda \to \infty$) these hierarchies are not relevant as already $z_{\eq} > z_{\mathrm{osc}}$ for BP1 and BP2, but due to the presence of the dim-6 operator there can be an intermediate value of $\Lambda$ for which $z_{\eq} < z_{\mathrm{osc}}$ but $z_{\eq,\mathrm{slow}} > z_{\mathrm{osc}}$. For these values of $\Lambda$ an overdamped evolution is possible which is indicated in green in Fig.~\ref{fig:final}. This overdamped evolution indeed does happen for BP2\footnote{For BP1 the hierarchy between $z_{\rm eq}$ and $z_{\rm eq, slow}$ is not large enough to be visible in Fig.~\ref{fig:final}.}, as we will see in Fig.~\ref{fig:final} in Section~\ref{sec:numerical_solution}.

\begin{figure}[t]
\begin{subfigure}[b]{0.5\textwidth}
    \includegraphics[width=\textwidth]{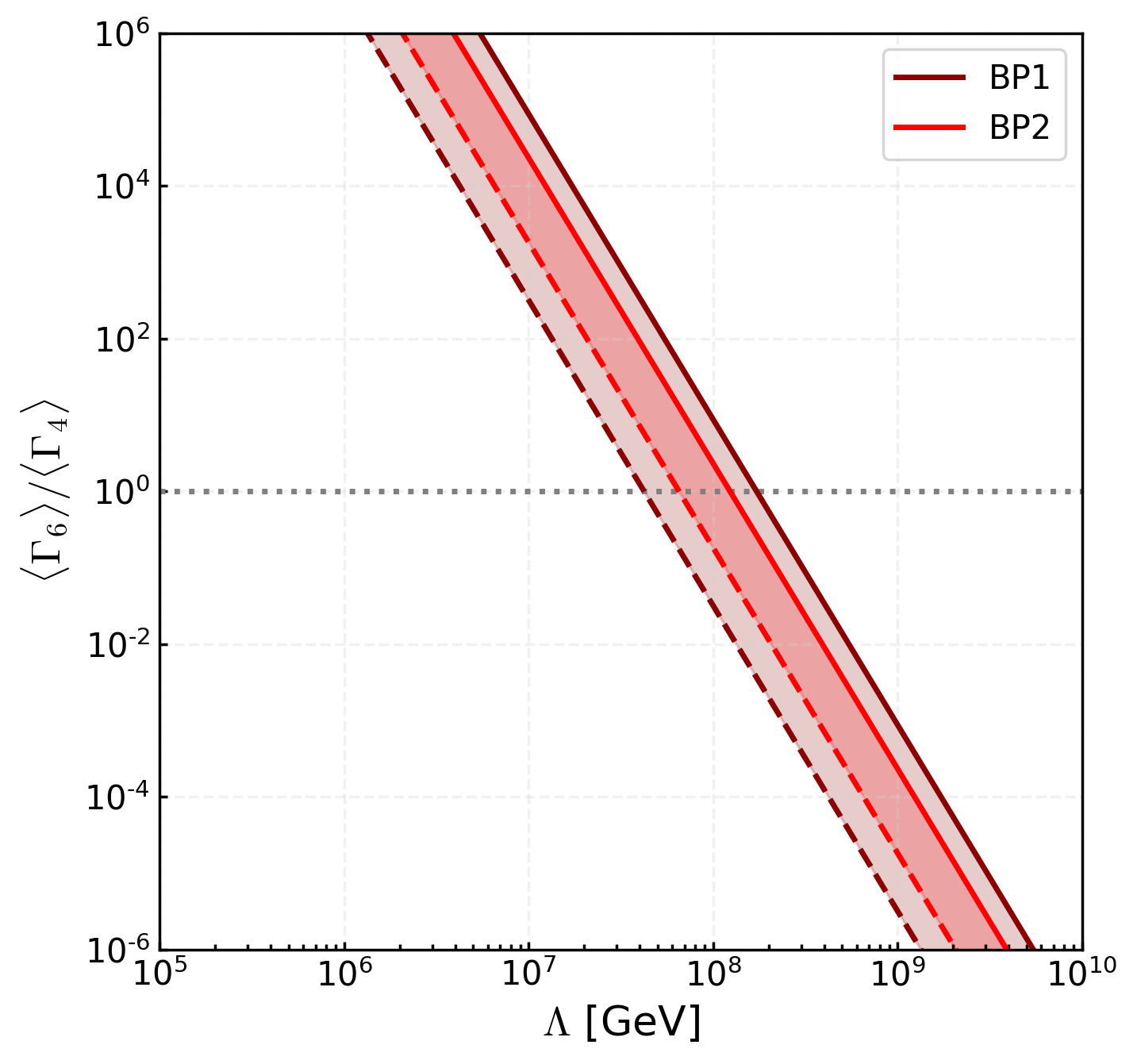}
\end{subfigure}
\hfill
\begin{subfigure}[b]{0.5\textwidth}
    \includegraphics[width=\textwidth]{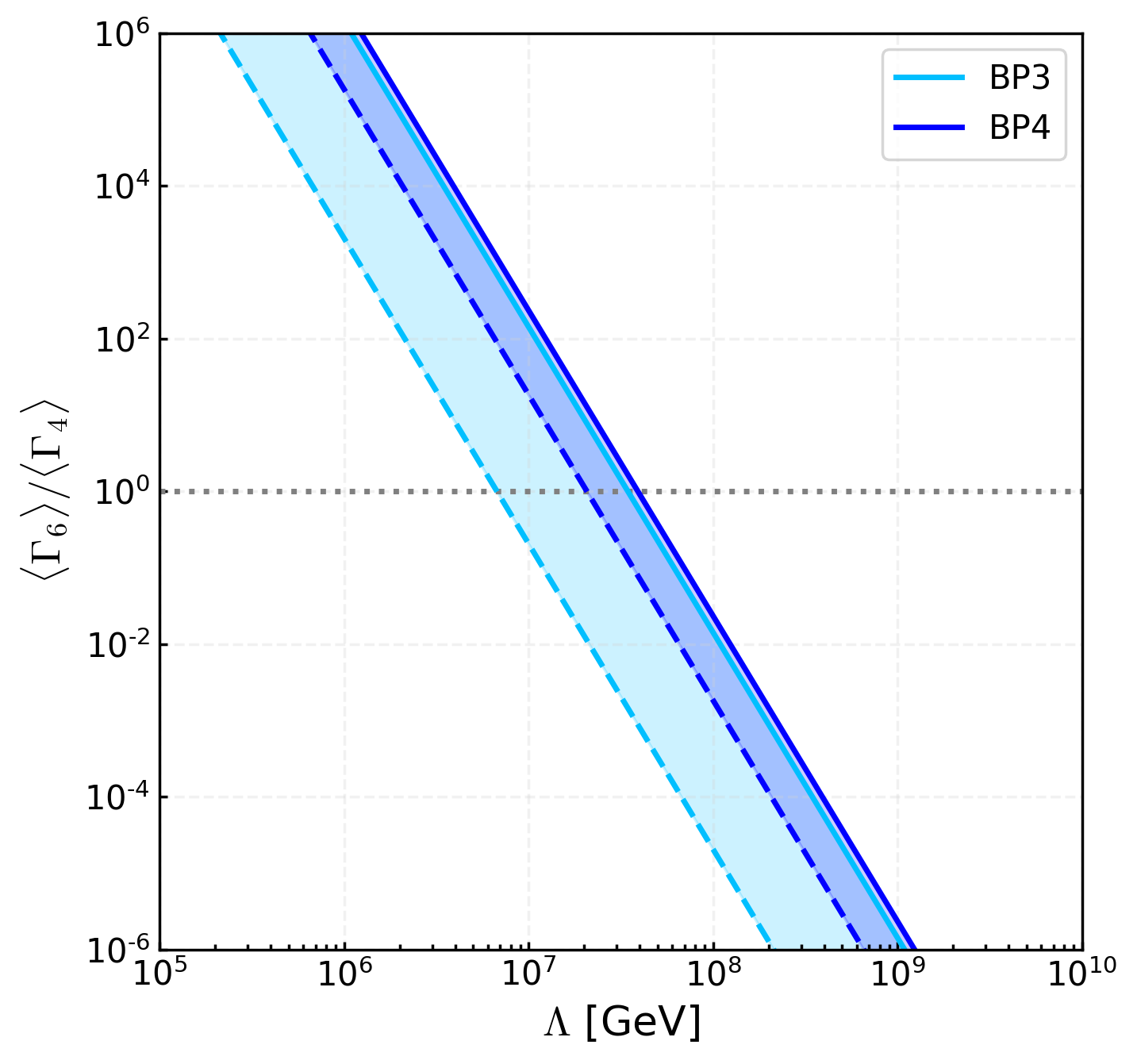}
\end{subfigure}
\caption{Ratios $\expval{\Gamma_6}/\expval{\Gamma_4}$ as a function of the dim-6 operator scale $\Lambda$ of the operator $\mathcal{O}_6$ in Eq.~\eqref{eq:our_dim6} with $G= F / \sqrt{\Tr(F^{\dagger}F)}$ for $z$ values between $z_{\mathrm{rh}}$ (solid) and $z_{\mathrm{osc}}$ (dashed). The gray dotted line corresponds to $\expval{\Gamma_6}/\expval{\Gamma_4}=1$. Left (right) for the BPs in the oscillatory (overdamped) regime.}
\label{fig:ratios}
\end{figure}

From the above discussion one might naively conclude that the only effect of the dim-6 operator is to suppress the BAU for $\Lambda$ below $\Lambda_{\rm crit}$. This is in fact not the case: while for larger values of $\Lambda$ the operator does not change the equilibration time $z_{\eq}$, it still produces additional RHNs compared to the dim-4 contribution only. This happens especially at early times, due to the UV dominance of the dim-6 operator rate, such that there are more RHNs in the thermal bath which can undergo oscillations, possibly increasing the asymmetry. To quantify this effect, we find it useful to compare the averaged rates directly via the ratio $\expval{\Gamma_6}/\expval{\Gamma_4}$. If this value is greater than one for a fixed $z$, the dim-6 operator dominates at that $z$. Due to the different scales involved in the problem, it is not clear at which value of $z$ the expression $\expval{\Gamma_6}/\expval{\Gamma_4}$ should be evaluated. Therefore, we show the ratio $\expval{\Gamma_6}/\expval{\Gamma_4}$ as a function of $\Lambda$ for $z$ values between $z_{\mathrm{rh}}$ and $z_{\mathrm{osc}}$ in Fig.~\ref{fig:ratios}. We find that values between $z_{\mathrm{rh}}$ and $z_{\mathrm{osc}}$ lead to reasonable indications of the dominance of the dim-6 operator in the following sections.
All lines in Fig.~\ref{fig:ratios} have the same slope due to $\expval{\Gamma_6} \propto \Lambda^{-4}$. Since we take the ratio of the rates and we use $G = F / \sqrt{\Tr(F^{\dagger} F)}$ for our BPs, the ratio is the same for each entry in flavor space. The difference between the different BPs arises then due to the difference in $\Tr(F^{\dagger} F) \simeq U^2M^2/v^2$ and the difference in $z_{\mathrm{osc}}$. For the BPs in the oscillatory regime, see left panel of Fig.~\ref{fig:ratios}, $\expval{\Gamma_6}/\expval{\Gamma_4}=1$ is reached for $\Lambda \approx \SI{1e8}{\giga\electronvolt}$. For the BPs in the overdamped regime, see right panel of Fig.~\ref{fig:ratios}, it is obtained at around $\Lambda \approx \SI{2e7}{\giga\electronvolt}$.
As we will see in Fig.~\ref{fig:final} in Section~\ref{sec:numerical_solution} the region in between the values of $\Lambda$ from Fig.~\ref{fig:ratios} and $\Lambda_{\rm crit}$ from Fig.~\ref{fig:eq_timescales} corresponds to an enhanced BAU. 

Thus with these results, we can find estimates for values of $\Lambda$, given in Fig.~\ref{fig:ratios}, below which the BAU will be enhanced up until a critical value $\Lambda_{\rm crit}$, given in Fig.~\ref{fig:eq_timescales}, below which the BAU will be exponentially suppressed. In the next section we solve the system numerically for the different BPs and compare with these estimates.

\subsection{Numerical Solution}\label{sec:numerical_solution}

\begin{figure}[t]
\centering
    \includegraphics[width=0.6\textwidth]{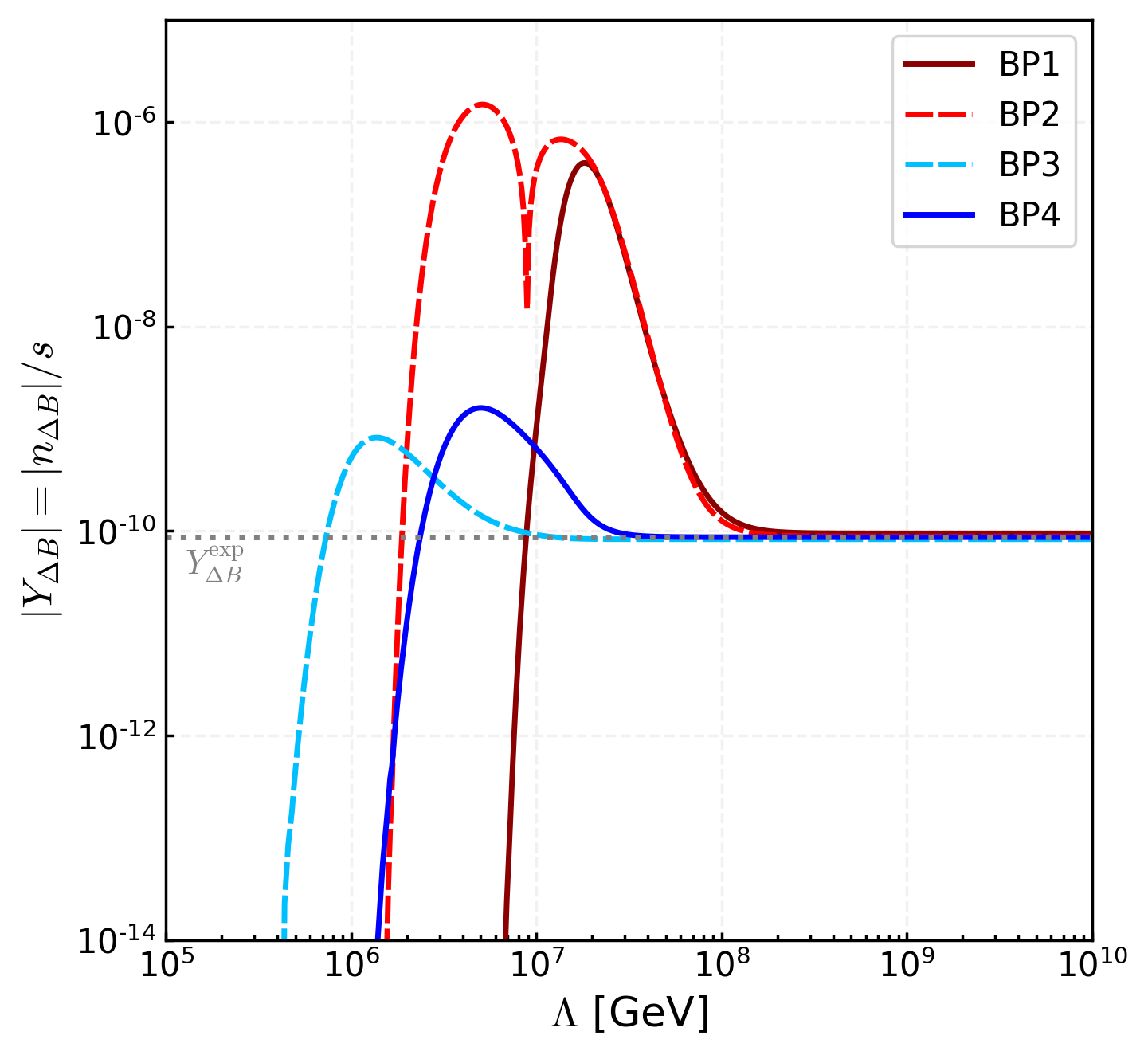}
    \caption{Final baryon asymmetry $Y_{\Delta B}$ as a function of the new physics scale $\Lambda$ of the operator $\mathcal{O}_6$ in Eq.~\eqref{eq:our_dim6} with $G= F / \sqrt{\Tr(F^{\dagger}F)}$ for different benchmark points (BPs) given in Figure~\ref{fig:BPs}. We fix $T_{\mathrm{rh}} = \SI{2e4}{\giga\electronvolt}$ and show the BPs in the oscillatory regime using the red colors and BPs in the overdamped regime using a blue color. Solid lines correspond to $M=\SI{10}{\giga\electronvolt}$ and dashed ones to $M=\SI{1}{\giga\electronvolt}$.}
    \label{fig:final}
\end{figure}

In this section, we solve the QKEs~\eqref{eq:QKE_LG_1}-\eqref{eq:QKE_LG_3} including the dim-6 operator numerically for the BPs listed in Table~\ref{tab:BPs}.\footnote{We solve these equations with a modified version of the ARS model of ULYSSES \cite{Granelli:2020pim,Granelli:2023vcm}, changing the equations to include our dim-6 operator as well as the other assumptions made in Section~\ref{sec:QKE}. Furthermore, we cross-check the result with an independent implementation in Mathematica \cite{Mathematica}.} In Fig.~\ref{fig:final} we show the final baryon yield $Y_{\Delta B}$ as a function of the new physics scale $\Lambda$ for the different BPs. 
We chose the benchmark points such that they reproduce the correct baryon asymmetry without the dim-6 operator, which we refer to as the standard case. In Fig.~\ref{fig:final} this corresponds to the decoupling limit $\Lambda \to \infty$. BP1 and BP2 lie in this limit in the oscillatory regime ($z_{\mathrm{osc}} < z_{\eq}^{\operatorname{dim-4}}$) and BP3 and BP4 in the overdamped regime ($z_{\mathrm{osc}} > z_{\eq}^{\operatorname{dim-4}}$) as given by the definitions in Section~\ref{sec:scales}.

Going from right to left in Fig.~\ref{fig:final} all curves start to rise at some value of $\Lambda$ and the BAU is enhanced. The position of this increase roughly corresponds to the values of $\Lambda$ for which $\expval{\Gamma_6}/\expval{\Gamma_4}=1$ in Fig.~\ref{fig:ratios}, e.g.\ for BP1 and BP2 around $\Lambda \approx \SI{1e8}{\giga\electronvolt}$.
Continuing from right to left in Fig.~\ref{fig:final}, we see that the enhancement can be multiple orders of magnitude until all curves start to be suppressed at some smaller value $\Lambda_{\rm crit}$ (see Section~\ref{sec:scales}). For BP1 and BP2, situated in the oscillatory regime, this decrease corresponds to a value of $ \Lambda \approx \SI{3e7}{\giga\electronvolt}$ in Fig.~\ref{fig:eq_timescales} at which $z_{\eq}$ rapidly decreases. As explained in Section~\ref{sec:scales}, due to the large hierarchies of the rates in BP2, the slow mode does not equilibrate before $z_{\mathrm{osc}}$ at this value of $\Lambda$ and the system switches from the oscillatory into the overdamped regime. Moreover, the BAU changes sign, leading to a second peak for BP2 in Fig.~\ref{fig:final}. For even smaller $\Lambda$ ($\Lambda < \Lambda_{\rm crit}$) also the slow modes equilibrate as seen by the rapid decrease of $z_{\eq,\mathrm{slow}}$ in Fig.~\ref{fig:eq_timescales}, which leads to an exponential decrease of the BAU starting around $\Lambda \approx \SI{2e6}{\giga\electronvolt}$. The same argument applies to BP3 and BP4, which already start in the overdamped regime, where only the equilibration of the slow mode matters. There are two cases in which the experimentally observed BAU can be reproduced. First, for $\Lambda\to\infty$ corresponding to the standard ARS case and second, for a particular value of $\Lambda$ for which the BAU changes from being enhanced to being exponentially suppressed.

\begin{figure}[th]
    \begin{subfigure}[b]{0.5\textwidth}
        \includegraphics[width=\textwidth]{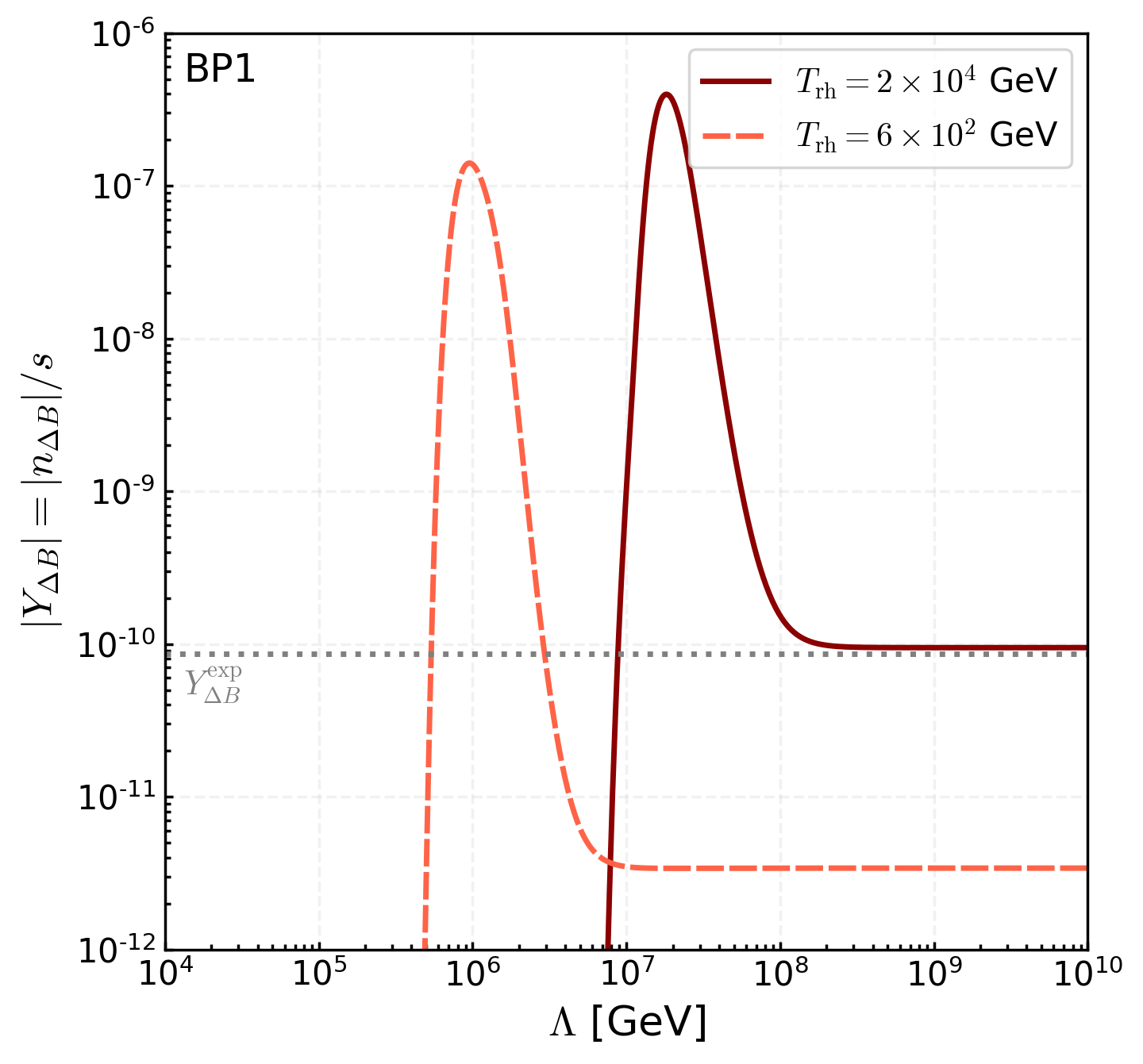}
    \end{subfigure}
    \hfill
    \begin{subfigure}[b]{0.5\textwidth}
        \includegraphics[width=\textwidth]{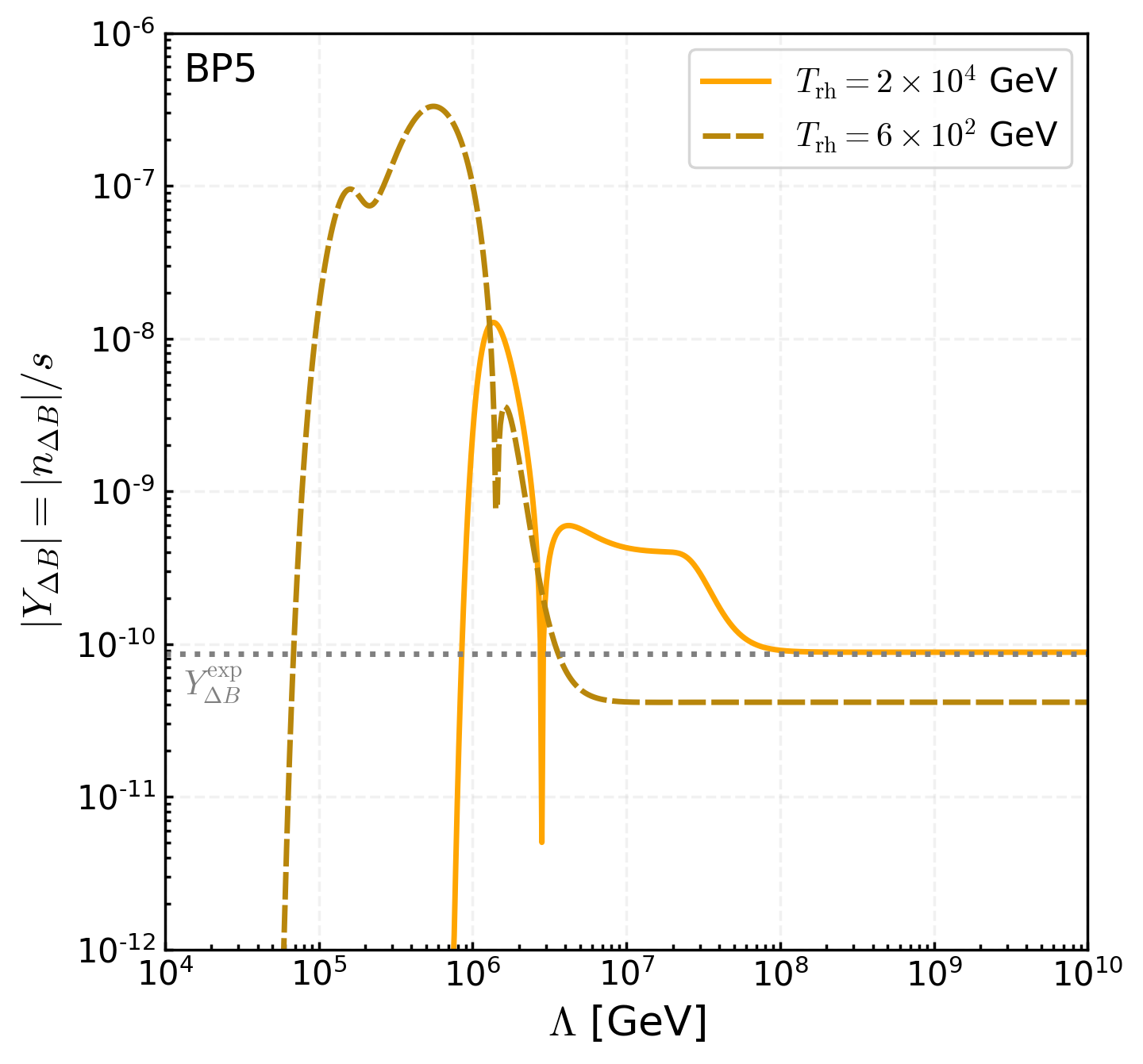}
    \end{subfigure}
    \caption{Final baryon asymmetry $Y_{\Delta B}$ as a function of the new physics scale $\Lambda$ of the operator $\mathcal{O}_6$ in Eq.~\eqref{eq:our_dim6} with $G= F / \sqrt{\Tr(F^{\dagger}F)}$ for different reheating temperatures. Left (right) for BP1 (BP5). Solid lines correspond to $T_{\mathrm{rh}} = \SI{2e4}{\giga\electronvolt}$ and dashed to $T_{\mathrm{rh}} = \SI{6e2}{\giga\electronvolt}$.}
    \label{fig:final_BP5}
\end{figure}

It is quite challenging to produce a sizable BAU and have lower values of $\Lambda$. This is because the rate of the dim-6 operator scales as $\expval{\Gamma_6} \sim T_{\mathrm{rh}}^5/\Lambda^4$, which is large for small $\Lambda$, and thus equilibrates the system erasing all asymmetries. One way to increase the BAU is to reduce the reheating temperature $T_{\mathrm{rh}}$, which is limited by $T_{\mathrm{EW}}$ from below to distribute the lepton asymmetry to the baryon sector via the sphaleron interactions (and from above by $T_{\mathrm{rh}} < \Lambda$). In the left panel of Fig.~\ref{fig:final_BP5} we show how BP1 changes when reducing $T_{\mathrm{rh}}=\SI{2e4}{\giga\electronvolt}$ (solid line) to $T_{\mathrm{rh}}=\SI{6e2}{\giga\electronvolt}$ (dashed line). Again, there are two cases for which the correct BAU can be found. While one of these values corresponds to $\Lambda \to \infty$ for $T_{\mathrm{rh}}=\SI{2e4}{\giga\electronvolt}$ there are two finite values for $\Lambda$ for $T_{\mathrm{rh}}=\SI{6e2}{\giga\electronvolt}$. This is due to the fact that the BPs are chosen in such a way to reproduce the correct BAU in the limit of $\Lambda\to\infty$ for $T_{\mathrm{rh}}=\SI{2e4}{\giga\electronvolt}$. While reducing $T_{\mathrm{rh}}$ lowers the scale $\Lambda$ below which a non-zero BAU can be achieved it also leads to $T_{\rm osc}\approx T_{\rm rh}$, and then there is only a small time window for the RHNs to oscillate coherently, which can produce most of the asymmetry. This can be seen in the $\Lambda\to\infty$ limit where the BAU for BP1 with lower $T_{\mathrm{rh}}$ is reduced. We will study the dependence on $T_{\mathrm{rh}}$ in more detail in the next section. To have large asymmetries at low values of $\Lambda$ one needs both small $T_{\mathrm{rh}}$ and sufficient separation until $T_{\mathrm{osc}}$. As an example of this, we show the additional BP5 in the right panel of Fig.~\ref{fig:final_BP5} which has in the limit $z_{\rm rh}\ll z_{\rm osc}$ for BP5 $z_{\mathrm{osc}} \approx 0.3$ compared to $z_{\mathrm{osc}} \approx 0.03$ for BP1. While the specific features of BP5 can be explained using the analysis based on Section~\ref{sec:scales} (see Appendix~\ref{sec:timescales_BP5}), we want to emphasize the fact that for relatively low values $\Lambda \sim \SI{1e5}{\giga\electronvolt}$ a large asymmetry can be obtained. 

We have shown in this section that the additional $\Tilde{L}$NC dim-6 interactions can change the resulting baryon asymmetry by multiple order of magnitudes compared to the dim-4 interactions usually considered. This can be an enhancement or a suppression depending on the scale of the dim-6 operator. Moreover, this change depends on the reheating temperature. In the following section we study the dependence on the reheating temperature in more detail and connect with the $0\nu\beta\beta$ decay results from Section~\ref{sec:0vbb}.

\section{Results}\label{sec:results}

In this section, we present how the reheating temperature affects the BAU and which conclusions we can draw for the observation of $0\nu\beta\beta$ decay. Moreover, we show that one can reduce the mass degeneracy of the RHNs while still producing the correct BAU.

\subsection{Dependence on the reheating temperature}
\begin{figure}[tb]
    \begin{subfigure}[b]{0.50\textwidth}
        \includegraphics[width=\textwidth]{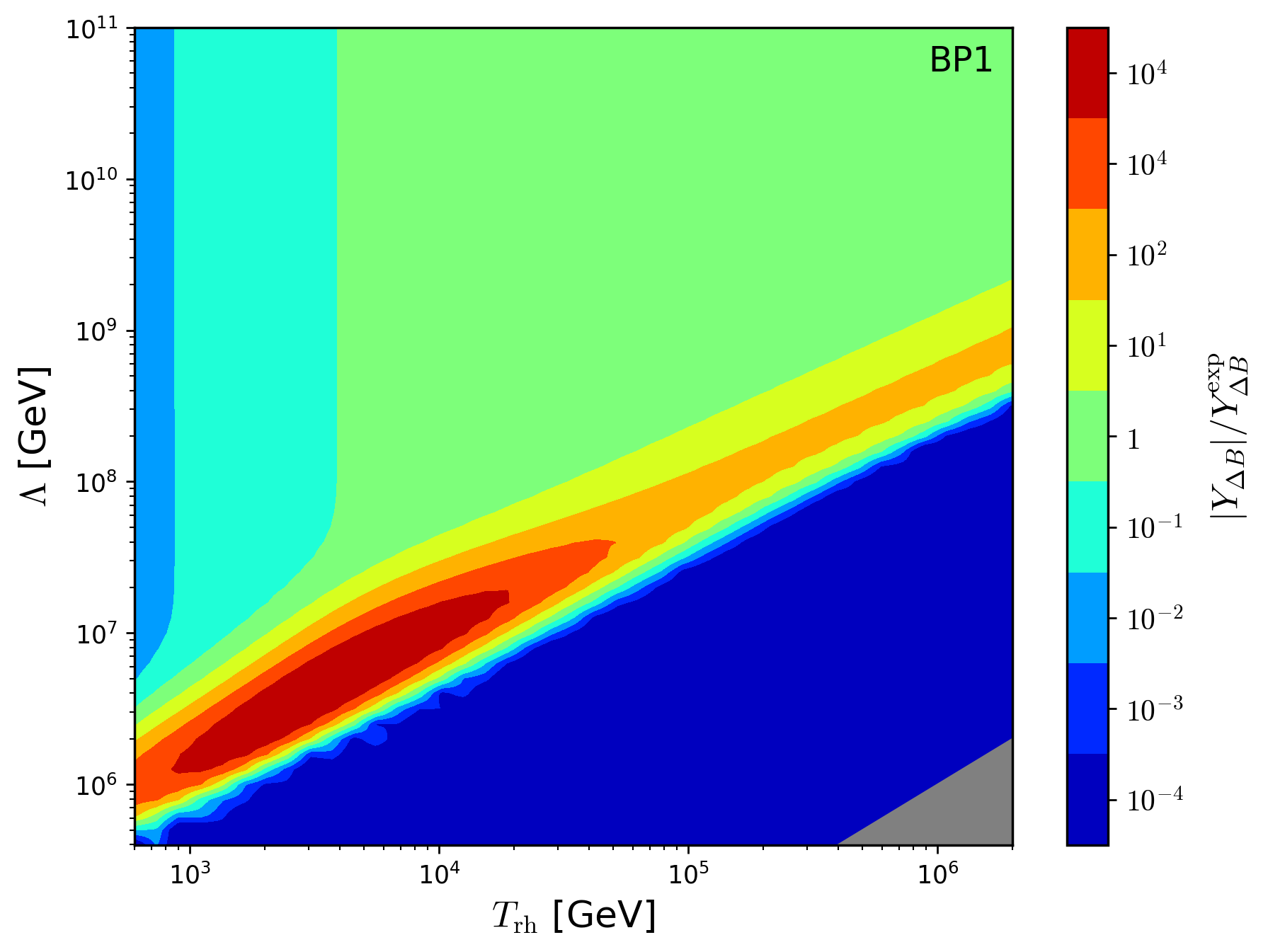}
    \end{subfigure}
    \hfill
    \begin{subfigure}[b]{0.5\textwidth}
        \includegraphics[width=\textwidth]{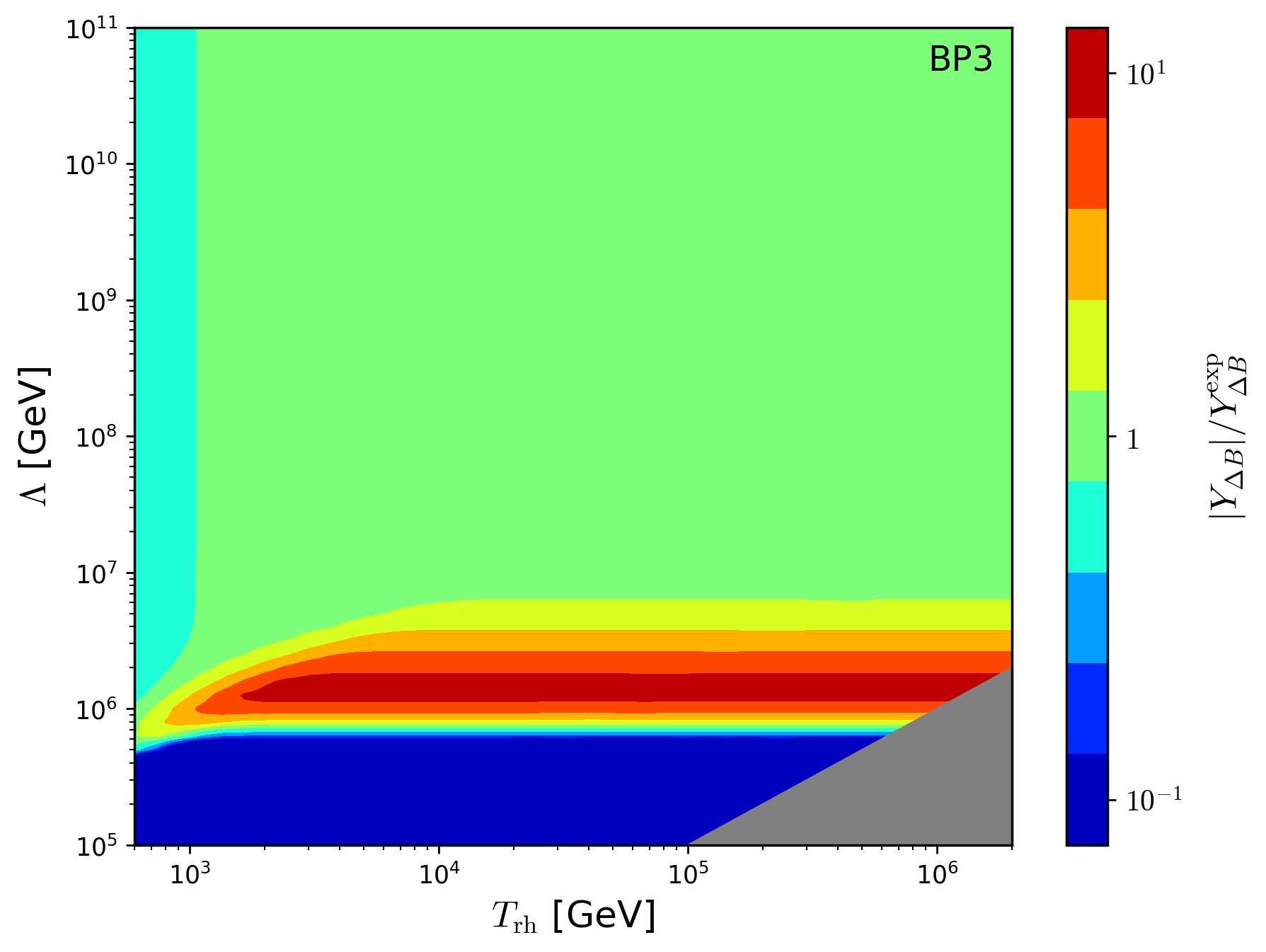}
    \end{subfigure}
    \vskip 0.5cm
    \centering
    \begin{subfigure}[b]{0.5\textwidth}
         \includegraphics[width=\textwidth]{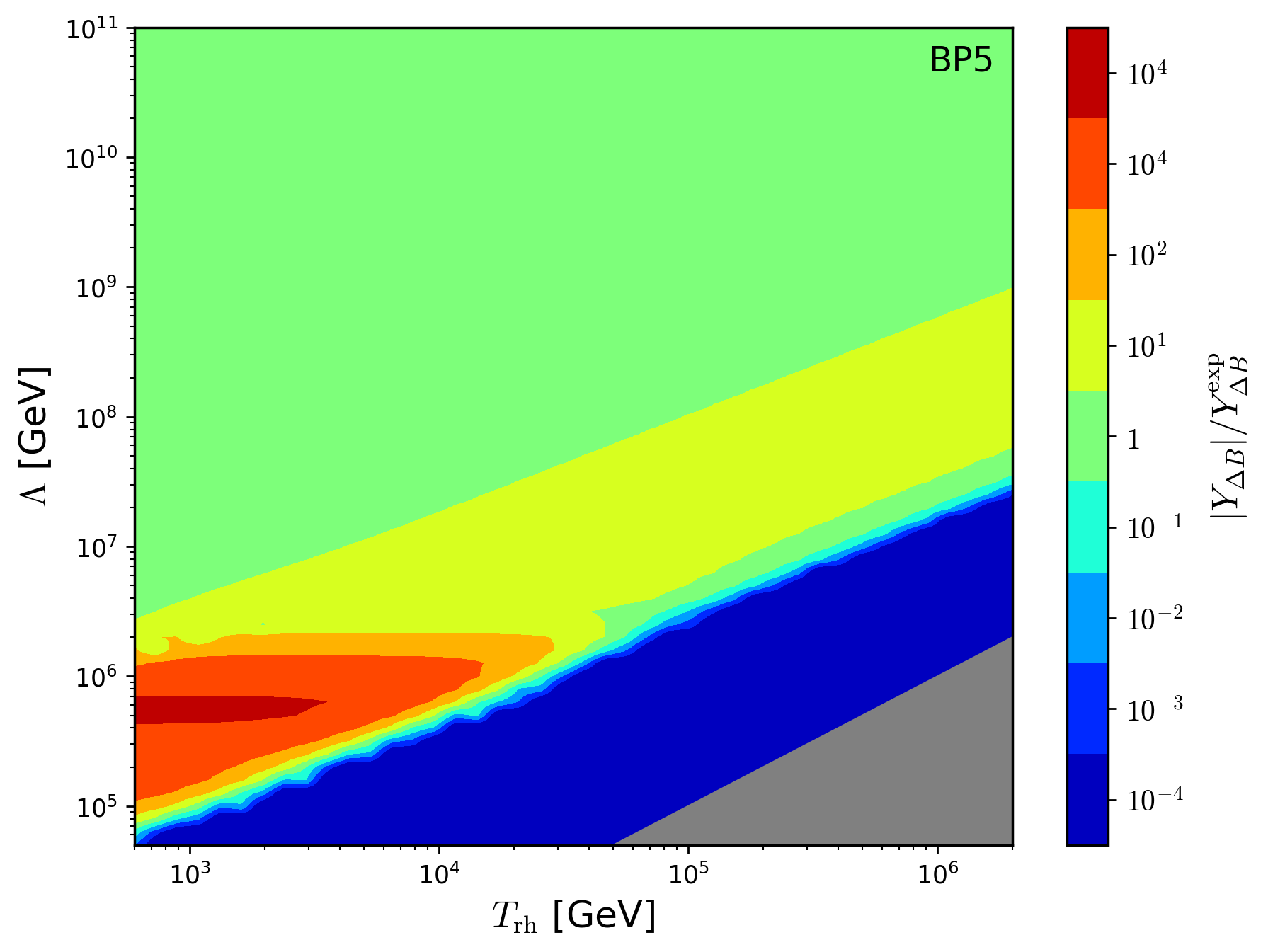}
    \end{subfigure}
    \caption{Resulting baryon yield normalized to the experimentally observed one $Y_{\Delta B}/Y^{\rm exp}_{\Delta B}$ (color) for different values of the reheating temperature $T_{\mathrm{rh}}$ and the dim-6 operator scale $\Lambda$, of the operator $\mathcal{O}_6$ in Eq.~\eqref{eq:our_dim6} with $G= F / \sqrt{\Tr(F^{\dagger}F)}$, for BP1 (left), BP3 (right) and BP5 (bottom) from Tab.~\ref{tab:BPs}. For future reference we also note that $\Delta M/M=10^{-8}$ for BP1, $\Delta M/M=\num{0.5e-6}$ for BP3 and $\Delta M/M=10^{-7}$ for BP5.}
    \label{fig:change_TRH}
\end{figure}

In Section~\ref{sec:numerical_solution} we observed that the resulting BAU depends on the reheating temperature $T_{\mathrm{rh}}$, see Fig.~\ref{fig:final_BP5}. To study this relationship we scan over $\Lambda$ and $T_{\mathrm{rh}}$ and show the resulting baryon asymmetry in Fig.~\ref{fig:change_TRH}. We choose BP1 as a representative BP for the oscillatory regime, BP3 as a representative BP in the overdamped regime and BP5 to demonstrate a point which can exhibit both an oscillatory and overdamped behavior depending on the values of $T_{\mathrm{rh}}$ and $\Lambda$.
The points which produce the experimentally observed BAU are colored in green, in red if the resulting BAU is larger and in blue if it is less. In the gray region the consistency requirement $T_{\mathrm{rh}} < \Lambda$ is not satisfied. For $\Lambda \to \infty$ the dim-6 operator decouples and the dynamics is mainly determined by the usual dim-4 Yukawa interaction. In this limit, and for sufficiently high reheating temperatures, the BAU is independent of the reheating temperature $T_{\mathrm{rh}}$, because the dim-4 interactions dominates in the IR. The baryon asymmetry starts to drop when the oscillation temperature approaches the reheating temperature, as then there is only a small time window for the RHNs to oscillate coherently, which can produce most of the asymmetry. For instance, BP1 features $T_{\mathrm{osc}}^{\mathrm{BP1}} \approx \SI{5e3}{\giga\electronvolt}$, BP3 $T_{\mathrm{osc}}^{\mathrm{BP3}} \approx \SI{4e3}{\giga\electronvolt}$ and BP5 $T_{\mathrm{osc}}^{\mathrm{BP5}} \approx \SI{5e2}{\giga\electronvolt}$ when $T_{\rm rh} \gg T_{\rm osc}$ and the BAU starts to drop when $T_{\rm rh}$ approaches $T_{\rm osc}$.

The dim-6 operator can enhance or reduce the baryon asymmetry, and the general behavior in Fig.~\ref{fig:change_TRH} is determined whether the evolution is in the oscillatory or in the overdamped regime. There  
is a linear relationship between $\Lambda$ and $T_{\mathrm{rh}}$ for BP1 
determined by $\overline{\Gamma}_6\propto M_P^{*}T_{\rm rh}^3/\Lambda^4$, which characterizes the production of RHNs which undergo the coherent oscillations in the oscillatory regime. In the overdamped regime, one RHN flavor reaches equilibrium early on, and the second flavor, mainly responsible for the BAU and corresponding to the slow mode, is produced via overdamped oscillations from the former around the oscillation timescale. This is similar to the production of sterile neutrinos in the Dodelson-Widrow mechanism. With increasing reheating temperature the first RHN reaches equilibrium earlier, however the second one will still be produced dominantly only at the oscillation timescale, which is largely independent of $T_{\rm rh}$. This explains why the resulting BAU is independent of the $T_{\rm rh}$ in the overdamped regime, i.e.~for BP3, at least unless $T_{\rm rh}$ is comparable to the oscillation timescale or below.
For BP5 (in contrast to BP1) there is a larger hierarchy between the overall equilibration timescale $z_{\rm eq}$ and the equilibration timescale of the slow mode $z_{\rm eq,slow}$ (see Fig.~\ref{fig:timescales_BP5}), such that --- due to the larger parameter range --- the behavior can either be oscillatory or overdamped depending on $(T_{\rm rh},\Lambda)$. Hence, BP5 shows a mixture of characteristic dependencies from both regimes.

Note also the different scales for BP1, BP3 and BP5: the enhancement for BP1 and BP5 can be multiple orders of magnitude while the enhancement is only about one order of magnitude for BP3. In all cases, for a fixed reheating temperature, one can find a value of $\Lambda$ which gives a maximum BAU (red region) and for smaller values of $\Lambda$ the asymmetry quickly reduces until it is exponentially suppressed (dark blue region). In the latter, the rate of the dim-6 operator is so large that the system exponentially reaches equilibrium and no asymmetry survives. This behavior was also illustrated in Fig~\ref{fig:final}. Since for BP1 the value of $\Lambda$, which gives the maximum BAU, depends on $T_{\rm rh}$ also the value $\Lambda_{\rm crit}$ below which the BAU becomes exponentially suppressed depends on $T_{\rm rh}$. For BP1 one can see that the baryon asymmetry is essentially zero for $\Lambda_{\rm crit} \lesssim \SI{1e6}{\giga\electronvolt}$ unless the reheating temperature is very low ($T_{\mathrm{rh}} \lesssim \SI{2e3}{\giga\electronvolt}$). For BP3 below $\Lambda_{\rm crit} \lesssim \SI{5e5}{\giga\electronvolt}$ no asymmetry survives, again with the exception of very low reheating temperatures. We see that only for BP5 a sizable BAU can exist at a scale of the order $\Lambda \sim \SI{1e5}{\giga\electronvolt}$ albeit for relatively small reheating temperatures. 

\subsection{Connecting leptogenesis and \texorpdfstring{$0\nu\beta\beta$}{0vbb} decay}
\begin{figure}[tb]
\centering
\includegraphics[width=7cm]{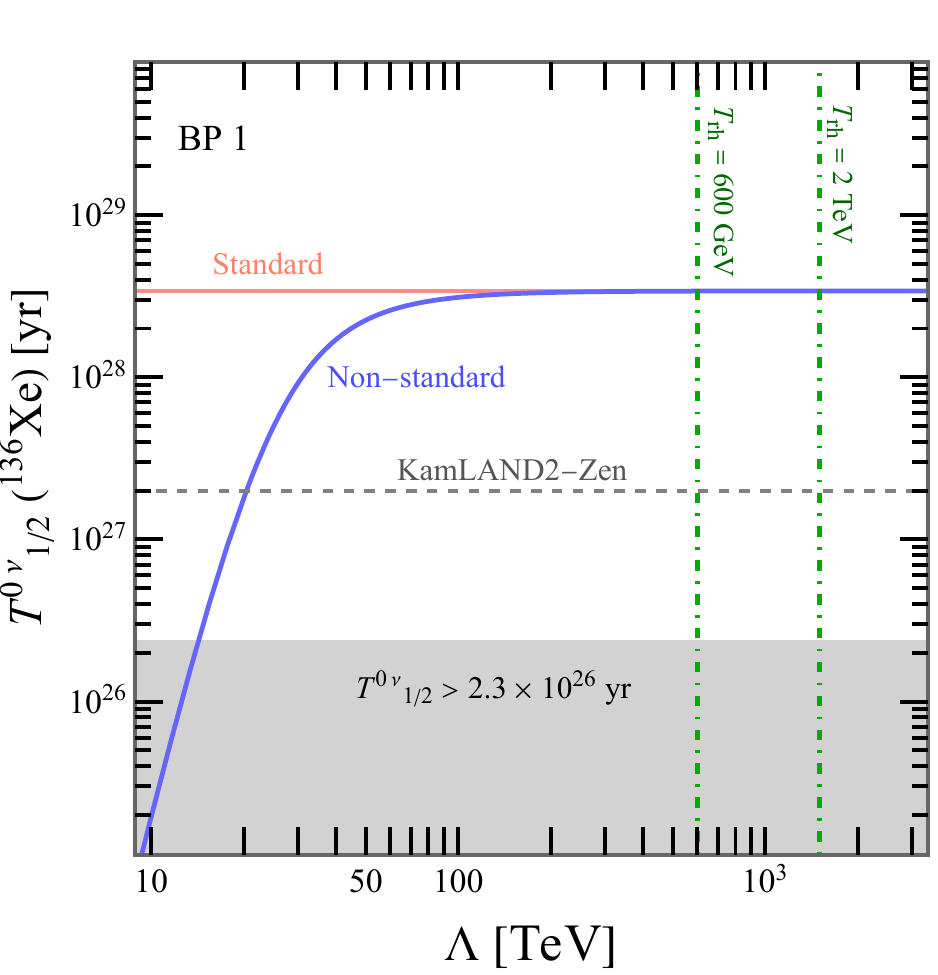}\hspace{0.5cm}
\includegraphics[width=7cm]{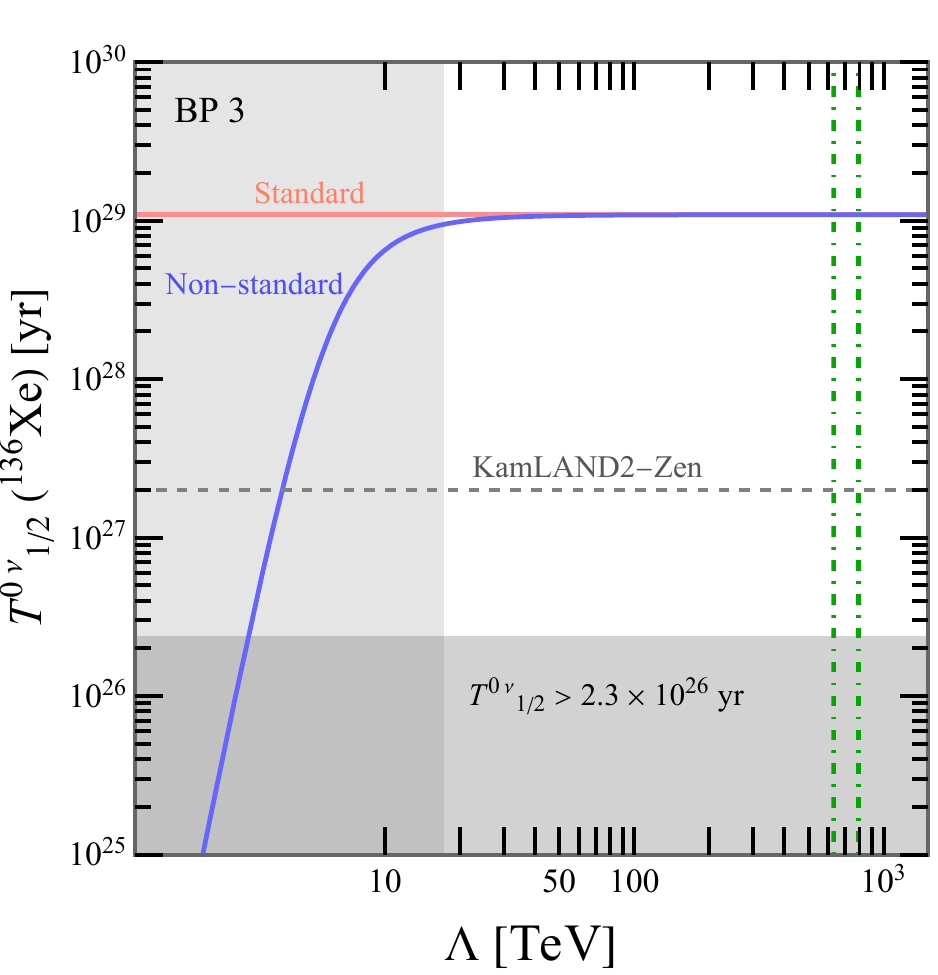}
\includegraphics[width=7cm]{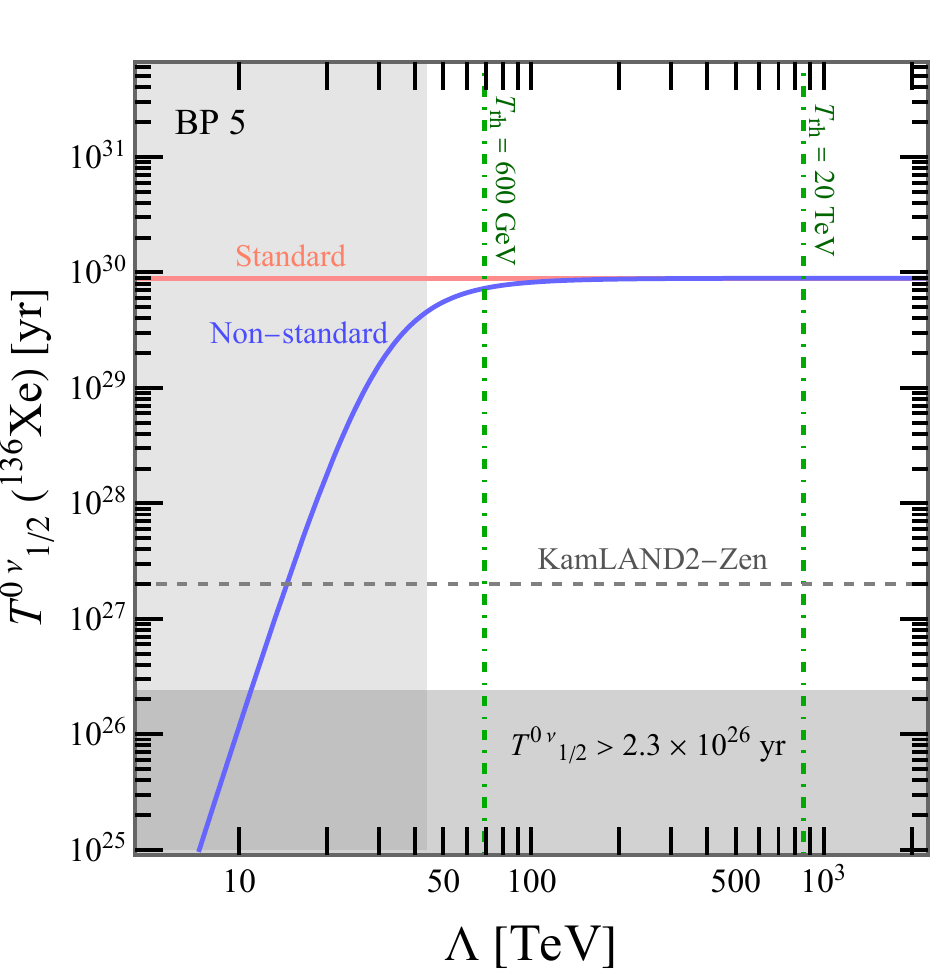}
\caption{$0\nu\beta\beta$ decay half-life for various benchmark points (BPs) as a function of the dim-6 operator scale $\Lambda$ of the operator $\mathcal{O}_6$ in Eq.~\eqref{eq:our_dim6} with $G= F / \sqrt{\Tr(F^{\dagger}F)}$. We show the contribution with (without) the dim-6 operator in blue (red) labeled ``non-standard" (``standard"). The excluded region is represented by the gray region, and the horizontal gray dashed line represents the expected sensitivity of $T^{0\nu}_{1/2}(^{136}{\rm Xe})=2.0\times 10^{27}~$yr at KamLAND2-Zen. The vertical green dash-dotted lines indicate the value of $\Lambda$, for different reheating temperatures {$T_{\mathrm{rh}}=600$ GeV~({\rm left}) and $20\,{\rm TeV}$ (right)}, at which the experimentally observed BAU can be reproduced. {The light gray region on the left represents the constraints from Tab.~\ref{tab:BP_bounds}.}}
\label{fig:DBD_halflife_lambda}
\end{figure}

In Section~\ref{sec:0vbb} we demonstrated that upcoming $0\nu\beta\beta$ decay experiments could potentially observe our BPs if the dim-6 operator scale is {{relatively low $\Lambda \sim {\cal O}(10)$ TeV}.} For BP1 and BP3, Fig.~\ref{fig:change_TRH} shows that for {such low} values of $\Lambda$ accessible at $0\nu\beta\beta$ decay experiments, the BAU is suppressed for reheating temperatures $T_{\rm rh} \gtrsim \SI{6e2}{\giga\electronvolt}$. The same is also true for BP2 and BP4. 
Compared to these BPs, BP5 can further go down to $\Lambda \sim \SI{1e5}{\giga\electronvolt}$, however, the energy scale is still somewhat high to see the dim-6 effect in $0\nu\beta\beta$ decay. Thus an observation of $0\nu\beta\beta$ decay mediated by the dim-6 operator in Eq.~\eqref{eq:our_dim6} and large reheating temperatures would imply that the ARS mechanism under consideration of the interactions in Eq.~\eqref{eq:lagrangian} is not able to produce the right BAU. To demonstrate this more explicitly, we present Fig.~\ref{fig:DBD_halflife_lambda} showing the $0\nu\beta\beta$ decay half-life as a function of $\Lambda$ together with the current exclusion limits (gray region) and targets of future experiments (gray dashed line). The green dash-dotted lines show the value of $\Lambda$ ({for two representative reheating temperature $T_{\mathrm{rh}}=600\,{\rm GeV}$ (left) and $20\,{\rm TeV}$ (right)}) for which the observed BAU can be explained in this model. {The region to the right (left) of each green dash-dotted line can successfully explain the correct BAU by increasing (decreasing) the reheating temperature $T_{\mathrm{rh}}$.} Note that the reheating temperature cannot be decreased arbitrarily since one needs EW sphalerons to convert the resulting lepton asymmetry to a baryon asymmetry. Within the current work we assume an instantaneous sphaleron shut-off such that they are only active above $T_{\mathrm{EW}} = \SI{131.7}{\giga\electronvolt}$. However, in reality this shut-off is a smooth process around this temperature. Similarly, we assume in this work an instantaneous reheating and neglect the short period of the broken phase of the EW theory. To reflect these assumptions we restrict ourselves therefore to reheating temperatures $T_{\mathrm{rh}}\geq\SI{600}{\giga\electronvolt}$ as a conservative bound.

In Fig.~\ref{fig:DBD_halflife_lambda}, the ``standard" line (red) is without the dim-6 operator and thus independent of $\Lambda$. The ``non-standard" lines (blue) show the resulting half-life with the inclusion of the dim-6 operator. In the limit $\Lambda \to \infty$ the dim-6 operator decouples and the blue line approaches the red one. The half-life can be suppressed or enhanced by several orders of magnitude depending on the scale $\Lambda$ (see Section~\ref{sec:0vbb}). {For BP3 and BP5, we also present the constraints given in Tab.~\ref{tab:BP_bounds} as light gray regions. The limit for BP1 is too weak to be visible.}

While the ``standard" case for all three BPs is well beyond experimental reach at KamLAND2-Zen experiment, the ``non-standard" case, i.e.\ in the presence of the dim-6 operator with a relatively low scale {$\Lambda \sim 10\, \si{\tera\electronvolt}$}  can be reached. Such a detection would have consequences for models explaining the BAU, including our set-up: {In all three BPs (BP1, 3, and 5), } 
the BAU cannot be explained through LG via RHN oscillation in the presence of the dim-6 operator $\mathcal{O}_6$, even for low reheating temperatures $T_{\mathrm{rh}} = \SI{600}{\giga\electronvolt}$, as indicated by the green dash-dotted lines. We expect a similar behavior for most of the RHN parameter space, given e.g.~by the parameters in the CI parametrization, as the rate of the dim-6 operator scales like $\expval{\Gamma_6} \sim T_{\mathrm{rh}}^5/\Lambda^4$, which is very large for small values of $\Lambda$ and high reheating temperatures. 
Note that the exclusion of most of the parameter space remains true if one includes other $\Tilde{L}$NC interactions like gauge scatterings or $1\leftrightarrow2$ scatterings, as they can only enhance the rate and drive the system closer to equilibrium. Their effect can change the size and position of the enhancement, which we leave for future work. Furthermore, similar conclusions should apply for other higher-dimensional operators or in specific UV completions as long as they do not have any additional $\Tilde{L}$NV interactions or an asymmetry generation mechanism of their own. Asymmetry generation can happen e.g.\ for the RHN self-interactions in \cite{Astros:2024yee}, for which the Lagrangian is $\Tilde{L}$NV, even at high temperatures.
 
\begin{figure}[tb]
\centering
\includegraphics[width=7cm]{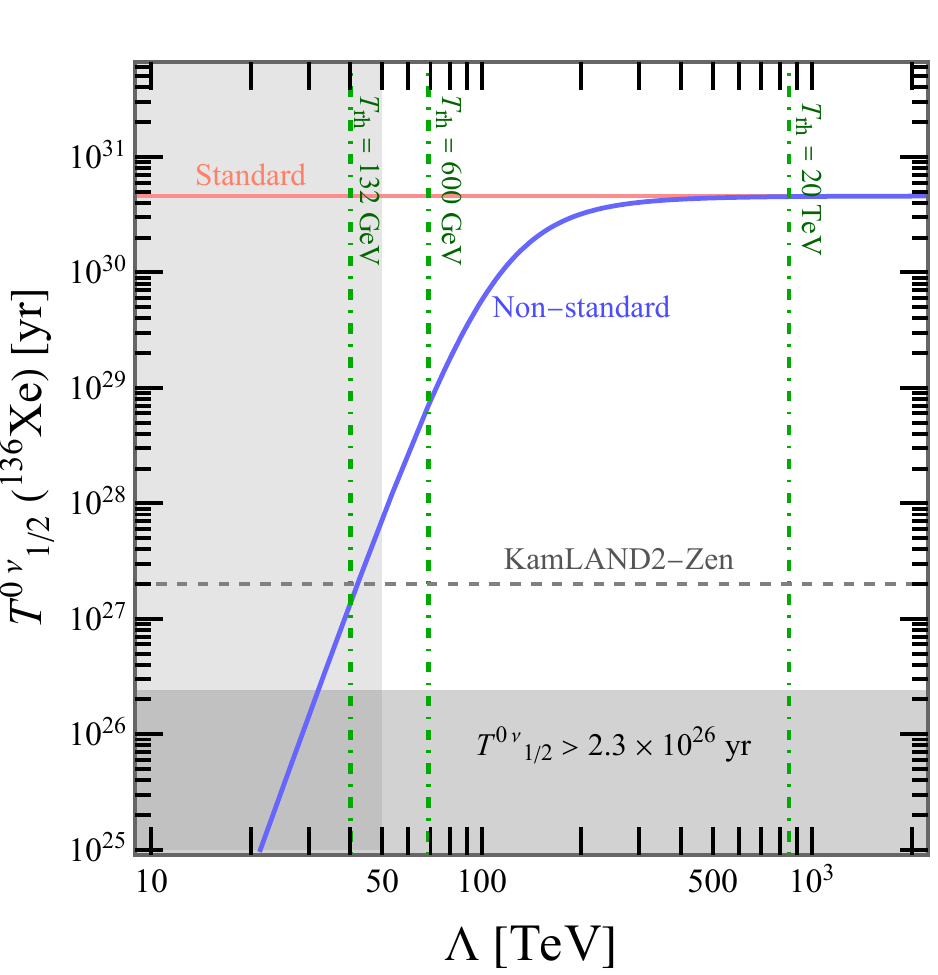}
\caption{$0\nu\beta\beta$ decay half-life as a function of $\Lambda$. We use $M=100 \,{\rm MeV},~\Delta M/M=10^{-6},~\alpha_{31}=3\pi, ~{\rm Re}(\omega)=3\pi/4, ~{\rm Im}(\omega)=0$. The vertical green dash-dotted lines indicate the value of $\Lambda$, for different reheating temperatures $T_{\mathrm{rh}}= 132\, {\rm GeV~(left)},~600$ GeV~({\rm middle}) and $20\,{\rm TeV}$ (right), at which the observed BAU can be reproduced. The vertical light gray region represents the constraint from low-energy laboratory experiments obtained by a recast of Ref.~\cite{Fernandez-Martinez:2023phj}.}
\label{fig:DBD_halflife_lambda_New}
\end{figure}

In order to demonstrate that there is still a possibility to have successful LG and an observation of $0\nu\beta\beta$ decay at the next-generation experiments, we consider another BP ($M=100 \,{\rm MeV},~\Delta M/M=10^{-6},~\alpha_{31}=3\pi, ~{\rm Re}(\omega)=3\pi/4, ~{\rm Im}(\omega)=0$) with a much lower reheating temperature, $T_{\rm rh}=132$ GeV.
If we take such a low reheating temperature, the period between reheating and the sphaleron shut-off is short. Therefore, the analysis indeed requires to include not only the effects from non-instantaneous reheating and non-instantaneous sphaleron shut-off but also an accurate determination of the scattering rates including effects from the broken phase of the EW theory. 
Keeping this caveat in mind, we confirm that this BP with a low reheating temperature can work for producing the observed BAU with lower $\Lambda$. Fig.~\ref{fig:DBD_halflife_lambda_New} shows that the predicted half-life can potentially be within the experimental reach at the next-generation $0\nu\beta\beta$ search, if we take $T_{\rm rh}=132$ GeV. It should be noted that this new BP is similarly subject to constraints from low-energy laboratory experiments \cite{Fernandez-Martinez:2023phj}. By recasting these results, we find that $\Lambda > 50.1$ TeV, which is shown as a light gray region in Fig.~\ref{fig:DBD_halflife_lambda_New}. 
In order to make a conclusive statement on whether the generation of the BAU is possible or impossible in the whole parameter space of RHNs for low $T_{\rm rh}$ and for values of $\Lambda$ in reach of $0\nu\beta\beta$ decay while not being excluded by low-energy constraints, it is required to perform a more detailed investigation, which we defer to a future publication, explicitly taking into account the aforementioned effects. 

Our findings suggest that if $0\nu\beta\beta$ decay is observed in next generation experiments, like KamLAND2-Zen, most of the RHN parameter space cannot explain the BAU through LG via RHN oscillations in the presence of the dim-6 operator $\mathcal{O}_6$ in Eq.~\eqref{eq:our_dim6}. If one assumes that the BAU is generated in this way, this would point towards 
a low reheating temperature $T_{\mathrm{rh}}$.

\subsection{Reducing the mass degeneracy}
\begin{figure}[tb]
    \begin{subfigure}[b]{0.5\textwidth}
        \includegraphics[width=\textwidth]{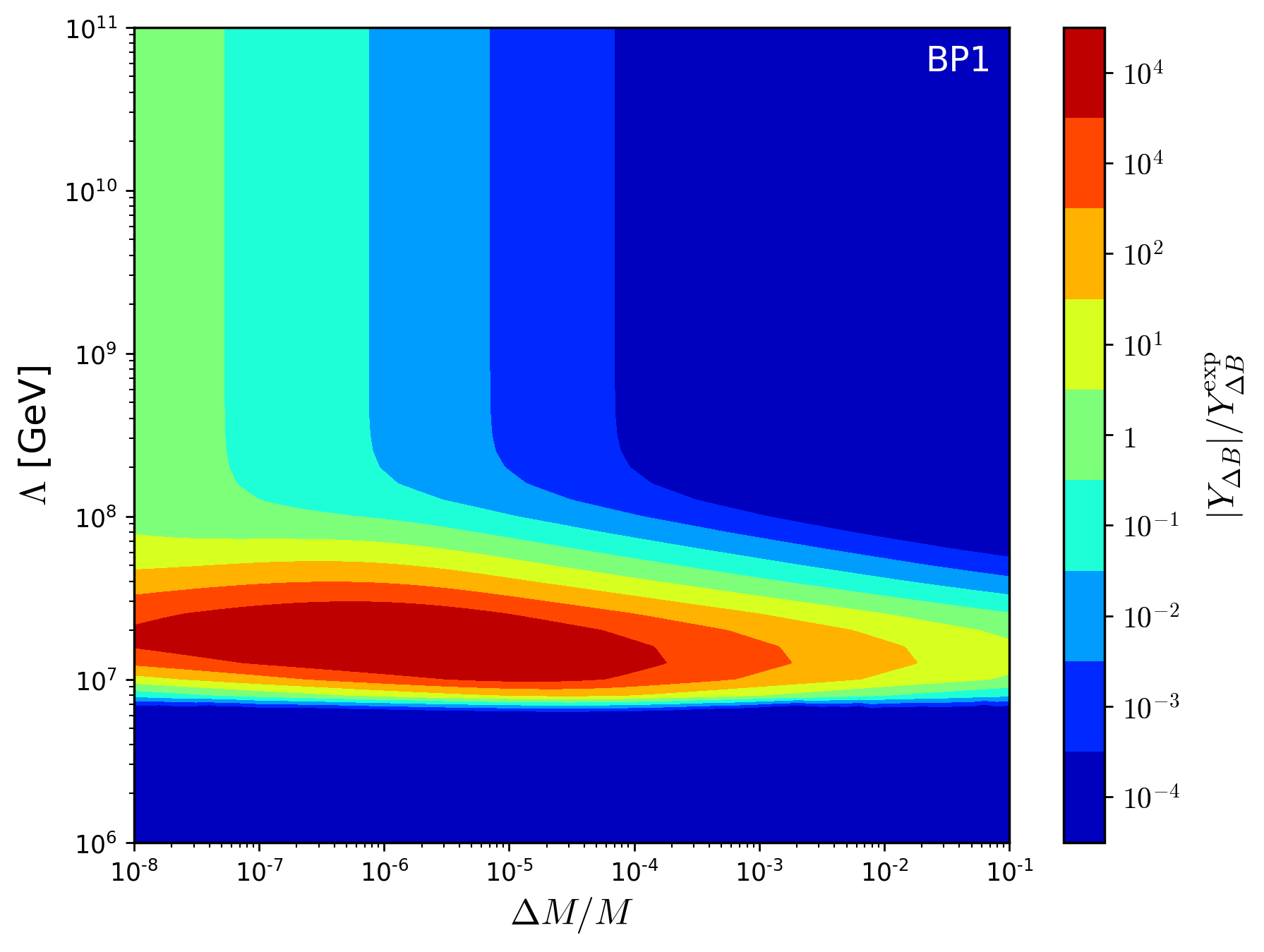}
    \end{subfigure}
    \hfill
    \begin{subfigure}[b]{0.5\textwidth}
        \includegraphics[width=\textwidth]{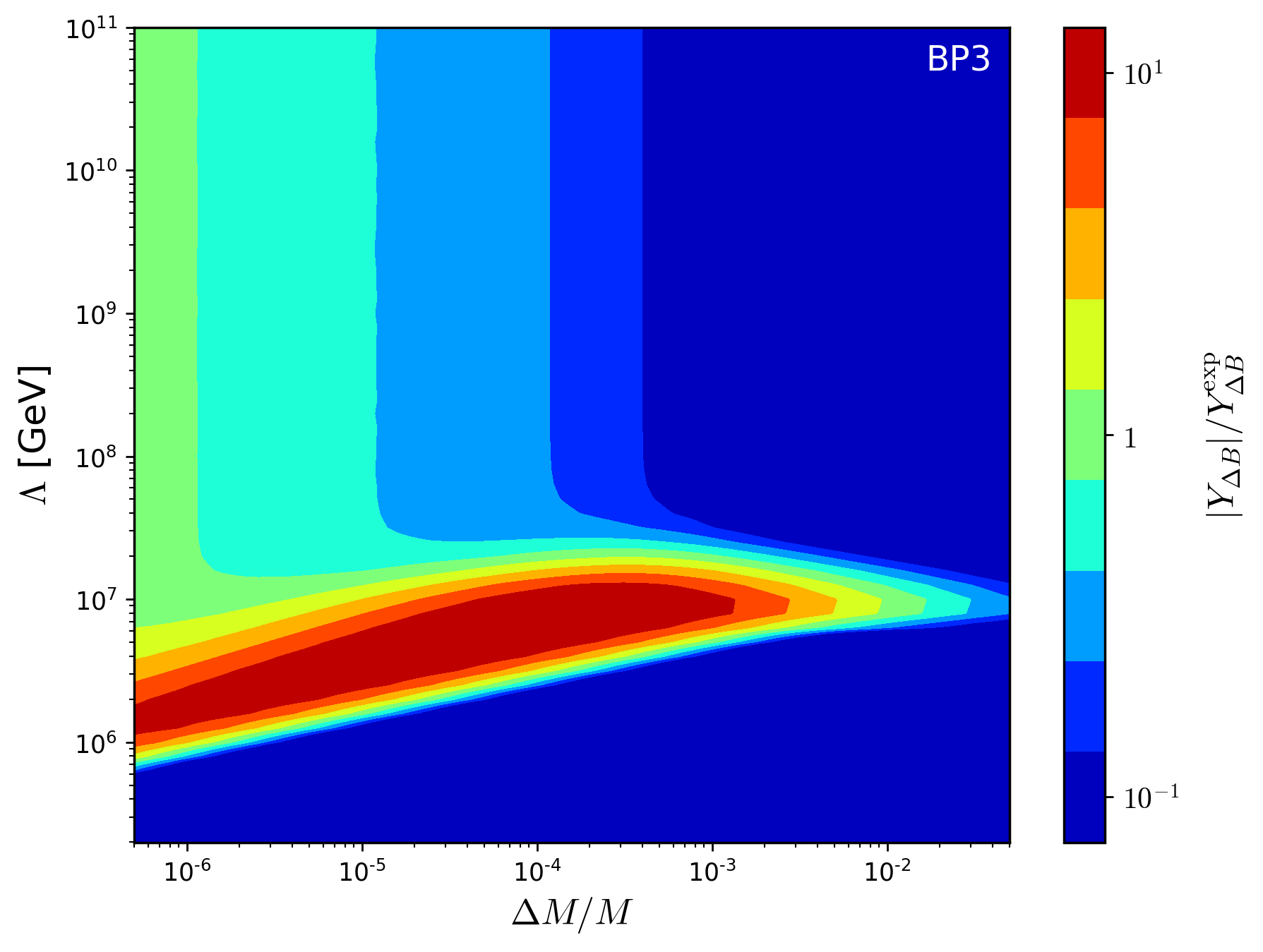}
    \end{subfigure}
    \caption{Resulting baryon yield normalized to the experimentally observed one (color) for different values of the mass degeneracy $\Delta M / M$ and dim-6 scale $\Lambda$, of the operator $\mathcal{O}_6$ in Eq.~\eqref{eq:our_dim6} with $G= F / \sqrt{\Tr(F^{\dagger}F)}$, for BP1 (left) and BP3 (right) from Tab.~\ref{tab:BPs}. We fix $T_{\mathrm{rh}}=\SI{2e4}{\giga\electronvolt}$. See text for details.}
    \label{fig:change_dM}
\end{figure}

Enhancing the BAU allows one to relax constraints on other parameters, e.g.\ one can reduce the mass degeneracy, i.e.~increasing the mass splitting $\Delta M$. Typically the mass splittings are chosen to be small, $\Delta M / M = \num{1e-8}$ for BP1 and $\Delta M / M = \num{5e-7}$ for BP3, to have small oscillation timescales $z_{\rm osc}$ and achieve the correct BAU. Increasing these mass splittings leads to a smaller BAU in the usual ARS case, but can result in the observed value with the inclusion of the interactions of the dim-6 operator. In Fig.~\ref{fig:change_dM} we show the resulting BAU for different values of $\Delta M$ and $\Lambda$. We fix the reheating temperature to $T_{\mathrm{rh}}=\SI{2e4}{\giga\electronvolt}$ such that $T_{\mathrm{osc}} < T_{\mathrm{rh}}$ for the smallest mass splittings in Fig.~\ref{fig:change_dM}.
Again, the observed BAU is reproduced for the parameter points colored in green, red corresponds to a larger BAU and blue to a smaller one. The limit $\Lambda\to\infty$ is the standard ARS case without the dim-6 operator. In this limit only small $\Delta M/M$ reproduces the correct BAU, while for increasing $\Delta M / M$ the asymmetry reduces. For both BPs the dim-6 operator can enhance the baryon asymmetry even for $\Delta M / M$ much larger than the initial BP values. For BP1 this happens around $\Lambda \approx \SI{2e7}{\giga\electronvolt}$ independent of the mass splitting. In this case one can change $\Delta M / M = \num{1e-8} $ to $\Delta M / M \approx \num{1e-1}$ to reproduce the correct baryon asymmetry. For BP3 the enhancement is maximal initially for $\Lambda \approx \SI{1.5e6}{\giga\electronvolt}$ and this value of $\Lambda$ increases with increasing $\Delta M / M$. Again one can change $\Delta M / M = \num{5e-7}$ to $\Delta M / M \approx \num{1e-2}$ to obtain the observed BAU.
In both cases, once $\Lambda$ is too low, the equilibration rate becomes too large leading to a reduced BAU. 
For a given set of parameters, larger mass splittings $\Delta M / M$ could potentially lead to new phenomenological features, like LNV signals at colliders due to RHN oscillations~\cite{Antusch:2017ebe} or allow for an increased range of parameter space in the context of an a$\Bar{L}$N symmetry in linear~\cite{Akhmedov:1995ip,Malinsky:2005bi} and inverse seesaw models~\cite{Mohapatra:1986aw,Mohapatra:1986bd} (see e.g.~\cite{Antusch:2017ebe,Drewes:2019byd,Fernandez-Martinez:2022gsu} for discussions in this context).\\

We showed in this section, that the dim-6 $\Tilde{L}$NC operator can enhance and reduce the resulting BAU significantly. 
Our result indicates that, if $0\nu\beta\beta$ decay is observed at next-generation searches, it could imply a presence of the dim-6 operator in Eq.~\eqref{eq:our_dim6} with relatively low scale $\Lambda$. However, such scales would not explain the observed BAU via low-scale LG within the $\nu$SMEFT. Such a situation would point towards low reheating temperature and small RHN masses if the BAU is generated via RHN oscillations including the dim-6 operator $\mathcal{O}_6$ in Eq.~\eqref{eq:our_dim6}. 
Furthermore, the mass degeneracy between the RHNs can be alleviated significantly via the inclusion of the additional interactions of the dim-6 $\Tilde{L}$NC operator.

\section{Conclusion}\label{sec:conclusion}

In this work, we have demonstrated that the dim-6, $\Tilde{L}$NC operator $\mathcal{O}_6 =(\overline{L_{\alpha}}\nu_{R,I})(\overline{u_R}Q)/\Lambda^2$ within the $\nu$SMEFT framework can have significant contributions to both low-scale LG and $0\nu\beta\beta$ decay. Although this operator preserves effective lepton number $\Tilde{L}$ and does not induce neutrino masses at tree-level, it modifies the RHN equilibration dynamics in the early Universe and enhances the $0\nu\beta\beta$ decay amplitude at low energies.

As discussed in \cite{Dekens:2020ttz}, we observe that the operator $\mathcal{O}_6$ can dominate $0\nu\beta\beta$ decay process over the dim-4 interactions, leading to shorter half-lives by several orders of magnitude. In particular, for RHN masses in the MeV-\si{\giga\electronvolt} range, next-generation $0\nu\beta\beta$ decay searches such as KamLAND2-Zen and LEGEND have great potential to probe the parameter space in the presence of the dim-6 operator. We focused on the case $G\propto F$, which greatly reduces the number of free parameters and allows for an approximate lepton number conservation in the system, which makes RHN parameter values with large active-sterile mixing angles and small active neutrino masses technically natural. However, we find that for parameter points obeying an approximate symmetry, cancellations in the rate stemming from the dim-6 operator can be present, suppressing $0\nu\beta\beta$ decay processes.

We presented the first detailed analysis of the effect of $\mathcal{O}_6$ on low-scale LG. We derived the relevant QKEs, incorporating the additional scattering processes mediated by this operator. From this, we showed how to estimate the resulting dynamics of the system analytically and computed the final BAU for several representative BPs numerically. Depending on the EFT cutoff scale $\Lambda$, the operator can both suppress or enhance the BAU by several orders of magnitude. We highlight that the enhancement of the BAU is possible even without introducing additional sources of lepton number violation. Furthermore, the operator $\mathcal{O}_6$ leads to UV freeze-in, i.e.\ it is highly active at large temperatures, making the resulting dynamics sensitive to the highest temperatures in the thermal history, particularly the reheating temperature. Explicitly, we showed the dependence of the BAU on the reheating temperatures and the cutoff scale $\Lambda$ for different benchmark points. While we focused on the case $G\propto F$, a similar behavior is expected also for other choices; in particular the regions where the BAU is exponentially suppressed should be the same for most flavor structures.

We further explored the interplay between our LG results and the $0\nu\beta\beta$ decay predictions. A potential detection of $0\nu\beta\beta$ decay, due to the dim-6 operator $\mathcal{O}_6$ considered in this work at scales corresponding to $\Lambda \sim \SI{10}{\tera\electronvolt}$, would imply large RHN equilibration rates, rendering low-scale LG scenarios with this dim-6 operator in significant tension with the observed BAU. 
This would effectively exclude large regions of the parameter space and in this case favor models with very low reheating temperatures. However, for a final statement on this, the inclusion of non-instantaneous reheating and non-instantaneous sphaleron shut-off as well as an accurate determination of the scattering rates including effects from the broken phase of the EW theory in future studies is crucial.
Moreover, as discussed in \cite{Fernandez-Martinez:2023phj}, the $\nu$SMEFT operators can also contribute to other low-energy observables. We find that the constraints point to a high scale $\Lambda$, leading to longer $0\nu\beta\beta$ half-lives that would be inaccessible to next-generation experiments. Their current estimation is based on a simplified rescaling procedure, therefore, a more detailed analysis would be needed to derive constraints relevant to our benchmark points.
Finally, we showed that in certain regions, the enhanced production of RHNs via $\mathcal{O}_6$ can alleviate the tight degeneracy requirements on RHN masses typically needed for successful low-scale LG.
Our EFT analysis shows that it is challenging to have a positive $0\nu\beta\beta$ decay signal at future experiments in agreement with other low-energy constraints and successful leptogenesis, in the presence of the studied dim-6 operator. A follow-up analysis investigating the multi-dimensional parameter space and the limitations of the EFT approach will allow us to give a conclusive statement.

While our analysis has focused on the single operator $\mathcal{O}_6$, the general framework and methodology are readily extendable to other ($\nu$)SMEFT operators. Furthermore, to derive QKEs which are simple to handle analytically and numerically we made some simplifying assumptions in Section~\ref{sec:LG} and focused on a few BPs. To make reliable predictions of the full available parameter space one can improve on these assumptions, for example by including additional terms in the equilibration rate coming from gauge scattering and $1\leftrightarrow2$ processes or studying the momentum dependent QKEs. Moreover, UV-complete models that naturally generate such operators and realize the observed BAU enhancement would be particularly illuminating.

Future studies including complementary low-energy observables will be crucial for sharpening these connections and further probe the nature of the physics responsible for both neutrino masses and the matter-antimatter asymmetry of the Universe.

\acknowledgments

We thank T.~Asaka, M.~Drewes, Y.~Georis, S.~Eijima, J.~Gargalionis, H.~Ishida, E.~Mereghetti, and S.~Sandner for valuable discussions.
We are grateful to the Mainz Institute for Theoretical Physics (MITP) of the Cluster of Excellence PRISMA+ (Project ID 390831469) and the workshops ``What's the Matter? A Cross-Frontier Pursuit of the Origin of Matter", and ``From the Cosmos to the Lab: Novel Links and Strategies", for its hospitality and support.
K.~F.~was supported by the US Department of Energy Office and by the Laboratory Directed Research and Development (LDRD) program of Los Alamos National Laboratory under project numbers 20210190ER, 20250164ER, and 20240078DR. Los Alamos National Laboratory is operated by Triad National Security, LLC, for the National Nuclear Security Administration of the U.S. Department of Energy (Contract No. 89233218CNA000001). K.~F. also acknowledges the support by the iTHEMS–KEK Theory Center Joint Research Program (under the Joint Research Agreement between RIKEN and KEK).  J.~H.~and S.~W.~acknowledge support by the Cluster of Excellence ``Precision Physics, Fundamental Interactions, and Structure of Matter" (PRISMA+ EXC 2118/1 and PRISMA++ EXC 2118/2) funded by the Deutsche Forschungsgemeinschaft (DFG, German Research Foundation) within the German Excellence Strategy (Project No. 390831469).

\appendix

\section{Mass diagonalization and Yukawa coupling parametrization}\label{app:yuakawa}

In the following, we review the diagonalization of the neutrino mass matrix and the parametrization used for the Yukawa couplings. We will only work with the renormalizable interactions in Eq.~\eqref{eq:lagrangian}. After EWSB the mass terms in the neutrino sector are given by
\begin{align}
    {\cal L}_m=-\frac{1}{2}\overline{N^c}M_{\nu}N
+\hc,\hspace{1cm}
M_{\nu}=
\begin{pmatrix}
    0 & M_D^*\\
    M_D^{\dagger} & M_R^{\dagger}
\end{pmatrix},
\end{align}
where $N=(\nu_{L\alpha},\nu_{RI}^c)^T$, $M_R = M$ and $M_D=vF$ with $v\simeq 174~$GeV.
The mass matrix can be diagonalized exactly even without the seesaw approximation $M_D \ll M_R$. The diagonalization is obtained by a $5\times 5$ unitary matrix $U$ as
\begin{align}\label{eq:neutrino_mass_diagonalization}
    {\cal L}_m=-\frac{1}{2}\overline{N^c}U^*\left(U^TM_{\nu}U\right)U^{\dagger}N+\hc = \frac{1}{2}\bar{\nu}M_{\nu}^{\rm diag}\nu,
\end{align}
with the Majorana mass eigenstates $\nu=N_m+N_m^c$, $N_m=U^{\dagger}N$ and
\begin{align}
    M^{\rm diag}_{\nu}={\rm diag}(m_1,m_2,m_3,m_4,m_5)\,.
\end{align}
The unitary matrix is described by \cite{Donini:2012tt, Huang:2013kma, Hernandez:2014fha}
\begin{align}
    U=
    \begin{pmatrix}
        U_{aa} & U_{as}\\
        U_{sa} & U_{ss}
    \end{pmatrix},
\end{align}
where for normal ordering (NO) and for a diagonal RHN mass matrix $M_R = \diag(M_1,M_2)$ one can parametrize the entries via \cite{Donini:2012tt, Huang:2013kma, Hernandez:2014fha}
\begin{align}
    U_{aa}&=U_{\rm PMNS}
    \begin{pmatrix}
        1 & 0 \\
        0 & H
    \end{pmatrix},&
    U_{as}&=iU_{\rm PMNS}
    \begin{pmatrix}
        0 \\
        Hm_{\ell}^{1/2}R^{\dagger}M_h^{-1/2}
    \end{pmatrix},\\
    U_{sa}&=i
    \begin{pmatrix}
        0 & \overline{H}M_{h}^{-1/2}Rm_{\ell}^{1/2}
    \end{pmatrix},&
    U_{ss}&=\overline{H},
\end{align}
with the Pontecorvo–Maki–Nakagawa–Sakata (PMNS) matrix $U_{\rm PMNS}$ and the $2\times 2$ mass matrices $m_{\ell}$ and $M_h$
\begin{align}
    m_{\ell}=
    \begin{pmatrix}
        m_2 & 0 \\
        0 & m_3
    \end{pmatrix},\hspace{1cm}
    M_{h}=
    \begin{pmatrix}
        m_4 & 0 \\
        0 & m_5
    \end{pmatrix}.
\end{align}
$R$ is a $2\times 2$ rotation matrix
\begin{align}
R&=\begin{pmatrix}
    \cos\left(\theta_{45}+i\gamma_{45}\right) &\sin\left(\theta_{45}+i\gamma_{45}\right)\\
    -\sin\left(\theta_{45}+i\gamma_{45}\right) & \cos\left(\theta_{45}+i\gamma_{45}\right)
\end{pmatrix}
\end{align}
and the $H$ and $\overline{H}$ matrices are described by
\begin{align}
    H&= \left[I+m^{1/2}_{\ell}R^{\dagger}M_h^{-1}Rm^{1/2}_{\ell} \right]^{-\frac{1}{2}},\hspace{1cm}
    \overline{H}=\left[I+M_h^{-1/2}Rm_{\ell}R^{\dagger}M_h^{-1/2} \right]^{-\frac{1}{2}}.
\end{align}

Using the relation $U_{as}M_h=M_D$ this parametrization of the unitary matrix $U$ directly leads to a parametrization of the Yukawa coupling. This parametrization agrees with the Casas-Ibarra (CI) parametrization \cite{Casas:2001sr} in the seesaw limit, as we show in the following. Taking the leading order in the expansion of $m_{\ell}/M_h$, one can see $H\simeq \overline{H}\simeq 1$ and one obtains \cite{Fuyuto:2024oii}
\begin{align}
    U\simeq
    \begin{pmatrix}
        U_{\mathrm{PMNS}} & \theta \\
        -\theta^{\dagger} U_{\mathrm{PMNS}} & 1
    \end{pmatrix},
\end{align}
where the active-sterile mixing matrix is 
\begin{equation}
    U_{as} \simeq \theta = i U_{\mathrm{PMNS}} (m_{\nu}^{\diag})^{1/2} {\cal R}^{\dagger} M_h^{-1/2} \, ,
\end{equation}
with the $3\times 3$ matrix of active neutrino masses $(m_{\nu}^{\diag})^{1/2} = \diag(m_1=0,m_2,m_3)$ and
\begin{align}
    {\cal R}^{\dagger}=
    \begin{pmatrix}
        0 & 0\\
        \cos\left(\omega\right) &-\sin\left(\omega\right)\\
    \sin\left(\omega\right) & \cos\left(\omega\right)
    \end{pmatrix},
\end{align}
with $\omega = \theta_{45}-i\gamma_{45}$. From the relation $U_{as}M_h=M_D$ one obtains the usual Casas-Ibarra parametrization 
\begin{align}
F_{\alpha I}=\frac{i}{v}U_{\rm PMNS}~(m_{\nu}^{\diag})^{1/2}{\cal R}^{\dagger}M_h^{1/2}\,.
\end{align}

The PMNS matrix and active neutrino masses are largely determined by neutrino oscillation experiments. The PMNS matrix reads \cite{ParticleDataGroup:2024cfk}
\begin{equation}
    U_{\nu} = \begin{pmatrix}
        1 & 0 & 0 \\
        0 & c_{23} & s_{23} \\
        0 & -s_{23} & c_{23}
    \end{pmatrix} \begin{pmatrix}
        c_{13} & 0 & s_{13} e^{-i\delta} \\
        0 & 1 & 0 \\
        -s_{13} e^{i\delta}& 0 & c_{13}
    \end{pmatrix} \begin{pmatrix}
        c_{12} & s_{12} & 0 \\
        -s_{12} & c_{12} & 0 \\
        0 & 0 & 1
    \end{pmatrix} \begin{pmatrix}
        1 & 0 & 0 \\
        0 & e^{i\alpha_{21}/2} & 0 \\
        0 & 0 & e^{i\alpha_{31}/2}\,,
    \end{pmatrix}
\end{equation}
where $c_{ij} = \cos \theta_{ij}, s_{ij} = \sin \theta_{ij}$ with the mixing angle $\theta_{ij}$, $\delta$ is the Dirac phase and $\alpha_{21,31}\in [0,4\pi]$ two Majorana phases.\footnote{We follow \cite{Granelli:2023vcm}, and explicitly chose a fixed orthogonal matrix $\mathcal{R}$, without any additional phase $\phi=\pm 1$. To allow for this degree of freedom one can extend the range of the Majorana phases from $[0,2\pi]$ to $\alpha_{21,31}\in[0,4\pi]$. Moreover for two RHNs and normal ordering (NO) we can set $\alpha_{21}=0$.} For two RHNs one of the active neutrinos will be massless and the other two masses are directly given by the neutrino oscillation data
\begin{equation}
    m_{\nu}^{\diag} = \diag\left(m_1 = 0, m_2 = \sqrt{\Delta m_{\mathrm{sol}}^2}, m_3 = \sqrt{\Delta m_{\mathrm{atm}}^2}\right) \qquad (\mathrm{NO}) .
\end{equation}
We will for simplicity always assume the normal ordering (NO) for the active neutrino masses in this work, the results for inverted ordering (IO) are expected to be similar. We use the latest \textit{NuFit-6.0}\footnote{See also \href{http://www.nu-fit.org/}{NuFIT 6.0 (2024), www.nu-fit.org}.} \cite{Esteban:2024eli} data for the oscillation parameters. We will slightly approximate the Dirac phase $\delta = \frac{59}{60}\pi$ for simplicity and set $\delta = \pi$. In Table~\ref{tab:oscillation_data} we show for convenience the values we use in this work.
\begin{table}[t]
    \centering
    \begin{tabular}{|cccccc|}
    \hline
        $\sin^2 \theta_{12}$ & $\sin^2 \theta_{23}$ & $\sin^2 \theta_{13}$ & $\delta$ & $\Delta m_{\mathrm{sol}}^2$ [$\si{\electronvolt}^2$] & $\Delta m_{\mathrm{atm}}^2$ [$\si{\electronvolt}^2$] \\
        \hline
        0.307 & 0.561 & 0.02195 & $\pi$ & \num{7.49e-5} & \num{2.534e-3}\\
        \hline
    \end{tabular}
    \caption{Neutrino oscillation data used in this work. Values taken from \cite{Esteban:2024eli}.}
    \label{tab:oscillation_data}
\end{table}
To quantify the overall active-sterile mixing one introduces 
\begin{equation}
    U^2 = \sum_{\alpha,I} (U_{as})_{\alpha I} (U_{as})_{\alpha I}^{*} = \Tr(U_{as} U_{as}^{\dagger}) \simeq \Tr(\theta^{\dagger}\theta) .
\end{equation}
which in the CI parametrization reads \cite{Drewes:2016gmt}
\begin{equation}
    U^2 \simeq \frac{\Delta M}{2 M_1 M_2} (m_2 - m_3) \cos(2\Re \omega) + \frac{M_1 + M_2}{2 M_1 M_2} (m_2 + m_3) \cosh(2\Im \omega ) \quad (\mathrm{NO})\,,
\end{equation}
For small mass splittings $U^2$ is almost entirely given by the second term, and since $\cosh(x) \geq 1$, we obtain the condition
\begin{equation}
     U^2 \gtrsim \frac{\sum_i m_i}{M} ,
\end{equation}
which is referred to as the seesaw line (see also Fig.~\ref{fig:BPs}). Another useful relation for small mass splittings reads
\begin{equation}
    \Tr(F^{\dagger} F) \simeq \frac{M^2}{v^2} U^2 .
\end{equation}
There are two interesting limits of the CI parametrization \cite{Drewes:2016gmt, Hernandez:2022ivz}. First, the \textit{naive seesaw} limit $\Im \omega \to 0$ which is characterized by small $U^2$, lying directly on the seesaw line. Second, the $\bar{L}$NC limit $\abs{\Im \omega} \gg 1$, for which larger values of $U^2$ can be achieved. Further note that we can maximize $\Re \omega$ to give the largest BAU with the analytic expressions given in \cite{Drewes:2016gmt,Hernandez:2022ivz}. This is achieved for $\abs{\sin(2\Re\omega)}=1$ or $\Re\omega \in \{\pi/4,3\pi/4\}$.

\section{Derivation of the Quantum Kinetic Equations}\label{app:rates}

One can calculate the matrix elements $\mathcal{M}$ for both the dim-4 quark scatterings as well as the scatterings mediated by the dim-6 operator using standard QFT techniques. The total matrix element is the sum of both contributions
\begin{equation}
    \mathcal{M}_{i} = \mathcal{M}_{i,\operatorname{dim-4}} + \mathcal{M}_{i,\operatorname{dim-6}} \ ,
\end{equation}
with $i\in\{A,B,C\}$ for each of the diagrams in Fig.~\ref{fig:quark_scatterings}. From this one can calculate the destruction and production rates $\Gamma^{d,p}$ for the RHNs $N$ and SM leptons $L$ via Eq.~\eqref{eq:destruction_rates_formula}. It is useful to explicitly factor out the flavor dependence of the rate. Since $\mathcal{M}_{i,\operatorname{dim-4}} \sim F$, $\mathcal{M}_{i,\operatorname{dim-6}} \sim G$ and $\Gamma \sim \abs{\mathcal{M}}^2$ one also obtains an interference term. One finds e.g
\begin{equation}
    \Gamma^{d}_{N,A} = \sum_{X,Y,\in \{F,G\}} \gamma_{N,A}^{XY} \left[X^{\dagger}Y\right] \ .
\end{equation}
Other rates can additionally involve the RHN or SM lepton density matrix $\rho_N$ and $\rho_L$ \cite{Asaka:2011wq} or --- after taking the thermal average 
\begin{equation}\label{eq:thermal_average}
   \expval{\gamma} = \frac{\int \mathrm{d}\vec{k} \, f^{\eq}(k) \gamma(k)}{\int \mathrm{d}\vec{k} \, f^{\eq}(k)}  \ ,
\end{equation}
with $f^{\eq}(k) = \exp(-E_k/T)$, using Eq.~\eqref{eq:density_matrix_RHN_momentum_ansatz} and expanding in small chemical potentials --- involve $R_N$ and $\mu$. We can then collect terms with the same factors of $R_N$ and $\mu$ resulting in the QKEs~\eqref{eq:QKE_LG_1}-\eqref{eq:QKE_LG_3}, where
\begin{equation}
\begin{array}{cccccccc}
    \expval{\gamma_N^{(0),XY}} &=& &\expval{\gamma_{N,A}^{XY}}& + &\expval{\gamma_{N,B}^{XY}}& + &\expval{\gamma_{N,C}^{XY}} \ ,  \\
    \expval{\gamma_N^{(1),XY}} &=& &\expval{\gamma_{N,A}^{XY}}& + &\expval{\gamma_{N,B}^{XY}}&  & \phantom{\expval{\gamma_{N,C}^{XY}}}\ ,  \\
    \expval{\gamma_N^{(3),XY}} &=& & & & & - &\expval{\gamma_{N,C}^{XY}} \ .
\end{array}
\end{equation}
Explicitly, we find the following rates
\begin{equation}\label{eq:explicit_rates}
\begin{array}{cccccccc}
    \expval{\gamma_{N,A}^{FF}}& = &\expval{\gamma_{N,B}^{FF}}& = &\expval{\gamma_{N,C}^{FF}}& = &\frac{N_D N_C h_t^2}{128\pi^3} T\ , \\
    \expval{\gamma_{N,A}^{GG}}& = &\expval{\gamma_{N,B}^{GG}}& & & = &\frac{N_D N_C }{2\pi^3} \frac{T^5}{\Lambda^4} \ , \\
    & && &\expval{\gamma_{N,C}^{GG}}& = &\frac{3 N_D N_C}{2\pi^3} \frac{T^5}{\Lambda^4} \ , \\
    \expval{\gamma_{N,A}^{FG}}& = &\expval{\gamma_{N,B}^{FG}}& & & = &-\frac{N_D N_C h_t}{32\pi^3} \frac{T^3}{\Lambda^2} \ , \\
    & && &\expval{\gamma_{N,C}^{FG}}& = &\frac{N_D N_C h_t}{16\pi^3} \frac{T^3}{\Lambda^2} \ ,
\end{array}
\end{equation}
with $\mathrm{SU(2)}_L$ ($\mathrm{SU(3)}_c$) factor $N_D = 2$ ($N_C = 3$), the top Yukawa coupling $h_t \simeq 1$ and $\expval{\gamma_{N,i}^{GF}} = \expval{\gamma_{N,i}^{FG}}$.

For the choice $G = F/\sqrt{\Tr(F^{\dagger}F)}$ the flavor structure can be factorized, e.g. 
\begin{equation}\label{eq:individual_contribution_to_rate}
    \expval{\Gamma^{d}_{N,A}} = \expval{\Gamma_{N,A}^{FF}} + \expval{\Gamma_{N,A}^{GG}} + 2\expval{\Gamma_{N,A}^{FG}} = \left(\expval{\Tilde{\gamma}_{N,A}^{FF}} + \expval{\Tilde{\gamma}_{N,A}^{GG}} + 2\expval{\Tilde{\gamma}_{N,A}^{FG}} \right)[G^{\dagger}G] 
\end{equation}
with
\begin{align}
    \expval{\Tilde{\gamma}_{N,A}^{FF}} &= \expval{\gamma_{N,A}^{FF}} \Tr(F^{\dagger}F) \ , \nonumber \\
    \expval{\Tilde{\gamma}_{N,A}^{FG}} &= \expval{\gamma_{N,A}^{FG}} \sqrt{\Tr(F^{\dagger}F)}  \ , \\
    \expval{\Tilde{\gamma}_{N,A}^{GG}} &= \expval{\gamma_{N,A}^{GG}} \ , \nonumber 
\end{align}
and similar definitions for the processes $(B)$ and $(C)$. The rates $\expval{\Tilde{\gamma}_{N,i}^{XY}}$ determine then the relative contribution of each term to the total rate. The rate $\expval{\Tilde{\gamma}_{N,i}^{FF}}$ is suppressed by the small Yukawa coupling $F$, the rate $\expval{\Tilde{\gamma}_{N,i}^{GG}}$ by $\Lambda > T$ and the rate $\expval{\Tilde{\gamma}_{N,i}^{FG}}$ by a combination of the two.

\section{Interference Terms}\label{sec:interference}

In the following, we study the interference terms between the dim-4 interactions and dim-6 operator contributions to the rates in the QKEs~\eqref{eq:QKE_LG_1}-\eqref{eq:QKE_LG_3} of LG.

\begin{figure}[t]
\begin{subfigure}[b]{0.5\textwidth}
    \includegraphics[width=\textwidth]{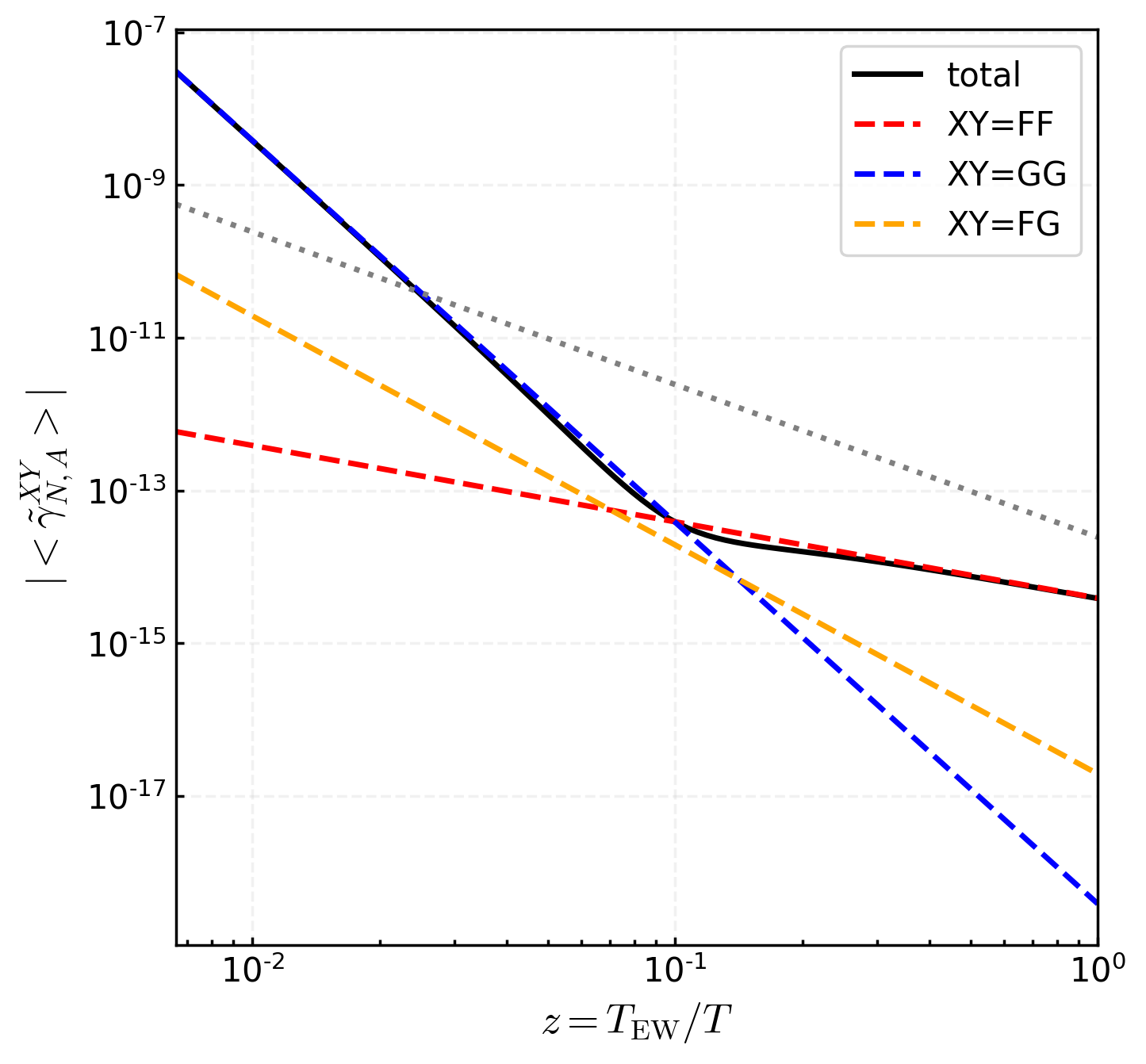}
\end{subfigure}
\hfill
\begin{subfigure}[b]{0.5\textwidth}
    \includegraphics[width=\textwidth]{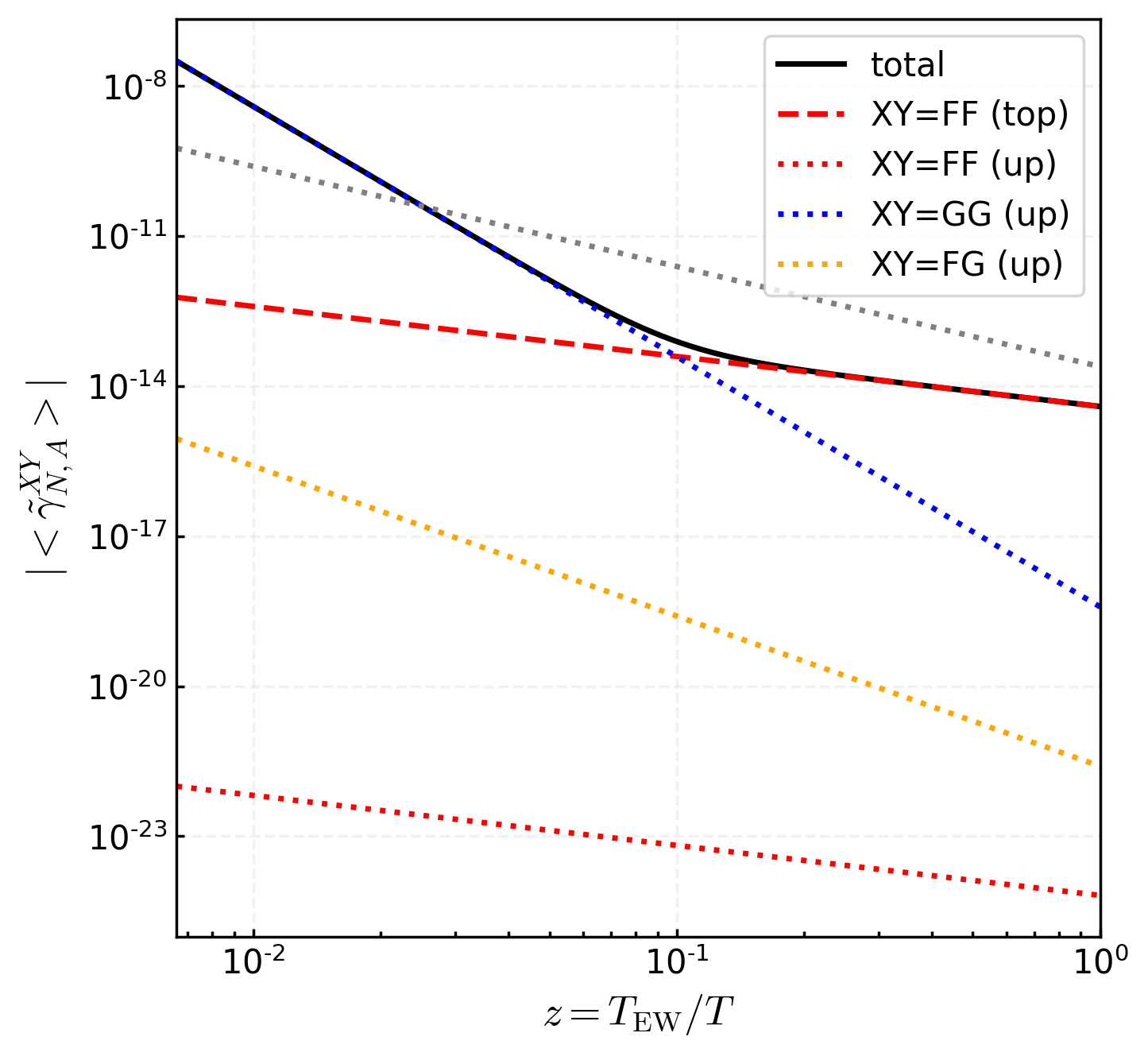}
\end{subfigure}
\caption{Individual rates as defined in Eq.~\eqref{eq:individual_contribution_to_rate} as a function of the timescale $z=T_{\mathrm{EW}}/T$ shown in color and the total rate in Eq.~\eqref{eq:individual_contribution_to_rate} shown in black for BP1 and $\Lambda = \SI{1e7}{\giga\electronvolt}$. We additionally show the Hubble rate $H=T^2/M_P^{*}$ as gray-dotted line. Left (right) corresponds to the case in which the dim-6 operator quarks are assumed to be only top-quarks (up-quarks).}
\label{fig:interference_example}
\end{figure}

First, note that the interference term for process $(A)$ and $(B)$ is negative, but we can show that the total rate of each particle is always positive. Since the rates for $(A)$ and $(B)$ are the same it suffices to show this for the process $(A)$. With the choice $G = F/\sqrt{\Tr(F^{\dagger}F)}$ one arrives, for the diagonal elements with $J=I\in\{1,2\}$, at
\begin{align}
    \expval{\Gamma_{N,A}}_{II} &= \expval{\Gamma_{N,A}^{FF}}_{II} + \expval{\Gamma_{N,A}^{GG}}_{II} + 2\expval{\Gamma_{N,A}^{FG}}_{II} \nonumber \\
    &= \left(\expval{\Tilde{\gamma}_{N,A}^{FF}} + \expval{\Tilde{\gamma}_{N,A}^{GG}} + 2\expval{\Tilde{\gamma}_{N,A}^{FG}} \right)[G^{\dagger}G]_{II} \nonumber \\
    &= \left[\frac{N_D N_C h_t^2}{128\pi^3} T \Tr(F^{\dagger}F) + \frac{N_D N_C }{2\pi^3} \frac{T^5}{\Lambda^4} -2\frac{N_D N_C h_t}{32\pi^3} \frac{T^3}{\Lambda^2} \sqrt{\Tr(F^{\dagger}F)} \right] [G^{\dagger}G]_{II} \nonumber \\
    &= \frac{N_D N_C T}{2\pi^3} \left[\frac{h_t^2}{64} \Tr(F^{\dagger}F) + \frac{T^4}{\Lambda^4} -2\frac{h_t}{16} \frac{T^2}{\Lambda^2} \sqrt{\Tr(F^{\dagger}F)} \right] [G^{\dagger}G]_{II}  \nonumber \\
    &= \frac{N_D N_C T}{2\pi^3} \left(\left[\frac{h_t}{8}\sqrt{\Tr(F^{\dagger}F)} - \frac{1}{2}\frac{T^2}{\Lambda^2}\right]^2 + \frac{3}{4}\frac{T^4}{\Lambda^4}\right) [G^{\dagger}G]_{II}  \nonumber \\
    &\geq 0 \ .
\end{align}
Moreover, the interference term is always subdominant in the total rate. This can be seen in the left panel of Fig.~\ref{fig:interference_example}, where we show the total rate and the individual rates given in Eq.~\eqref{eq:individual_contribution_to_rate} for the process $(A)$ for BP1 and $\Lambda = \SI{1e7}{\giga\electronvolt}$. 
The rate of the dim-6 operator $\expval{\Tilde{\gamma}_{N,A}^{GG}}$ is UV dominated and thus the largest rate for small $z$. The rate of the dim-4 interactions $\expval{\Tilde{\gamma}_{N,A}^{FF}}$, however, is IR dominated and thus the largest rate for large $z$. In between the interference term $\expval{\Tilde{\gamma}_{N,A}^{FG}}$ could be relevant, but from the left panel of Fig.~\ref{fig:interference_example} we see that the rate is always smaller than the total rate for the chosen BP and value of $\Lambda$ in this BP. The same is true for all BPs and all values of $\Lambda$ considered in this work. We can quantify this behaviour by defining
\begin{equation}
   r(z) = \left[ \frac{\abs{\expval{\Tilde{\gamma}_{N,A}^{FG}}}}{\expval{\Tilde{\gamma}_{N,A}^{FF}} + \expval{\Tilde{\gamma}_{N,A}^{GG}} + 2\expval{\Tilde{\gamma}_{N,A}^{FG}} } \right] \ .
\end{equation}
For BP1 and $\Lambda = \SI{1e7}{\giga\electronvolt}$, shown in the left panel of Fig.~\ref{fig:interference_example}, one finds $r(z)\ll1$ for most of the time evolution. For all BPs and $\Lambda$ considered in this work, we have explicitly checked that this holds, implying that the interference terms are subdominant. The same conclusion applies to processes of type (B) and (C).
Furthermore, we have explicitly confirmed that solving the QKEs~\eqref{eq:QKE_LG_1}-\eqref{eq:QKE_LG_3} without the interference term leads to indistinguishable result with respect to those shown in Fig.~\ref{fig:final}.

The above considerations are limited to the top-quark scattering used in the LG calculations in this work. In practice there will also be a contribution of the up-quark to the LG rate. While for the dim-4 contribution this effect is negligible as $h_u \simeq \num{1.3e-5}$ compared to ${h_t} \simeq \num{1}$, this is no longer true for the dim-6 operator, which can have a different flavor structure for the quarks independent of the Yukawa couplings in the dim-4 interactions. In the right panel of Fig.~\ref{fig:interference_example} we show the LG rates given in Eq.~\eqref{eq:individual_contribution_to_rate} for the process $(A)$ for BP1 and $\Lambda = \SI{1e7}{\giga\electronvolt}$ when considering up-quarks instead of top-quarks in the dim-6 operator. Note that the top-quarks are present in the thermal plasma such that the dim-4 quark scatterings between them and the RHNs cannot be turned off (red dashed line in Fig.~\ref{fig:interference_example}). Moreover, here we assumed that the coupling of the dim-6 operator to the quark flavor remains the same in both panels of Fig.~\ref{fig:interference_example}, such that the blue-dashed line in the left panel and the blue-dotted line in the right panel are identical. In the right panel of Fig.~\ref{fig:interference_example} it still remains true, that the interference term is always subdominant compared to the dim-4 and dim-6 interactions of the same flavor. Furthermore, the dim-4 top quark rate exceeds the up-quark interference term as well as the up-quark dim-4 rate such that, effectively, the total rate is given by the sum of the dim-4 top-quark rate and the dim-6 up-quark rate.

The total rate is the same in the left and right panels of Fig.~\ref{fig:interference_example} since the dim-6 operator rate in Eq.~\eqref{eq:explicit_rates} does not depend on the quark Yukawa coupling. Thus our results can easily be mapped to an arbitrary composition of quark flavor with the same total rate. Explicitly, the dim-6 operator coupling we use in this work is the sum over all up-type quark flavors $u,c,t$
\begin{equation}
   \frac{G^{\dagger}G}{\Lambda^4} = \frac{1}{\Lambda_0^4} \left([G^{\dagger} G]_{uu} + [G^{\dagger} G]_{cc} + [G^{\dagger} G]_{tt}\right) \ .
\end{equation}
For example in a flavor democratic assignment ($[G^{\dagger} G]_{uu}=[G^{\dagger} G]_{cc}=[G^{\dagger} G]_{tt}$) the operator scales are related by $\Lambda_0 = \sqrt[4]{3}\Lambda$.

\section{A more detailed Look at BP5}\label{sec:timescales_BP5}

\begin{figure}[t]
\begin{subfigure}[b]{0.5\textwidth}
    \includegraphics[width=\textwidth]{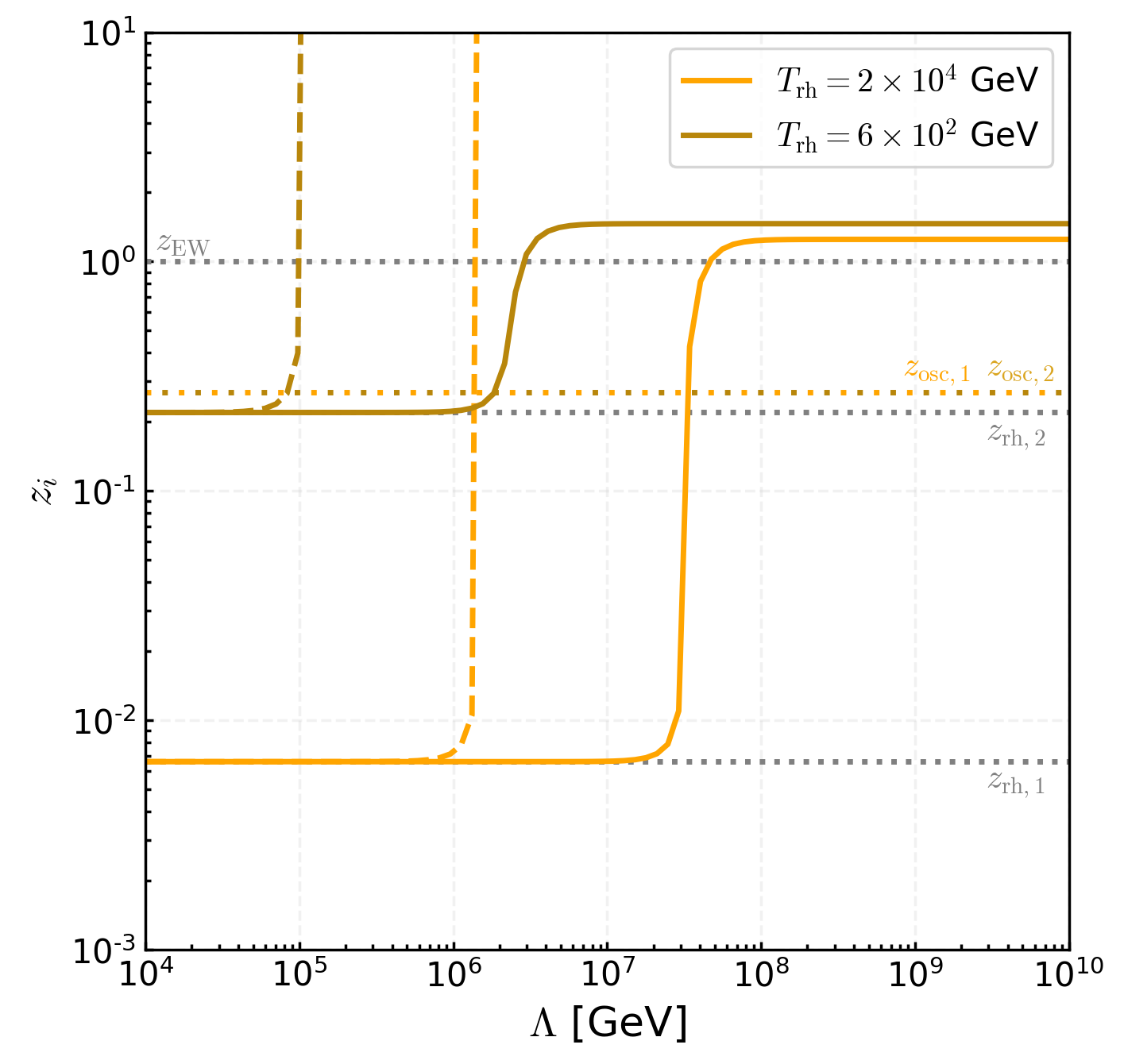}
\end{subfigure}
\hfill
\begin{subfigure}[b]{0.5\textwidth}
    \includegraphics[width=\textwidth]{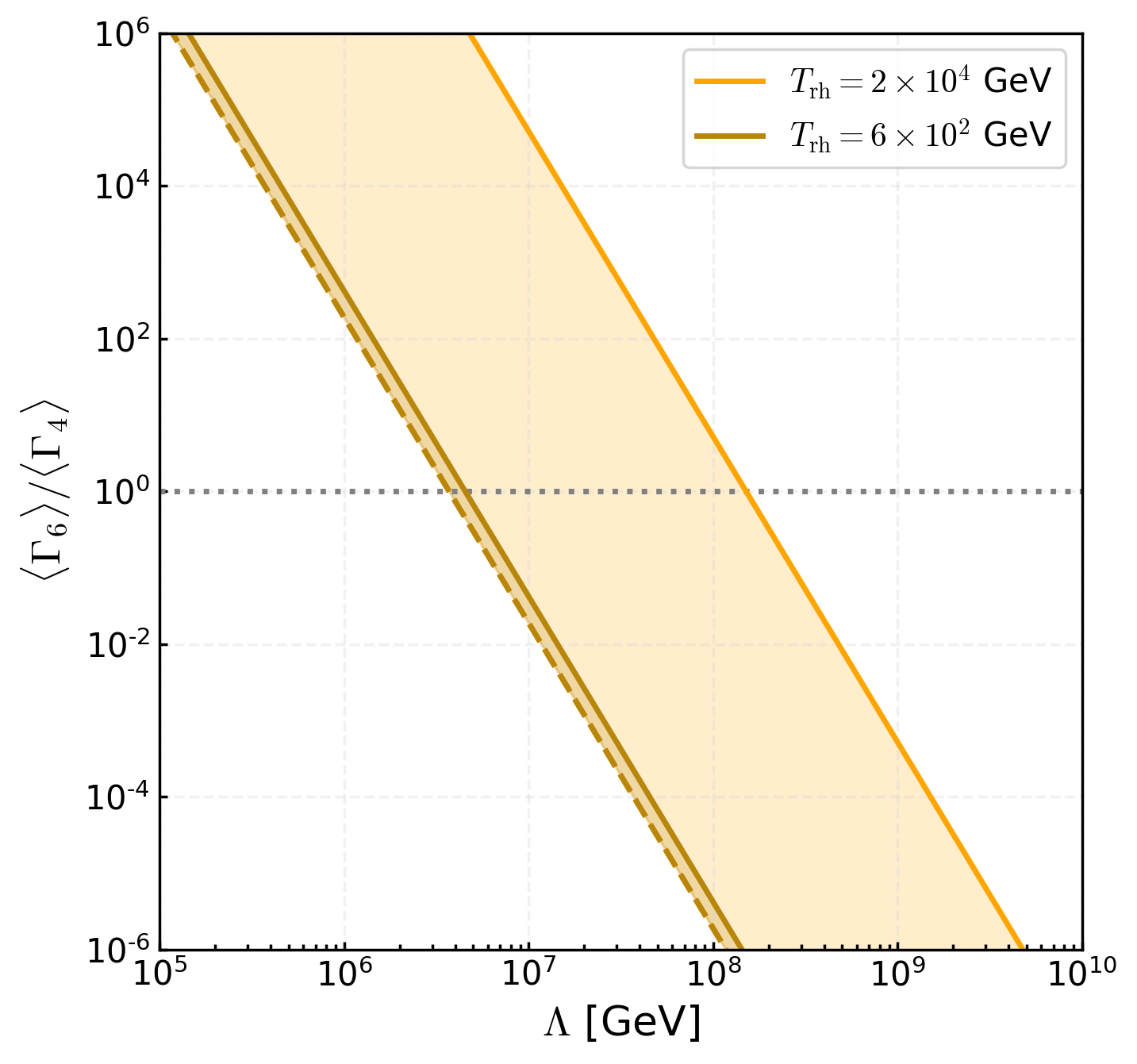}
\end{subfigure}
\caption{Left: Various timescales $z_i$ as a function of the dim-6 operator scale $\Lambda$ of the operator $\mathcal{O}_6$ in Eq.~\eqref{eq:our_dim6} with $G= F / \sqrt{\Tr(F^{\dagger}F)}$ for different reheating temperatures for BP5 given in Fig.~\ref{fig:BPs}. Solid lines are the overall equilibration timescales $z_{\eq}$ and dashed lines the equilibration timescales $z_{\eq,\mathrm{slow}}$ for the slow mode. In colored-dotted lines we show the corresponding oscillation timescales $z_{\mathrm{osc}, \, 1}$ ($z_{\mathrm{osc}, \, 2}$) and in gray-dotted the reheating timescale $z_{\mathrm{rh}, \, 1}$ ($z_{\mathrm{rh},\, 2}$) for $T_{\mathrm{rh}} = \SI{2e4}{\giga\electronvolt}$ ($T_{\mathrm{rh}} = \SI{6e2}{\giga\electronvolt}$) and the EW timescale $z_{\rm EW}$. Right: Ratios $\expval{\Gamma_6}/\expval{\Gamma_4}$ as a function of the dim-6 operator scale $\Lambda$ of the operator $\mathcal{O}_6$ in Eq.~\eqref{eq:our_dim6} with $G= F / \sqrt{\Tr(F^{\dagger}F)}$ for $z$ values between $z_{\mathrm{rh}}$ (solid) and $z_{\mathrm{osc}}$ (dashed). The gray dotted line corresponds to $\expval{\Gamma_6}/\expval{\Gamma_4}=1$.}
\label{fig:timescales_BP5}
\end{figure}

In this section we explain the features in the right panel of Fig.~\ref{fig:final_BP5}. For this we show in Fig.~\ref{fig:timescales_BP5} the timescales described in Section~\ref{sec:scales} for BP5. The same conclusions as in Section~\ref{sec:scales} and Section~\ref{sec:numerical_solution} apply. When $\expval{\Gamma_6}/\expval{\Gamma_4} = 1$, as shown in the right panel of Fig.~\ref{fig:timescales_BP5}, the BAU starts to increase. As seen in the left panel of Fig.~\ref{fig:timescales_BP5}, when $z_{\rm eq}$ becomes less than $z_{\rm osc}$, the system switches from the oscillatory regime into the overdamped regime and when $z_{\rm eq, slow}$ becomes less than $z_{\rm osc}$ the BAU becomes exponentially suppressed. In this way BP5 is very similar to BP2.

However, Fig.~\ref{fig:timescales_BP5} shows an additional feature at $\Lambda \approx\SI{3e6}{\giga\electronvolt}$ for $T_{\rm rh} = \SI{2e4}{\giga\electronvolt}$ and $\Lambda \approx\SI{1e6}{\giga\electronvolt}$ for $T_{\rm rh} = \SI{6e2}{\giga\electronvolt}$, where the BAU changes sign. Note that the BAU is proportional to $Y_{\Delta B} \propto \sum_{\alpha} \mu_{\alpha}$. Some of the $\mu_{\alpha}$ are positive and some negative and which one dominates the sum depends on the flavor dependent rates $\Gamma_4$ and $\Gamma_6$. Due to the flavor dependent rate $\Gamma_6$, it also depends on the scale $\Lambda$, whether the positive or negative values of $\mu_{\alpha}$ dominate the sum. In principle, the value of $\Lambda$ for which this happens is independent of the scales in Fig.~\ref{fig:timescales_BP5} and lies for BP5 somewhere in between, as can be seen in Fig.~\ref{fig:final_BP5}. For BP2 the sign change roughly coincides with the change to the overdamped regime as can be seen in Fig.~\ref{fig:final}.

\clearpage

\bibliographystyle{./JHEP}
\bibliography{ref_v2}

\end{document}